\documentclass[12pt]{iopart}
\usepackage{iopams}
\usepackage{rotating}
\usepackage{epsfig}
\bibstyle{plain}
\newcommand{\be}{\begin{equation}}
\newcommand{\ee}{\end{equation}}
\newcommand{\bea}{\begin{eqnarray}}
\newcommand{\eea}{\end{eqnarray}}
\newcommand{\nn}{\nonumber\\ }

\newcommand{\pp}{p^{\prime} }

\newcommand{\ppi}{p^{\prime}_{1} }
\newcommand{\ppii}{p^{\prime}_{2} }
\newcommand{\talpha}{\tilde{\alpha}}
\newcommand{\tmu}{\tilde{\mu}}
\newcommand{\fag}{\hspace{.1cm}^{\ast}F}
\newcommand{\jg}{\hskip .05cm ^{\ast}\hskip -.07cm j}
\newcommand{\Jg}{\hskip .05cm ^{\ast}\hskip -.07cm J}
\newcommand{\Gg}{\hspace{.1cm}^{\ast}G}
\newcommand{\dslash}{\gamma\partial}

\newcommand{\Aslash}{\gamma A}
\newcommand{\Bslash}{\gamma B}
\newcommand{\bAslash}{\gamma \bar{A}}
\newcommand{\bBslash}{\gamma \bar{B}}
\newcommand{\dx}{(\rmd x)}
\newcommand{\dxp}{(\rmd x^{\prime})}
\newcommand{\dxpp}{(\rmd x^{\prime\prime})}

\newcommand{\dy}{(\rmd y)}

\newcommand{\dzi}{(\rmd z_1)}
\newcommand{\dzii}{(\rmd z_2)}
\newcommand{\dxi}{(\rmd x_1)}
\newcommand{\dxii}{(\rmd x_2)}
\newcommand{\dyi}{(\rmd y_1)}
\newcommand{\dyii}{(\rmd y_2)}
\newcommand{\cq}{{{\mathsf{q}}_1\cdot{\mathsf{q}}_2}}
\newcommand{\wq}{{{\mathsf{q}}_1{\times}{\mathsf{q}}_2}}

\newcommand{\xp}{x^\prime}
\newcommand{\xpp}{x^{\prime\prime}}

\newcommand{\A}{{\cal A}}

\newcommand{\bA}{{\bar{A}}}

\newcommand{\tD}{{\tilde{D}}}
\newcommand{\tA}{\tilde{A}}
\newcommand{\tB}{\tilde{B}}
\newcommand{\D}{{\cal D}}
\newcommand{\N}{{\cal N}}
\newcommand{\J}{{\cal J}}
\newcommand{\K}{{\cal K}}

\newfont{\fib}{cmfi10 at 10pt}
\newcommand{\ps}{\mbox{\fib{s}}}
\newcommand{\pat}{\mbox{\fib{t}}}

\eqnobysec

\begin{document}
\review[Magnetic Monopoles]{Theoretical and experimental status
of  magnetic monopoles}
\author{Kimball A Milton\footnote{Permanent address:
Oklahoma Center for High Energy Physics and Homer L. Dodge
Department of Physics and Astronomy, The University of
Oklahoma, Norman, OK 73019 USA}}
\address{Department of Physics, Washington University, St.~Louis,
MO 63130 USA}\ead{milton@nhn.ou.edu}

\date\today

\begin{abstract}
The Tevatron has inspired new interest in the subject of magnetic monopoles.
First there was the 1998 D0 limit on the virtual production of monopoles,
based on the theory of Ginzburg and collaborators. In 2000 and 2004
results from an experiment {(Fermilab E882)} searching 
for real magnetically
charged particles bound to elements from the CDF and D0 detectors were
reported. The strongest
direct experimental limits, from the CDF collaboration,
have been reported in 2005. Less strong, but complementary, limits from the
H1 collaboration at HERA were reported in the same year. Interpretation of 
these experiments also require new developments in theory. 
Earlier experimental and observational constraints on 
point-like (Dirac) and non-Abelian monopoles were given from the
1970s through the 1990s, with occasional short-lived positive evidence
for such exotic particles reported. The status
of the experimental limits on monopole masses will be reported,
as well as the limitation of the theory of magnetic charge at present.
\end{abstract}
\pacs{14.80.Hv, 12.20.-m, 11.15.Kc, 11.80.Fv}
\maketitle

\section{Introduction}
\label{sec:intro}
The origin of the concept of magnetic charge, if not the name, goes back to
antiquity.  Certain stones in Magnesia, in Anatolia (Asia Minor), 
were found to exhibit
an attractive force on iron particles, and thus was discovered magnetism.
(Actually there is an ancient confusion about the origin of the name,
for it may refer to Magnesia, a prefecture in Thessaly, Greece, from
whence came the settlers (``Magnets'') of the city 
(or the ruins of another city) which is now in Turkey. 
For a recent discussion on the etymology see \cite{online}.)
Electricity likewise was apparent to the ancients, but without any evident
connection to magnetism.   Franklin eventually posited that there were two
kinds of electricity, positive and negative poles or charge; there were 
likewise two types of magnetism, north and south poles, but experience showed
that those poles were necessarily always associated in pairs.  Cutting a 
magnet, a dipole,  in two did not isolate a single pole, but resulted in
two dipoles with parallel orientation;  
the north and south poles so created were
bound to the opposite poles already existing \cite{peregrinus}.  
This eventually was formalized
in {\em Amp\`ere's hypothesis} (1820): Magnetism has its source in the 
motion of
electric charge.  That is, there are no intrinsic magnetic poles, but rather
magnetic dipoles are created by circulating electrical currents, 
macroscopically or at the atomic level.

Evidently, the latter realization built upon the emerging recognition of the
connection between electricity and magnetism.  Some notable landmarks along
the way were Oersted's discovery (1819) that an electrical current produced 
magnetic forces in its vicinity; Faraday's visualization of lines of force
as a physical picture of electric and magnetic fields; his discovery that
a changing magnetic field produces a electric field (Faraday's law
of magnetic induction, 1831); and Maxwell's crowning achievement 
in recognizing that a changing electric field must 
produce a magnetic field, which permitted him to write his equations
describing electromagnetism (1873).  
The latter accomplishment, built on the work of
many others, was the most important development in the 19th Century.
It is most remarkable that Maxwell's equations, written down in a less than
succinct form in 1873, have withstood the revolutions of the 20th Century,
relativity and quantum mechanics, and they still hold forth unchanged 
as the governing field equations of quantum electrodynamics, by far the
most successful physical theory ever discovered.

The symmetry of Maxwell's equations was spoiled, however, by the absence
of magnetic charge, and it was obvious to many, including
Poincar\'e \cite{poincare:1896} and Thomson 
\cite{thomson:1904,thomson:1904a}, 
the discoverer of the electron, that the concept of 
magnetic charge had utility, and
its introduction into the theory results in significant simplifications.
(Faraday \cite{faraday}
had already demonstrated the heuristic value of magnetic charge.)
But at that time, the consensus was clearly that magnetic charge had no
independent reality, 
and its introduction into the theory was for computational convenience
only \cite{heaviside}, although Pierre Curie \cite{curie} did suggest
that free magnetic poles might exist.
It was only well after the birth of quantum mechanics that a
serious proposal was made by Dirac \cite{Dirac:1931kp} that particles
carrying magnetic charge, or magnetic monopoles, should exist.
This was based on his observation that the phase unobservability in
quantum mechanics permits singularities manifested as sources of magnetic 
fields, just as point electric monopoles are sources of electric fields.
This was only possible if the product of electric and magnetic charges was
quantized.
This prediction was an example of what Gell-Mann would later call the
``totalitarian principle'' -- that anything which is not forbidden is
compulsory \cite{gm-tot}.  Dirac eventually became disillusioned with the 
lack of experimental evidence for magnetic charge, but Schwinger,
who became enamored of the subject around 1965, never gave up hope.
This, in spite of his failure to construct a computationally useful field 
theory of magnetically charged monopoles, or more generally particles carrying
both electric and magnetic charge, which he dubbed dyons (for his
musing on the naming of such hypothetical particles, see 
\cite{Schwinger:1969ib}).  Schwinger's failure to construct a manifestly
consistent theory caused many, including Sidney Coleman, to suspect that
magnetic charge could not exist. The subject of magnetic charge really took
off with the discovery of extended classical monopole solutions of
non-Abelian gauge theories by Wu and Yang, 't Hooft, Polyakov, Nambu, and 
others \cite{wuyang,'tHooft:1974qc,Polyakov:1974ek,nambu,Julia:1975ff,%
Wu:1975vq}.  With the advent of grand unified theories, this implied that
monopoles should have been produced in the early universe, and therefore
should be present in cosmic rays.  (The history of magnetic monopoles up
to 1990 is succinctly summarized with extensive
references in the Resource Letter of Goldhaber and Trower \cite{rl}.)

So starting in the late 1960s there was a burst of activity both in
trying to develop the theory of magnetically charged particles and in
attempting to find their signature either in the laboratory or in the cosmos.
As we will detail, the former development was only partially successful,
while no evidence at all of magnetic monopoles has survived.  Nevertheless,
the last few years, with many years of running of the Tevatron, and on
the eve  of the opening of the LHC, have witnessed new interest in the
subject, and new limits on monopole masses have emerged.  However,
the mass ranges where monopoles might most likely be found are yet
well beyond the reach of earth-bound laboratories, while cosmological
limits depend on monopole fluxes, which are subject to large uncertainties.
It is the purpose of this review to summarize the state of knowledge at
the present moment on the subject of magnetic charge, with the hope of
focusing attention on the unsettled issues with the aim of laying the 
groundwork for the eventual discovery of this exciting new state of matter.

A word about my own interest in this subject.  I was a student of Julian
Schwinger, and co-authored an important paper on the subject with him
in the 1970s \cite{Schwinger:1976fr}.  Many years later my colleague
in Oklahoma, George Kalbfleisch, asked me to join him in a new experiment
to set limits on monopole masses  based on Fermilab experiments
\cite{Kalbfleisch:2000iz,Kalbfleisch:2003yt}.  His interest grew out of
that of his mentor Luis Alvarez, who had set one of the best earlier limits on
low-mass monopoles \cite{Alvarez:1963zp,Alvarez:1970zu,Eberhard:1971re,%
Ross:1973it,Eberhard:1975en}.
Thus, I believe I possess the bona fides to present this review.

Finally, I offer a guide to the reading of this review.  Since the issues
are technical, encompassing both theory and experiment, not all parts of this
review will be equally interesting or relevant to all readers.  I have
organized the review so that the main material is contained in sections and
subsections, while the third level, subsubsections, contains material which is
more technical, and may be omitted without loss of continuity at a first 
reading.
Thus in \sref{sec:qt}, sections \ref{yang}--\ref{app} constitute a detailed
proof of the quantization condition, while \sref{sec:nrelham} describes
the quantum mechanical cross section.
 
In this review we use Gaussian units, so, for example, the fine-structure
constant is $\alpha=e^2/\hbar c$.  We will usually, particularly in
field theoretic contexts, choose natural units where $\hbar=c=1$.

\section{Classical theory}
\subsection{Dual symmetry}
The most obvious virtue of introducing magnetic charge is the
symmetry thereby imparted to Maxwell's equations in vacuum,
\begin{eqnarray}
\bnabla\cdot{\bf E}=4\pi\rho_e,\qquad \bnabla\cdot{\bf B}=4\pi\rho_m,
\nonumber\\
\bnabla\times{\bf B}={1\over c}{\partial\over\partial t}{\bf E}+
{4\pi\over c}{\bf j}_e,\qquad
-\bnabla\times{\bf E}={1\over c}{\partial\over\partial t}{\bf B}+
{4\pi\over c}{\bf j}_m.
\end{eqnarray}
Here $\rho_e$, ${\bf j}_e$ are the electric charge and current densities,
and $\rho_m$, ${\bf j}_m$ are the magnetic charge and current densities,
respectively.
These equations are invariant under a global  {\em duality \/}
transformation.  If $\cal E$ denotes any electric quantity, such as 
$\bf E$, $\rho_e$, or ${\bf j}_e$, while $\cal M$ denotes any magnetic
quantity, such as $\bf B$, $\rho_m$, or ${\bf j}_m$,
the dual Maxwell equations are invariant under
\numparts
\be
{\cal E}\to{\cal M},\quad {\cal M}\to-{\cal E},\label{duality}
\ee
or more generally
\begin{eqnarray}
{\cal E}\to{\cal E}\cos\theta+{\cal M}\sin\theta,\label{dualtransf0}\qquad
{\cal M}\to{\cal M}\cos\theta-{\cal E}\sin\theta,\label{dualtransf}
\end{eqnarray}
\endnumparts
where $\theta$ is a constant.

Exploitation of this dual symmetry is useful in practical calculations, even
if there is no such thing as magnetic charge.  For example, 
its appearance may
be used to facilitate an elementary derivation of the laws of energy and
momentum conservation in classical electrodynamics \cite{CE}.  A more 
elaborate
example is the use of fictitious magnetic currents in calculate diffraction
from apertures \cite{bethe44}. (See also \cite{er}.)

\subsection{Angular momentum}
\label{sec:angmom}
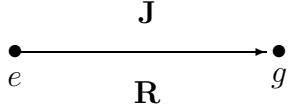
\begin{figure}
\centering
\begin{picture}(200,50)
\put(50,25){\makebox(0,0){$\bullet$}}
\put(50,15){\makebox(0,0){$e$}}
\put(150,15){\makebox(0,0){$g$}}
\put(150,25){\makebox(0,0){$\bullet$}}
\put(50,25){\vector(1,0){95}}
\put(100,10){\makebox(0,0){$\bf R$}}
\put(100,40){\makebox(0,0){$\bf J$}}
\end{picture}
\caption{Static configuration of an electric change and a magnetic 
monopole.}
\label{fig1}
\end{figure}

J. J. Thomson observed in 1904 
\cite{thomson:1904,thomson:1904a,thomson:1909,thomson:1937} 
the remarkable fact that a {\it static\/}
system of an electric ($e$) and a magnetic ($g$) charge
separated by a distance $\bf R$ possesses an angular momentum,
see \fref{fig1}.
The angular momentum is obtained by integrating the moment of the momentum
density of the static fields:
\begin{eqnarray}
{\bf J}=\int (\rmd{\bf r})\,{\bf r\times G}=\int (\rmd{\bf r})\,{\bf r}
\times\frac{{\bf E\times B}}{4\pi c}\nonumber\\
=\frac{1}{4\pi c}\int (\rmd{\bf r})\,{\bf r}\times\left[\frac{e{\bf r}}{r^3}
\times\frac{g({\bf r-R})}{({\bf r-R})^3}\right]
=\frac{eg}{c}{\bf\hat R},
\label{tang}
\end{eqnarray}
which follows from symmetry 
(the integral can only supply a numerical factor,
which turns out to be $4\pi$ \cite{CE}).  The quantization of charge 
follows by applying semiclassical quantization of angular
momentum:
\numparts
\begin{equation}
{\bf J\cdot\hat R}=\frac{eg}{c}=n\frac{\hbar}{2},
\qquad n=0,\,\pm1,\,\pm2,\,\dots,
\end{equation}
or
\be
eg=m'\hbar c,\qquad m'=\frac{n}2.
\ee\endnumparts
(Here, and in the following, we use $m'$ to designate this ``magnetic
quantum number.''  The prime will serve to distinguish this quantity from
an orbital angular momentum quantum number, or even from a particle mass.)

\subsection{Classical scattering}
Actually, earlier in 1896, Poincar\'e \cite{poincare:1896}
 investigated the motion of an electron
in the presence of a magnetic pole. This was inspired by a slightly
earlier report of anomalous motion of cathode rays in the presence
of a magnetized needle \cite{birkeland}.
 Let us generalize the analysis to two dyons
(a term coined by Schwinger in 1969 \cite{Schwinger:1969ib}) 
with charges $e_1$, $g_1$, and $e_2$, $g_2$, respectively.
There are two charge combinations
\be
q=e_1e_2+g_1g_2,\qquad
\kappa=-{e_1g_2-e_2g_1\over c}.\label{qandkappa}
\ee
Then the classical equation of relative motion is 
($\mu$ is the reduced mass
and $\bf v$ is the relative velocity)
\begin{equation}
\mu{\rmd^2\over \rmd t^2}{\bf r}
=q{{\bf r}\over r^3}-\kappa{\bf v}\times{{\bf r}
\over r^3}.
\end{equation}
The constants of the motion are the energy and the angular momentum,
\begin{eqnarray}
E={1\over2}\mu v^2+{q\over r},\qquad
{\bf J}={\bf r\times\mu v}+\kappa{\bf \hat r}.
\label{enandang}
\end{eqnarray}
Note that Thomson's angular momentum (\ref{tang}) is prefigured here.

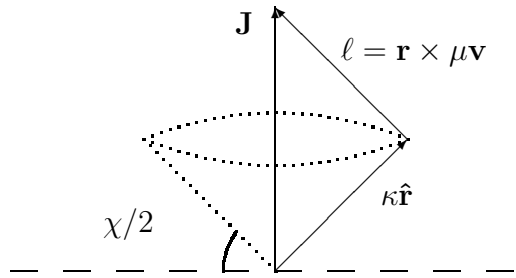
\begin{figure}
\centering
\begin{picture}(200,100)
\multiput(0,0)(20,0){10}{\line(1,0){10}}
\put(100,0){\vector(0,1){100}}
\put(85,90){$\bf J$}
\put(100,0){\vector(1,1){50}}
\put(140,25){$\kappa{\bf\hat r}$}
\put(125,80){$\bf \ell=r\times\mu v$}
\put(150,50){\vector(-1,1){50}}
\thicklines
\qbezier[20](100,0)(75,25)(50,50)
\qbezier[30](50,50)(100,70)(150,50)
\qbezier[30](50,50)(100,30)(150,50)
\put(35,15){$\chi/2$}
\qbezier(80,0)(80,7.5)(85,15)
\end{picture}
\vspace*{13pt}
\caption{The relative motion of two dyons is confined to the surface of
a cone about the direction of the angular momentum.}
\label{fig3}
\end{figure}

\begin{figure}[t]
\centering
\includegraphics*[height=10cm,angle=1,scale=1]{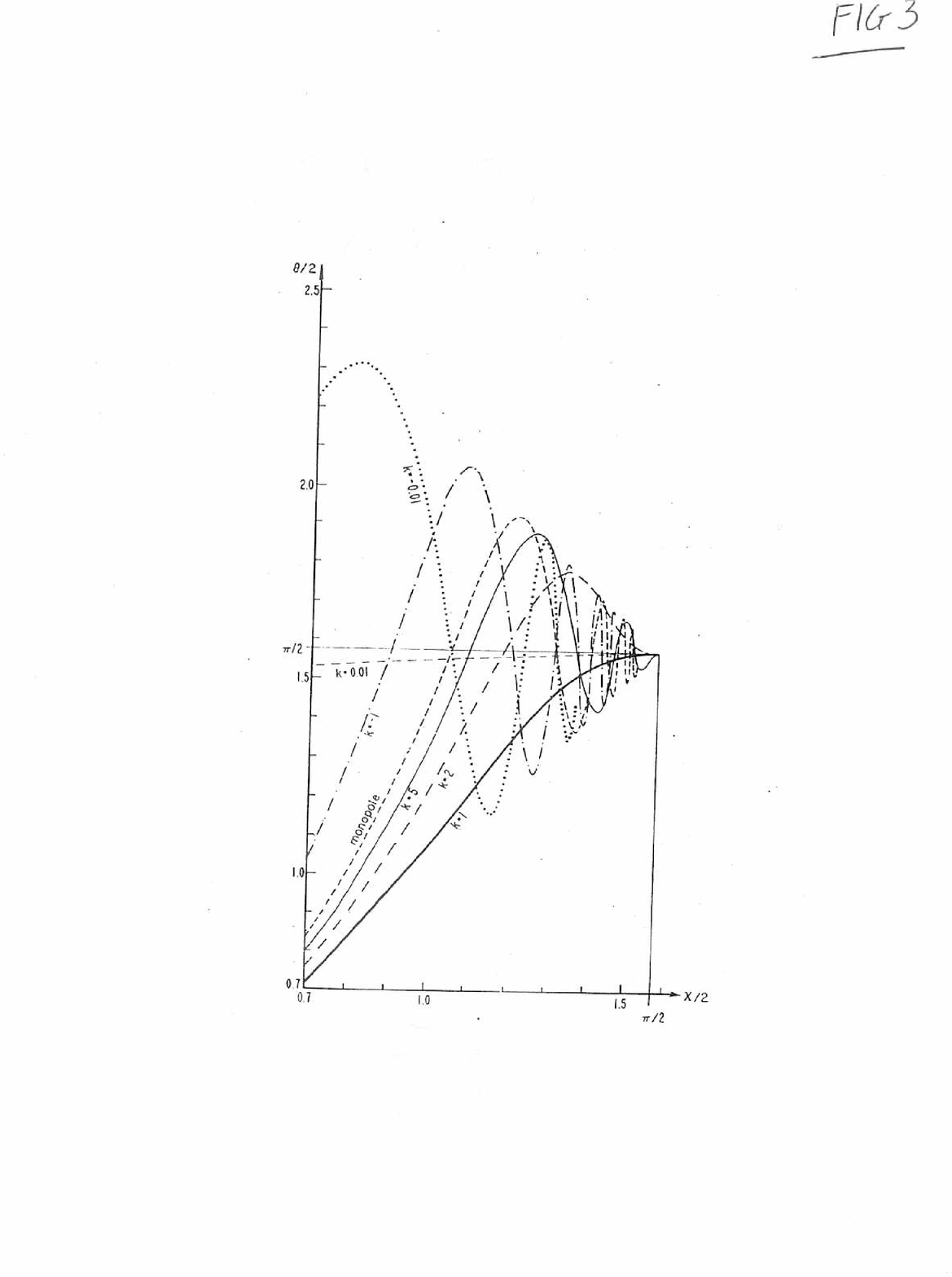}
\caption{Scattering angle $\theta$ as a function of the impact parameter
variable $\chi$.  Here $k\pi=|\kappa|v_0/q$.}  
\label{fig4}
\end{figure}

\begin{figure}
\centering
\includegraphics*[height=10cm, angle=01, scale=1.3]{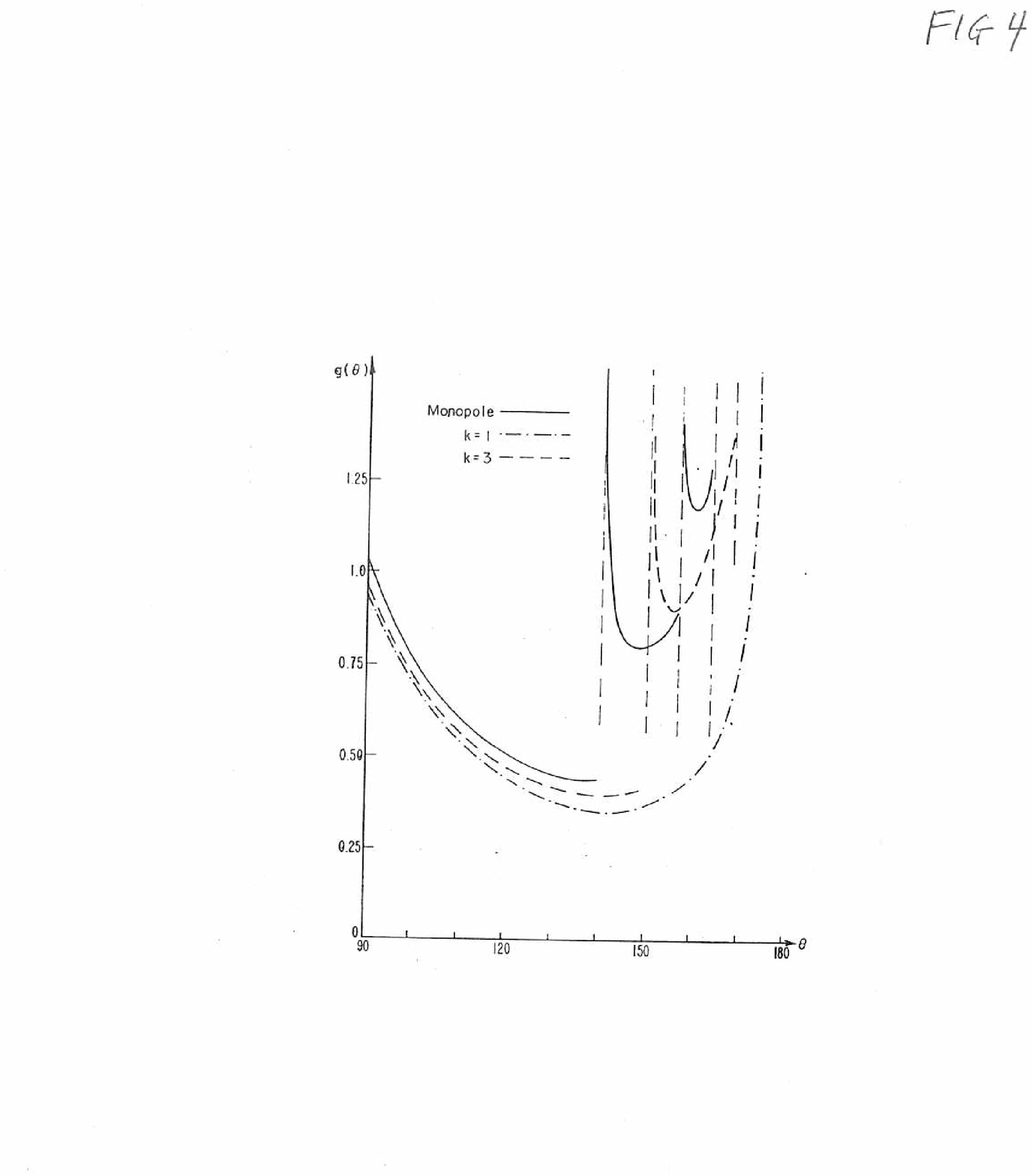}
\caption{Classical cross section for monopole-electron and 
dyon-dyon scattering. Again, $k\pi=|\kappa|v_0/q$, while $g(\theta)=
(\mu v_0/\kappa)^2(\rmd \sigma/\rmd\Omega)$.}
\label{fig5}
\end{figure}

Because $\bf J\cdot \hat r=\kappa$, the motion is confined to a cone,
as shown in \fref{fig3}.
Here the angle of the cone is given by
\begin{eqnarray}
\cot{\chi\over2}&=&{l\over|\kappa|},\qquad l=\mu v_0b,
\end{eqnarray}
where $v_0$ is the relative speed at infinity, 
and $b$ is the impact parameter. The scattering angle $\theta$ is given by
\numparts
\begin{eqnarray}\cos\frac{\theta}{2}=\cos\frac{\chi}{2}\left|\sin
\left(\frac{\xi/2}{\cos\chi/2}\right)\right|,
\end{eqnarray}
where
\begin{equation}
\frac{\xi}{2}=\left\{\begin{array}{cc}
\arctan\left(\frac{|\kappa|v_0}{q}\cot\frac{\chi}{2}\right),&q>0,\\
\pi-\arctan\left(\frac{|\kappa|v_0}{|q|}\cot\frac{\chi}{2}\right),&q<0.
\end{array}\right.
\end{equation}
\endnumparts
When $q=0$ (monopole-electron scattering), $\xi=\pi$.
The impact parameter $b(\theta)$ is a multiple-valued function of
$\theta$, as illustrated in \fref{fig4}.
The differential cross section is therefore
\begin{eqnarray}
\frac{\rmd\sigma}{\rmd\Omega}=\left|\frac{b\,\rmd b}
{\rmd(\cos\theta)}\right|
=\left(\frac{\kappa}{\mu v_0}\right)^2
\underbrace{\sum_\chi \frac{1}{4\sin^4
\frac{\chi}{2}}
\left|\frac{\sin\chi\,\rmd\chi}{\sin\theta\,
\rmd\theta}\right|}_{g(\theta)}.
\end{eqnarray}
Representative results are given in \cite{Schwinger:1976fr}, 
and reproduced here in \fref{fig5}.

The cross section becomes infinite in two circumstances; first, when
\begin{equation}
\sin\theta=0 \quad(\sin\chi\ne0), \qquad \theta=\pi,
\end{equation}
we have what is called a {\it glory}.
For monopole-electron scattering this occurs for 
\begin{equation}
{\chi_g\over2}=1.047,\,1.318,\,1.403,\,\dots.
\end{equation}
The other case in which the cross section diverges is when
\begin{equation}
{\rmd\theta\over \rmd\chi}=0.
\end{equation}
This is called a {\it rainbow}.
For monopole-electron scattering this occurs at
\begin{equation}
\theta_r=140.1^\circ,\,156.7^\circ,\,163.5^\circ,\,\dots.
\end{equation}

For small scattering angles we have the generalization of the
Rutherford formula
\begin{eqnarray}
\fl\frac{\rmd\sigma}{\rmd\Omega}=\frac{1}{(2\mu v_0)^2}\left\{\left(
\frac{e_1g_2-e_2g_1}c
\right)^2+\left(\frac{e_1e_2+g_1g_2}{v_0}\right)^2\right\}
\frac{1}{(\theta/2)^4},\qquad\theta\ll1.\label{ruth}
\end{eqnarray}
Note that for electron-monopole scattering, $e_1=e$, $e_2=0$,
$g_1=0$, $g_2=g$, this cross section differs from the 
Rutherford one for electron-electron scattering by the replacement
\be
\frac{e_2}{v}\to \frac{g}c.\label{velsup}
\ee  This is a universal feature
which we (and others) used in our experimental analyses.

\section{Quantum theory}
\label{sec:qt}
Dirac showed in 1931 \cite{Dirac:1931kp}
 that quantum mechanics was consistent with the existence
of magnetic monopoles provided the quantization condition holds,
\be
eg=m'\hbar c,\label{quant}
\ee
where $m'$ is an integer or an integer plus 1/2,
which explains the quantization of electric charge.
This was generalized by Schwinger to dyons:
\be
e_1g_2-e_2g_1=-m'\hbar c.\label{dyon}
\ee
(Schwinger sometimes argued \cite{Schwinger:1975ww}
that $m'$ was an integer, or perhaps an even integer.)
We will demonstrate these quantization conditions in the following.
Henceforth in this section we shall set $\hbar=c=1$.

\subsection{Vector potential}
One can see where charge
quantization  comes from by considering quantum mechanical scattering.
To define the Hamiltonian, one must introduce a vector potential, 
which must be singular because
\be
\bnabla\cdot\mathbf{B}\ne0\Rightarrow
{\bf B}\ne\bnabla\times{\bf A}.\label{novec}
\ee
For example, a potential singular along the entire line $\bf \hat n$ is
\begin{equation}
{\bf A(r)}=-\frac{g}{r}\frac{1}{2}\left(\frac{{\bf \hat n\times r}}{
r-{\bf \hat n\cdot r}}-\frac{{\bf \hat n\times r}}{
r+{\bf \hat n\cdot r}}\right)
=-\frac{g}{r}\cot\theta\mbox{\boldmath{$\hat\phi$}}\label{3.4}
\end{equation}
where the latter form applies if 
${\bf \hat n=\hat z}$,
which corresponds to the magnetic field produced by a magnetic monopole at
the origin,
\be
{\bf B(r)}=g\frac{{\bf r}}{r^3}.
\ee
In view of \eref{novec}, we can write
\numparts
\be
\mathbf{B(r)}=\bnabla\times\mathbf{A(r)}+g\mathbf{f(r)},
\label{string0a}
\ee
where
\be
\bnabla\cdot\mathbf{f(r)}=4\pi\delta(\mathbf{r}),
\label{string0b}
\ee
\endnumparts
in which $\mathbf{f}$ has support only along the line $\mathbf{\hat n}$
passing through the origin. 
The line of singularities is called the {\em string}, and $\mathbf{f}$
is called the string function.
Invariance of the theory (wavefunctions must be single-valued) under
string rotations implies the charge quantization condition (\ref{quant}).
  This is a nonperturbative statement, which is proved in \sref{sec:strings}.

\subsubsection{Yang's approach}
\label{yang}
Yang offered another approach, which is fundamentally equivalent
\cite{Wu:1976qk,Wu:1976ge,Kazama:1976sr,
Wu:1977qk,Yang:1977qv,Kazama:1977fm,Yang:1978td,Yang:1979tx}.
He insisted that there be no singularities, 
but rather different potentials in different but overlapping regions:
\numparts
\begin{eqnarray}
A^a_\phi={g\over r\sin\theta}(1-\cos\theta)
={g\over r}\tan{\theta\over2},\qquad\theta<\pi,\label{3.6a}\\
A^b_\phi=-{g\over r\sin\theta}(1+\cos\theta)
=-{g\over r}\cot{\theta\over2},\label{3.6b}
\qquad \theta>0.
\end{eqnarray}
\endnumparts
These correspond to the same magnetic field, so must differ by a gradient:
\be
A^a_\mu-A^b_\mu={2g\over r\sin\theta}\mbox{\boldmath{$\hat\phi$}}
=\partial_\mu\lambda,
\ee where $\lambda=2g\phi.$
Requiring now that $\rme^{\rmi e\lambda}$ be single valued leads to the
quantization condition,
$eg=m'$, $m'$ a half integer.

\subsubsection{Spin approach}
There is also a intrinsic spin formulation, pioneered by
Goldhaber  \cite{gold,Goldhaber:1977xw}.
The energy (\ref{enandang}),
\be
E={1\over2}\mu v^2+{q\over r},\qquad q=e_1e_2+g_1g_2,
\ee
differs by a gauge transformation from
\be
{\cal H}={1\over2\mu}\left(p_r^2+{J^2
-({\bf J\cdot\hat r})^2\over r^2}\right)
+{q\over r},
\ee
where
\numparts
\begin{eqnarray}
{\bf J}={\bf r\times p+S},\label{3.10a}
\\
\mu{\bf v}={\bf p}+{{\bf S\times r}\over r^2}.\label{3.10b}
\end{eqnarray}
\endnumparts
The quantization condition appears as
\be
{\bf S\cdot\hat r}=m'.\label{3.11}
\ee
The elaboration of this \cite{Milton:1976jq} is given in \sref{sec:spin}.

\subsubsection{Strings}
\label{sec:strings}
Let us now discuss in detail the nonrelativistic, quantum scattering of two
dyons, with electric and magnetic charges $e_1,g_1$ and $e_2,g_2$,
respectively.  The Hamiltonian for the system is
\be
\mathcal{H}=\frac12 m_1v_1^2+\frac12 m_2 v_2^2+\frac{q}{|\mathbf{
r}_1-\mathbf{r}_2|},
\ee
where, in terms of the canonical momenta, the velocities are given by
\numparts
\bea
m_1\mathbf{v}_1=\mathbf{p}_1-e_1\mathbf{A}_{e2}(\mathbf{r}_1,t)-g_1
\mathbf{A}_{m2}(\mathbf{r}_1,t),\\
m_2\mathbf{v}_2=\mathbf{p}_2-e_2\mathbf{A}_{e1}(\mathbf{r}_2,t)-g_2
\mathbf{A}_{m1}(\mathbf{r}_2,t).
\eea\endnumparts
The electric ($e$) and magnetic ($m$) vector potentials are
\numparts
\bea
4\pi\mathbf{A}_e(\mathbf{r},t)=4\pi\bnabla\lambda_e(\mathbf{r},t)
-\int(\rmd \mathbf{r}')\mathbf{f(r-r')}\times\mathbf{B}(\mathbf{r}',t),
\label{3.14a}\\
4\pi\mathbf{A}_m(\mathbf{r},t)=4\pi\bnabla\lambda_m(\mathbf{r},t)
+\int(\rmd \mathbf{r}'){}^*\mathbf{f(r-r')}\times\mathbf{E}(\mathbf{r}',t),
\label{3.14b}
\eea
\endnumparts
with
\numparts
\bea
\lambda_e(\mathbf{r},t)=\int(\rmd\mathbf{r}')\mathbf{f(r-r')}\cdot
\mathbf{A}_e(\mathbf{r}',t),\label{lambdae}\\
\lambda_m(\mathbf{r},t)=\int(\rmd\mathbf{r}'){}^*\mathbf{f(r-r')}\cdot
\mathbf{A}_m(\mathbf{r}',t).\label{lambdam}
\eea
\endnumparts
Here, the functions $\mathbf{f}$ and ${}^*\mathbf{f}$ represent the strings
and must satisfy
\be
\bnabla\cdot{}^{(*)}\mathbf{f(r-r')}=4\pi\delta(\mathbf{r-r'}).\label{2.6}
\ee
A priori, $\mathbf{f}$ and ${}^*\mathbf{f}$ need not be related, and could be
different for each source.  So, for the case of dyon-dyon scattering, it would
seem that four independent strings are possible.

The first condition we impose on the Schr\"odinger equation,
\be
\mathcal{H}\Psi=E\Psi,
\ee
is that it separates when center-of-mass and relative coordinates are
employed, which implies
\numparts
\bea
e_1\mathbf{A}_{e2}(\mathbf{r}_1,t)=-g_2 \mathbf{A}_{m1}(\mathbf{r}_2,t)
\equiv e_1g_2 \boldsymbol{\mathcal{A}}(\mathbf{r}),\\
e_2\mathbf{A}_{e1}(\mathbf{r}_2,t)=-g_1 \mathbf{A}_{m2}(\mathbf{r}_1,t)
\equiv e_2g_1 \boldsymbol{\mathcal{A}}'(\mathbf{r}),
\eea
\endnumparts
where $\mathbf{r}=\mathbf{r}_1-\mathbf{r}_2$.  Correspondingly, there are
relations between the various string functions,
\be
{}^*\mathbf{f}_1(\mathbf{x})=-\mathbf{f}_2(-\mathbf{x}),\qquad
{}^*\mathbf{f}_2(\mathbf{x})=-\mathbf{f}_1(-\mathbf{x}),
\ee
leaving only two independent ones.  The Hamiltonian for the relative
coordinates now reads
\be
\mathcal{H}=\frac1{2\mu}[\mathbf{p}
-e_1g_2 \boldsymbol{\mathcal{A}}(\mathbf{r})
+e_2g_1 \boldsymbol{\mathcal{A}}'(\mathbf{r})]^2+\frac{q}{r},
\label{nrham2pot}
\ee
where $\mu$ is the reduced mass.

If we further require that only one vector potential be present,
$\boldsymbol{\mathcal{A}}=\boldsymbol{\mathcal{A}}'$, so that only
the antisymmetric
combination of electric and magnetic charges occurring in (\ref{dyon})
appears, one more relation is obtained between the two $\mathbf{f}$ functions,
\be
\mathbf{f}_2(\mathbf{x})=-\mathbf{f}_1(-\mathbf{x}).\label{strcon1}
\ee
Notice that (\ref{strcon1}) possesses two types of solutions.
\begin{itemize}
\item There is a single string, necessarily infinite, satisfying
\be
\mathbf{f}(\mathbf{x})=-\mathbf{f}(-\mathbf{x}).\label{strcon2}
\ee
Then it is easily seen that the vector potential transforms the
same way as charges and currents do under duality transformations 
(\ref{dualtransf}).  This is the so-called {\em symmetric\/} case.
\item There are two strings, necessarily semi-infinite, which are negative
reflections of each other.  
\end{itemize}

If identical semi-infinite strings are employed,
so that $\boldsymbol{\mathcal{A}}\ne\boldsymbol{\mathcal{A}}'$, the 
individual charge products $e_1g_2$ and $e_2g_1$ occur in the
dynamics.  The singularities of $\boldsymbol{\mathcal{A}}$ and
$\boldsymbol{\mathcal{A}}'$ lie on lines parallel and antiparallel to the
strings, respectively. 
 We will see the consequences for the charge quantization condition of these
different choices in the following.

For now, we return to the general situation embodied in (\ref{nrham2pot}).
For simplicity, we choose the string associated with $\boldsymbol{\mathcal{A}}$
to be a straight line lying along the direction $\bf \hat n$,
\be
\boldsymbol{\mathcal{A}}=\cases{-\frac1r\frac{\mathbf{\hat n\times r}}
{r-(\mathbf{\hat n\cdot r})},&\mbox{semi-infinite}\\
-\frac1r\frac12\left(\frac{\mathbf{\hat n\times r}}{r-(\mathbf
{\hat n\cdot r})}-\frac{\mathbf{\hat n\times r}}{r+(\mathbf{\hat n\cdot r})}
\right),&\mbox{infinite}}.\label{potential}
\ee
This result is valid in the gauge in which $\lambda_{e(m)}$ [(\ref{lambdae}),
(\ref{lambdam})] is equal to zero.  Without loss of generality, we will take
$\boldsymbol{\mathcal{A}}'$ to be given by (\ref{potential}) with $\mathbf{\hat n
\to\hat z}$, which corresponds to taking the string associated with 
$\boldsymbol{\mathcal{A}}'$ to point along the $-z$ axis, $\mathbf{f}_1
\propto -\mathbf{\hat z}$.

We now wish to convert the resulting Hamiltonian, $\mathcal{H}$, into a form
$\mathcal{H}'$ in which all the singularities lie along the $z$ axis.  It
was in that case that the Schr\"odinger equation was solved in
\cite{Schwinger:1976fr}, as described in  \sref{sec:nrelham}, yielding
the quantization condition (\ref{dyon}).  This conversion is effected
by a unitary transformation \cite{zumino} (essentially a gauge transformation),
\be
\mathcal{H}'=\rme^{\rmi \Lambda}\mathcal{H}\rme^{-\rmi \Lambda}.
\label{2.14}
\ee
The differential equation determining $\Lambda$ is
\be
\bnabla \Lambda=e_1g_2[\boldsymbol{\mathcal{A}}'(\mathbf{r})-
\boldsymbol{\mathcal{A}}(\mathbf{r})].\label{diffeqlam}
\ee
We take $\mathbf{\hat n}$ to be given by
\be
\mathbf{\hat n}=\sin\chi\cos\psi\mathbf{\hat x}+\sin\chi\sin\psi\mathbf{\hat y}
+\cos\chi\mathbf{\hat z},\label{2.16}
\ee
and use spherical coordinates [$\mathbf{r}=(r,\theta,\phi)$], to find
\be
\Lambda=-e_1g_2\beta(\mathbf{\hat n,r}),\label{Lambdaasbeta}
\ee
where, for the semi-infinite string (Dirac)
\numparts
\be
\beta_D=\phi-\psi+(\cos\theta-\cos\chi)F_-(\theta,\phi-\psi,\chi)-2\pi
\eta(\chi-\eta),\label{betaD}
\ee
and for the infinite string (Schwinger)
\bea
\fl\beta_S=\frac12\left[(\cos\theta-\cos\chi)F_-(\theta,\phi-\psi,\chi)
+(\cos\theta+\cos\chi)F_+(\theta,\phi-\psi,\chi)-2\pi\eta(\chi-\theta)\right].
\nn \label{betaS}
\eea
\endnumparts
The functions occurring here are
\bea
F_\pm(\theta,\alpha,\chi)=\int_0^\alpha\frac{\rmd \phi'}{1\pm
\cos\chi\cos\theta\pm \sin\chi\sin\theta\cos\phi'}\nn
=\frac{2\epsilon(\alpha)}{|\cos\theta\pm\cos\chi|}\arctan
\left[\left(\frac{1\pm\cos(\chi+\theta)}{1\pm\cos(\chi-\theta)}\right)^{1/2}
\tan\frac{|\alpha|}2\right],
\eea
where the arctangent is not defined on the principal branch, but is chosen
such that $F_\pm(\theta,\alpha,\chi)$ is a monotone increasing function
of $\alpha$.  The step functions occurring here are defined by
\numparts
\bea
\eta(\xi)=\cases{1,&$\xi>0$,\\
0,&$\xi<0$,}\label{stepfunction}\\
\epsilon(\xi)=\cases{1,&$\xi>0$,\\
-1,&$\xi<0$.}\label{epsilon}
\eea
\endnumparts
The phases, $\beta_D$ and $\beta_S$, satisfy the appropriate differential
equation (\ref{diffeqlam}), for $\theta\ne\chi$ (as well as
$\theta\ne\pi-\chi$ for $\beta_S$) and are determined up to constants.
The step functions $\eta$ are introduced here in order to make
$\rme^{\rmi\Lambda}$ continuous at $\theta=\chi$ and $\pi-\chi$, as will
be explained below.  We now observe that
\be
F_\pm(\theta,2\pi+\alpha,\chi)-F_\pm(\theta,\alpha,\chi)=\frac{2\pi}{|\cos
\theta\pm\cos\chi|},
\ee
so that the wavefunction
\be
\Psi=\rme^{-\rmi\Lambda}\Psi',
\ee
where $\Psi'$ is the solution to the Schr\"odinger equation with the
singularity on the $z$ axis, is single-valued under the substitution
$\phi\to\phi+2\pi$ when the quantization condition (\ref{quant}) is
satisfied.

Notice that integer quantization follows when an infinite string is used
while a semi-infinite string leads to half-integer quantization, since
$\beta_S$ changes by a multiple of $2\pi$ when $\phi\to\phi+2\pi$, while
$\beta_D$ changes by an integer multiple of $4\pi$.  Notice that $\beta_D$
possesses a discontinuity, which is a multiple of $4\pi$, at $\theta=\chi$,
while $\beta_S$ possesses discontinuities, which are multiples of $2\pi$,
at $\theta=\chi$, $\pi-\chi$. In virtue of the above-derived quantization
conditions, $\rme^{\rmi\Lambda}$ is continuous everywhere.  Correspondingly,
the unitary operator $\rme^{\rmi\Lambda}$, which relates solutions of 
Schr\"odinger equations with different vector potentials, is alternatively
viewed as a gauge transformation relating physically equivalent
descriptions of the same system, since it converts one string into another.
[Identical arguments applied to the case when only one vector potential is
present leads to the condition (\ref{dyon}),
where $m'$ is an integer, or an integer plus one-half,
 for infinite and semi-infinite strings, respectively.]

It is now a simple application of the above results to transform a system
characterized by a single vector potential with an infinite string along the
direction $\bf \hat n$ into one in which the singularity line is semi-infinite
and lies along the $+z$ axis.  This can be done in a variety of ways;
particularly easy is to break the string at the origin and transform the
singularities to the $z$ axis.  Making use of (\ref{Lambdaasbeta}) with
$e_1g_2\to-m'/2$ and (\ref{betaD}) for $\bf \hat n$ and $-\mathbf{\hat n}$, 
we find
\be
\Lambda=m'\beta'_S(\mathbf{\hat n,r})\qquad\mbox{with}\qquad \beta_S'=\phi
-\psi+\beta_S.\label{infinitytosemi}
\ee
In particular, we can relate the wavefunctions for infinite and semi-infinite
singularity lines on the $z$ axis by setting $\chi=0$ in 
(\ref{infinitytosemi}),
\be
\beta'_S=\phi-\psi,\label{betaS'}
\ee
so
\be
\Psi(\mbox{infinite})=\rme^{-\rmi m'(\phi-\psi)}\Psi(\mbox{semi-infinite}).
\label{rotwf}
\ee
Note that (\ref{rotwf}) or (\ref{infinitytosemi}) reiterates that an infinite
string requires integer quantization.

\subsubsection{Scattering}

In the above subsection, we related the wavefunction when the string
lies along the direction $\mathbf{\hat n}$ with that when the string lies
along the $z$ axis.  When there is only a single vector potential (which,
for simplicity, we will assume throughout the following), this relation is
\be
\Psi_{\bf \hat n}=\rme^{-\rmi m'\beta(\mathbf{\hat n,r})}\Psi',\label{psi} 
\ee
where $\beta$ is given by (\ref{betaD}), (\ref{betaS}), or 
(\ref{infinitytosemi})
for the various cases.  For concreteness, if we take $\Psi'$ to be a
state corresponding to a semi-infinite singularity line along the $+z$
axis, then $\beta$ is either $\beta_D$ (\ref{betaD}) or $\beta_S'$
(\ref{infinitytosemi}) depending on whether the singularity characterized by 
$\bf\hat n$
is semi-infinite or infinite.  By means of (\ref{psi}), we can easily build
up the relation between solutions corresponding to two arbitrarily oriented
strings, with $\bf \hat n$ and $\bf \hat n'$ say,
\be
\Psi_{\bf \hat n'}=\rme^{-\rmi m'(\beta(\mathbf{\hat n',r})
-\beta(\mathbf{\hat n,r}))}\Psi_{\bf \hat n},
\ee
which expresses the gauge covariance properties of the wavefunctions.

For scattering, we require a solution that consists of an incoming plane
wave and an outgoing spherical wave.  We will consider an eigenstate of 
$\mathbf{J\cdot\hat k}$ where $\mathbf{J}$ is the total angular
momentum
\be
\mathbf{J}=\mathbf{r}\times(\mathbf{p}+m'\boldsymbol{\mathcal{A}}_{\bf\hat n})
+m'\mathbf{\hat r},\label{3.38}
\ee
and $\mathbf{\hat k}$ is the unit vector in the direction of propagation
of the incoming wave (not necessarily the $z$ axis).  This state cannot be
an eigenstate of $\mathbf{\hat k\cdot(r\times p)}$, since this operator does
not commute with the Hamiltonian.  However, since
\numparts
\be
\rme^{\rmi \Lambda}\mathbf{\hat k\cdot J}\rme^{-\rmi \Lambda}
=\mathbf{\hat k\cdot(r\times p)}-m',\label{2.12}
\ee
for a reorientation of the string from $\bf \hat n$ to $\bf \hat k$,
\be
\Lambda=m'[\beta(\mathbf{\hat n,r})-\beta_D(\mathbf{\hat k,r})],\label{3.5}
\ee
\endnumparts
because
\be
\frac{\mathbf{\hat k\cdot(\hat r\times(\hat k\times \hat r))}}{1-
\mathbf{\hat k\cdot\hat r}}=1+\mathbf{\hat k\cdot\hat r},
\ee 
the incoming state with eigenvalue \cite{Boulware:1976tv}
\be
(\mathbf{\hat k\cdot J})'=-m'
\ee
is simply related to an ordinary modified plane wave [$\eta$ is defined below
in \eref{3.46}]
\be
\Psi_{\rm in}=\rme^{-\rmi \Lambda}\exp\left\{\rmi\left[\mathbf{k\cdot r}+
\eta\ln(kr-\mathbf{k\cdot r})\right]\right\}.  
\ee
This state exhibits the proper gauge covariance under reorientation of the
string.

The asymptotic form of the wavefunction is
\be
\fl \Psi\sim \rme^{-\rmi m'\beta(\mathbf{\hat n,r})}\sum_{j\bar m}A_{kj\bar m}
\mathcal{Y}^{m'}_{j\bar m}(\mathbf{\hat r})\rme^{\rmi m'\phi}\frac1{kr}
\sin\left(kr-\eta\ln 2kr-\frac\pi2 L+\delta_L\right),\qquad
r\to\infty.\label{3.8}
\ee
The summation in (\ref{3.8}) is the general from of the solution when the
singularity line is semi-infinite, extending along the $+z$ axis.  In
particular, $\mathcal{Y}^{m'}_{j\bar m}$ is a generalized spherical
harmonic, which is another name for the rotation matrices in quantum
mechanics [$\mathbf{\hat r}=(\theta,\phi)$],
\be
\langle jm'|\rme^{\rmi \psi J_3}\rme^{\rmi \theta J_2}\rme^{\rmi\phi J_3}|jm
\rangle=\rme^{\rmi m'\psi}\frac1{\sqrt{2j+1}}\mathcal{Y}^{m'}_{jm}(\mathbf{
\hat r})=\rme^{\rmi m'\psi}U_{m'm}^{(j)}(\theta)\rme^{\rmi m\phi},
\ee
$\delta_L$ is the Coulomb phase shift for noninteger $L$,
\be
\delta_L=\mbox{arg}\,\Gamma(L+1+\rmi \eta),\label{coulombph}
\ee
and
\be
L+\frac12=\sqrt{\left(j+\frac12\right)^2-m^{\prime2}},\qquad
\eta=\frac{\mu q}{k},\qquad q=e_1e_2+g_1g_2.\label{3.46}
\ee
Upon defining the outgoing wave by
\be
\Psi\sim\rme^{-\rmi \Lambda}\left(\rme^{\rmi[\mathbf{k\cdot r}+\eta
\ln(kr-\mathbf{k\cdot r})]}+\Psi_{\rm out}\right),
\ee
where $\Lambda$ is given by (\ref{3.5}), we find that 
\be
\Psi_{\rm out}=\frac1r\rme^{\rmi (kr-\eta\ln kr)}\rme^{\rmi m'\gamma}
f(\bar\theta).\label{3.13}
\ee
In terms of the scattering angle, $\bar\theta$, which is the 
angle between $\mathbf{k}$ and $\mathbf{r}$, the scattering amplitude is
\be
2\rmi k f(\bar\theta)=\sum_{j=|m'|}^\infty \sqrt{2j+1}
\mathcal{Y}^{m'}_{jm'}(\pi-\bar\theta,0)\rme^{-\rmi (\pi L-2\delta_L)}.
\label{3.49}
\ee
The extra phase in (\ref{3.13}) is given by (where $\mathbf{\hat k}$ 
characterized by $\theta'$, $\phi'$ and $-\mathbf{\hat k}$ by $\pi-\theta'$,
$\phi'\pm \pi$)
\be
\gamma=\beta_D(\mathbf{\hat k,-\hat k})+\phi-\phi'\mp \pi-\beta_D(\mathbf{
\hat k,r})+\bar\phi,
\ee
where
\be
\arctan\frac12\bar\phi=\frac{\cos\left(\frac{\theta+\pi-\theta'}2\right)
\sin\left(\frac{\phi-\phi'\mp\pi}2\right)}
{\cos\left(\frac{\theta-\pi+\theta'}2\right)
\cos\left(\frac{\phi-\phi'\mp\pi}2\right)}.\ee
Straightforward evaluation shows that
\be
\frac\gamma2=0\,\, (\mbox{mod}\, 2\pi),
\ee
so that there is no additional phase factor in the outgoing wave.  

\subsubsection{Spin}\label{sec:spin}
Classically, the electromagnetic field due to two dyons at rest carries
angular momentum, as in \sref{sec:angmom},
\be
\mathbf{S}_{\rm classical}=m'\mathbf{\hat r}.\label{4.1}
\ee
A quantum-mechanical transcription of this fact allows us to replace
the nonrelativistic description explored above, in which the interaction
is through the vector potentials (apart from the Coulomb term), by one in 
which the particles interact with an intrinsic spin.  The derivation of the
magnetic charge problem from this point of view
seems first to have carried out by Goldhaber
\cite{gold} in a simplified context, and was revived in the context of 
't Hooft-Polyakov monopoles 
\cite{'tHooft:1974qc,Polyakov:1974ek,wu75,yang76,nambu,Boulware:1976tv}, 
where the spin is called ``isospin.''

Before introducing the notion of spin, we first consider the angular
momentum of the actual dyon problem.  For simplicity we will describe
the interaction between two dyons in terms of a single vector potential
$\boldsymbol{\mathcal{A}}$, and an infinite string satisfying (\ref{strcon2}).
(The other cases are simple variations on what we do here, and the consequences
for charge quantization are the same as found in \sref{sec:strings}.)
Then the relative momentum of the system is
\be
\mathbf{p}=\mu\mathbf{v}-m'\boldsymbol{\mathcal{A}}.
\ee
Since from \eref{3.14a}, \eref{3.14b}, we have the result in \eref{string0a}, or
\be
\bnabla\times\boldsymbol{\mathcal{A}}=\frac{\mathbf{r}}{r^3}-\mathbf{f(r)},
\label{3.55}
\ee
we have the following commutation property valid everywhere,
\be
\mu\mathbf{v}\times\mu\mathbf{v}=-\rmi m'\left[\frac{\mathbf{r}}{r^3}-\mathbf
{f(r)}\right].
\ee
Motivated by the classical situation, we assert that the total angular
momentum operator is \eref{3.38}, or
\be
\mathbf{J}=\mathbf{r}\times\mu\mathbf{v}+m'\mathbf{\hat r}.\label{4.5}
\ee
This is confirmed \cite{Schwinger:1969ib} 
by noting that, almost everywhere,
$\mathbf{J}$ is the generator of rotations:
\numparts
\bea
\frac1\rmi[\mathbf{r,J}\cdot\delta\bomega]=\delta\bomega\times\mathbf{r},\\
\frac1\rmi[\mu\mathbf{v},\mathbf{J}\cdot\delta\bomega]=\delta\bomega\times
\mu\mathbf{v}-m'\mathbf{f(r)}\times(\delta\bomega\times\mathbf{r}),
\label{4.7}
\eea
\endnumparts
where $\delta\bomega$ stands for an infinitesimal rotation.  The presence of 
the extra term in (\ref{4.7}) is consistent only because of the quantization
condition \cite{Schwinger:1975ww}. For example, consider the effect of
a rotation on the time evolution operator,
\be
\rme^{-\rmi\mathbf{J}\cdot\delta\bomega}\exp\left[-\rmi\int\rmd t\,\mathcal{H}
\right]\rme^{\rmi\mathbf{J}\cdot\delta\bomega}
=\exp\left[-\rmi\int\rmd t\left(\mathcal{H}+\delta\mathcal{H}\right)\right],
\ee
where
\be
\delta\mathcal{H}=\rmi[\mathcal{H},\mathbf{J}\cdot\delta\bomega]
=m'\mathbf{v}\cdot[\mathbf{f(r)\times\delta r}],
\qquad
\delta\mathbf{r}=\delta\bomega\times\mathbf{r}.
\ee
Using the representation for the string function,
\be
\mathbf{f(r)}=4\pi\int_C\rmd \mathbf{x}\frac12\left[
\delta(\mathbf{r-x})-\delta(\mathbf{r+x})\right],
\ee
where $C$ is any contour starting at the origin and extending to
infinity, and the notation $\rmd t\mathbf{v}=\rmd \mathbf{r}$, we have
\be
-\rmi\int\rmd t\,\delta\mathcal{H}=-\rmi m'4\pi\int\rmd\mathbf{r}\cdot
(\rmd \mathbf{x}\times\delta\mathbf{r})\frac12\left[\delta(\mathbf{r-x})
-\delta(\mathbf{r+x})\right].
\ee
Since the possible values of the integral are 0, $\pm\frac12$, $\pm1$, the
unitary time development operator is unaltered by a rotation only if $m'$
is an integer.  (Evidently, half-integer quantization results from the use
of a semi-infinite string.)

Effectively, then, $\mathbf{J}$ satisfies the canonical angular momentum
commutation relations (see also \sref{app}) 
\be
\frac1\rmi\mathbf{J\times J=J},
\ee
and is a constant of the motion,
\be
\frac{\rmd}{\rmd t}\mathbf{J}=\frac1\rmi[\mathcal{H},\mathbf{J}]=0.\ee
And, corresponding to the classical field angular momentum (\ref{4.1}),
the component of $\mathbf{J}$ along the line connecting the two dyons, $m'$,
should be an integer.

The identification of $m'$ as an angular momentum component invites us to
introduce an independent spin operator $\mathbf S$.  We do this by first
writing \cite{Schwinger:1969ib}, as anticipated in \eref{3.11}, \eref{3.10b}, 
\numparts
\be
m'=\mathbf{S\cdot \hat r},\label{4.15}
\ee
and
\be
\mu\mathbf{v}=\mathbf{p}+\frac{\mathbf{S\times r}}{r^2},\label{mechmom0}
\ee
\endnumparts
which, when substituted into (\ref{4.5}) yields \eref{3.10a},
\be
\mathbf{J=r\times p+S}.\label{4.17}
\ee
We now ascribe independent canonical commutation relations to $\mathbf{S}$,
and regard (\ref{4.15}) as an eigenvalue statement.  The consistency of this
assignment is verified by noting that the commutation property
\be
\mu\mathbf{v}\times \mu\mathbf{v}=-\rmi m'\frac{\mathbf{r}}{r^3}
\ee
holds true, and that $\mathbf{S\cdot \hat r}$ is a constant of the motion,
\be
[\mathbf{S\cdot\hat r},\mu\mathbf{v}]=0.
\ee
In this angular momentum description, the Hamiltonian, (\ref{nrham2pot}),
can be written in the form
\be
\mathcal{H}=\frac1{2\mu}\left[p^2+\frac{2\mathbf{S\cdot L}}{r^2}+
\frac{\mathbf{S}^2-(\mathbf{S\cdot \hat r})^2}{r^2}\right]+\frac{q}r,
\label{4.19}
\ee
in terms of the orbital angular momentum,
\be
\mathbf{L=r\times p}.
\ee
The total angular momentum $\mathbf{J}$ appears when the operator
\be
p_r^2=\frac1{r^2}\left[(\mathbf{r\cdot p})^2
+\frac1\rmi \mathbf{r\cdot p}\right],\label{4.21}
\ee
is introduced into the Hamiltonian
\be
\mathcal{H}=\frac1{2\mu}\left[p_r^2+\frac{\mathbf{J}^2-(\mathbf{J\cdot \hat r}
)^2}{r^2}\right]+\frac{q}r.\label{4.22}
\ee
In an eigenstate of $\mathbf{J}^2$ and $\mathbf{J\cdot \hat r}$.
\be
(\mathbf{J}^2)'=j(j+1),\qquad (\mathbf{J\cdot \hat r})'=m',
\ee
(\ref{4.22}) yields the radial Schr\"odinger equation \eref{3.129a} solved 
in \sref{sec:nrelham}.  This modified formulation, only formally
equivalent to our starting point, makes no reference to a vector potential
or string.

We now proceed to diagonalize the $\mathbf{S}$ dependence of the Hamiltonian,
(\ref{4.19}) or (\ref{4.22}), subject to the eigenvalue constraint
\be
(\mathbf{S\cdot\hat r})'=m'.\label{4.24}
\ee
This is most easily done by diagonalizing \cite{Boulware:1976tv}
the angular momentum operator
(\ref{4.17}).  In order to operate in a framework sufficiently general to
include our original symmetrical starting point, we first write $\mathbf{S}$
as the sum of two independent spins
\be
\mathbf{S}=\mathbf{S}_a+\mathbf{S}_b.
\ee
We then subject $\mathbf{J}$ to a suitable unitary transformation \cite{gold}
\be
\mathbf{J}'=U\mathbf{J}U^{-1},
\ee
where
\be
U=\exp[\rmi(\mathbf{S}_a\cdot \hat{\bphi})\theta]\exp[\rmi(\mathbf{S}_b\cdot
\hat{\bphi})(\theta-\pi)],\label{4.27}
\ee
which rotates $\mathbf{S}_{a,b}\cdot\mathbf{\hat r}$ into $\pm(\mathbf{S}_{a,b}
)_3$.  This transformation is easily carried out by making use of the
representation in terms of Euler angles,
\be
\exp(\rmi \mathbf{S}\cdot\bphi\,\theta)=\exp(-\rmi\phi S_3)\exp(\rmi\theta S_2)
\exp(\rmi \phi S_3).\label{4.28}
\ee
The general form of the transformed angular momentum,
\be\mathbf{J}'=\mathbf{r}\times\left[\mathbf{p}+\frac{\hat{\bphi}}r
\sin\theta\left(\frac{S_{a3}}{1+\cos\theta}+\frac{S_{b3}}{1-\cos\theta}
\right)\right]+\mathbf{\hat r}(\mathbf{S}_a-\mathbf{S}_b)'_3,
\ee
is subject, a priori, only to the constraint (\ref{4.24}), or
\be
(\mathbf{S}_a-\mathbf{S}_b)'_3=m'.\label{4.30}
\ee
We recover the unsymmetrical and symmetrical formulations by imposing the
following supplementary eigenvalue conditions:
\numparts
\bea
(1):\qquad S_{a3}'=0,\label{4.31a}\\
(2):\qquad (\mathbf{S}_a+\mathbf{S}_b)'_3=0.\label{4.31b}
\eea
\endnumparts
These yield the angular momentum in the form (\ref{4.5}) or (\ref{3.38}), 
the vector potential appearing there being, respectively,
\numparts
\bea
(1):\qquad \boldsymbol{\mathcal{A}}=-\frac{\hat{\bphi}}r\cot\frac\theta2,
\label{4.32a}\\
(2):\qquad \boldsymbol{\mathcal{A}}=-\frac{\hat{\bphi}}r\cot\theta,
\label{4.32b}
\eea
\endnumparts
which are (\ref{potential}) with $\mathbf{\hat n=\hat z}$. [See also
\eref{3.6b}, \eref{3.4}.]

The effect of this transformation on the Hamiltonian is most easily seen
from the form (\ref{4.22}),
\be
U\left[p_r^2+\frac{J^2-(\mathbf{J\cdot{\hat r}})^2}{r^2}\right]U^{-1}
=p_r^2+\frac1{r^2}(\mathbf{r\times\mu v})^2=(\mu\mathbf{v})^2,
\ee
making use of (\ref{4.21}), or
\be
\mathcal{H}'=U\mathcal{H}U^{-1}=\frac12\mu v^2+\frac{q}r.
\label{4.33}
\ee

So by means of the transformation given in (\ref{4.27}) we have derived
the explicit magnetic charge problem, expressed in terms of $\mathbf{J}'$
and $\mathcal{H}'$, from the implicit formulation in terms of spin.  These
transformations are not really gauge transformations, because the physical
dyon theory is defined only after the eigenvalue conditions (\ref{4.30})
and (\ref{4.31a})--(\ref{4.31b})  are imposed.  The unsymmetrical
condition (1), (\ref{4.31a}), gives rise to the Dirac formulation of 
magnetic charge, with a semi-infinite singularity line, and, from (\ref{4.30}),
$m'$ either integer or half-integer.   The symmetrical condition (2),
(\ref{4.31b}) gives the Schwinger formulation: An infinite singularity
line [with (\ref{strcon2}) holding], and integer quantization of $m'$.
These correlations, which follow directly from the commutation properties
of angular momentum (the group structure), are precisely the conditions
required for the consistency of the magnetic charge theory, as we have
seen in \sref{sec:strings}.

Even though the individual unitary operators $U$ are not gauge
transformations, a sequence of them, which serves to reorient the
string direction, is equivalent to such a transformation.  For example,
if we formally set $\mathbf{S}_a=0$ in (\ref{4.27}),
\be
U_{(1)}=\exp(\rmi \mathbf{S}\cdot\hat{\bphi}(\theta-\pi)],\label{4.35}
\ee
we have the transformation which generates a vector potential 
with singularity along the positive $z$ axis, (\ref{4.32a}), while
\be
U_{(2)}=\exp[\rmi\mathbf{S\cdot \hat u}_2(\Theta-\pi)]\label{4.36}
\ee
generates a vector potential with singularity along $\mathbf{\hat n}$,
the first form in (\ref{potential}), where $\Theta$ is the angle
between $\mathbf{\hat n}$ and $\mathbf{r}$,
\be
\cos\Theta=\cos\theta\cos\chi+\sin\theta\sin\chi\cos(\phi-\psi)
\ee
[the coordinates of $\mathbf{\hat n}$ are given by (\ref{2.16})], and
\be
\mathbf{\hat u}_2=\frac{\mathbf{\hat n\times r}}{|\mathbf{\hat n\times r}|}.
\ee
The transformation which carries (\ref{4.32a}) into the first form
in (\ref{potential}) is
\be
U_{(12)}=U^{\vphantom{1}}_{(2)}U^{-1}_{(1)}.
\ee
Since $U_{(12)}$ reorients the string from the direction $\mathbf{\hat z}$
to the direction $\mathbf{\hat n}$, it must have the form
\be
U_{(12)}=\exp(\rmi\mathbf{S\cdot\hat n}\Phi)\exp(-\rmi \mathbf{S}\cdot
\hat{\bpsi}\chi).\label{4.40}
\ee
The angle of rotation about the $\mathbf{n}$ axis, $\Phi$, is most
easily determined by considering the spin-1/2 case $\mathbf{S}=\frac12\bsigma$,
and introducing a right-handed basis,
\be
\mathbf{\hat u}_1=\mathbf{\hat n},\qquad \mathbf{\hat u}_2=
\frac{\mathbf{\hat n\times r}}{|\mathbf{\hat n\times r}|},\qquad
\mathbf{\hat u}_3=\mathbf{\hat n\times \hat u}_2.
\ee
Then straightforward algebra yields
\numparts
\bea
\cos\frac12\Phi=\frac{\sin\frac12\theta\cos\frac12\chi-\cos\frac12\theta
\sin\frac12\chi\cos(\phi-\psi)}{\sin\frac12\Theta},\label{4.42a}\\
\sin\frac12\Phi=\frac{-\cos\frac12\theta
\sin\frac12\chi\sin(\phi-\psi)}{\sin\frac12\Theta}.\label{4.42b}
\eea
\endnumparts
The corresponding transformation carrying the vector potential with 
singularities along the negative $z$ axis [(\ref{4.32a}) with $\theta
\to\theta-\pi$], into the vector potential with singularities along the
direction of $-\mathbf{\hat n}$ [the first form in (\ref{potential}) with 
$\mathbf{\hat n\to-\hat n}$], are obtained from (\ref{4.40}) and 
(\ref{4.42a})--(\ref{4.42b}) [see also (\ref{4.35}) and (\ref{4.36})]
by the substitutions
\be
\theta\to\theta+\pi,\qquad \Theta\to\Theta+\pi.
\ee  
The combination of these two cases gives the transformation of the infinite
string, of which (\ref{4.27}) is the prototype.

Since the effect of $\exp(-\rmi \mathbf{S}\cdot\hat{\bpsi}\chi)$ is
completely given by
\be
\exp(-\rmi \mathbf{S}\cdot\hat{\bpsi}\chi)S_3
\exp(\rmi\mathbf{S}\cdot\hat{\bpsi}\chi)=\mathbf{S\cdot \hat n},
\ee
that is, for the transformation (\ref{4.40}),
\bea
U^{\vphantom{1}}_{(12)}\left[\mathbf{r}\times\left(\mathbf{p}+\frac{\hat{\bphi}}r
\cot\frac\theta2 S_3\right)-\mathbf{\hat r}S_3\right]U^{-1}_{(12)}\nn
=\exp(\rmi\mathbf{S\cdot \hat n}\Phi)\left[\mathbf{r}\times\left(\mathbf{p}
+\frac{\hat{\bphi}}r \cot\frac\theta2\mathbf{S\cdot \hat n}\right)
-\mathbf{\hat r
S\cdot \hat n}\right]\exp(-\rmi\mathbf{S\cdot\hat n}\Phi),
\eea
in a state when $\mathbf{S\cdot \hat n}$ has a definite eigenvalue $-m'$,
$U_{(12)}$ is effectively just the gauge transformation which reorients
the string from the $z$ axis to the direction $\mathbf{\hat n}$.  And, indeed,
in this case,
\be
\frac12\Phi=\frac12\beta_D \,\,(\mbox{mod}\,2\pi),
\ee
where $\beta_D$ is given by (\ref{betaD}) as determined by the differential
equation method.

\subsubsection{Singular gauge transformations}
\label{app}
We now make the observation that it is precisely the
singular nature of the gauge transformations (\ref{2.14}) and (\ref{4.33})
which is required for the consistency of the theory, that is, the 
nonobservability of the string.  To illustrate this, we will consider
a simpler context, that of an electron moving in the field of a static
magnetic charge of strength $g$, which produces the magnetic field
\be
\mathbf{B}=g\frac{\mathbf{\hat r}}{r^2}.\label{a1}
\ee
The string appears in the relation of $\mathbf{B}$ to the vector potential,
\eref{3.55}, \eref{string0a}, or
\be
\mathbf{B}=\bnabla\times \mathbf{A}+g\mathbf{f(r)},\label{a2} 
\ee
where the string function $\bf f$ satisfies (\ref{2.6}).
Reorienting the string consequently changes $\mathbf{A}$,
\be
\mathbf{A\to A'},
\ee
which induces a phase change in the wavefunction,
\be
\Psi\to\Psi'=\rme^{\rmi \Lambda}\Psi.\label{a4}
\ee
The equation determining $\Lambda$ is (\ref{diffeqlam}), or
\be
\bnabla\Lambda=e(\mathbf{A'-A}),\label{a5}
\ee
which makes manifest that this is a gauge transformation of a singular
type, since
\be
\bnabla\times\bnabla\Lambda\ne0.
\ee

Recognition of this fact is essential in understanding the commutation
properties of the mechanical momentum (called $\mu\mathbf{v}$ above),
\be
\bpi=\mathbf{p}-e\mathbf{A},\label{3.104}
\ee
since
\be
\bpi\times\bpi=-\bnabla\times\bnabla+\rmi e(\bnabla\times \mathbf{A}).
\label{a8}
\ee
(Here, the parentheses indicate that $\bnabla$ acts only on $\mathbf{A}$,
and not on anything else to the right.)  Consider the action of the operator
(\ref{a8}) on an energy eigenstate $\Psi$.  Certainly $\bnabla\times\bnabla
\Psi=0$ away from the string; on the string, we isolate the singular term
by making a gauge transformation reorienting the string,
\be
\Psi=\rme^{-\rmi\Lambda}\Psi',
\ee
where $\Psi'$ is regular on the string associated with $\mathbf{A}$.
Hence
\be
-\bnabla\times\bnabla\Psi=\cases{0&\mbox{off string},\\
\rmi(\bnabla\times\bnabla\Lambda)\Psi&\mbox{on string},}
\ee
so by (\ref{a5}) and (\ref{a2}),
\be
-\bnabla\times\bnabla\Psi(\mathbf{r})=\rmi eg\mathbf{f(r)}\Psi(\mathbf{r}).
\ee
Thus, when acting on an energy eigenstate [which transforms like (\ref{a4})
under a string reorientation], (\ref{a8}) becomes
\be
\bpi\times\bpi\to\rmi e[(\bnabla\times\mathbf{A})+g\mathbf{f(r)}]=\rmi e
\mathbf{B}.\label{a12}
\ee
This means that, under these conditions, the commutation properties of the 
angular momentum operator (\ref{4.5}),
\be
\mathbf{J}=\mathbf{r\times \bpi}-eg\mathbf{\hat r},
\ee
are precisely the canonical ones
\numparts
\bea
\frac1\rmi[\mathbf{r,J}\cdot\delta\bomega]\to\delta\bomega\times\mathbf{r},
\label{a14a}\\
\frac1\rmi[\mathbf{\bpi,J}\cdot\delta\bomega]\to\delta\bomega\times\bpi.
\label{a14b}
\eea
\endnumparts
In \sref{sec:spin}, we considered the operator properties of $\mathbf{J}$ on
the class of states for which $\bnabla\times\bnabla=\mathbf{0}$, so an
additional string term appears in the commutator (\ref{4.7}).
Nevertheless, in this space, $\mathbf{J}$ is consistently recognized as
the angular momentum, because the time evolution operator is invariant under
the rotation generated by $\mathbf{J}$.  Here, we have considered the
complementary space, which includes the energy eigenstates, in which case
the angular momentum attribution of $\mathbf{J}$ is immediate, from 
(\ref{a14a})--(\ref{a14b}).

Incidentally, note that the replacement (\ref{a12}) is necessary to
correctly reduce the Dirac equation describing an electron moving in the
presence of a static magnetic charge,
\be
(\gamma\pi+m)\Psi=0,\label{a15a}
\ee
to nonrelativistic form, since the second-order version of (\ref{a15a})
is
\be
(\pi^2+m^2-e\bsigma\cdot\mathbf{B})\Psi=0,
\ee
where $\mathbf{B}$ is the fully gauge invariant, string independent, field
strength (\ref{a1}), rather than $(\bnabla\times\mathbf{A})$, as might be
naively anticipated.  This form validates the consideration of the magnetic
dipole moment interaction, including the anomalous magnetic moment coupling,
which we will consider numerically below, both in connection with scattering
(\sref{sec:magdipole}) and binding (\sref{sec:binding}).

Similar remarks apply to the non-Abelian, spin, formulation of the theory,
given by (\ref{4.19}).  If we define the non-Abelian vector potential by
\be
e\mathbf{A}=-\frac{\mathbf{S\times r}}{r^2},\label{spinvp}
\ee
the mechanical momentum \eref{mechmom0} of a point charge moving in this field 
is again given by \eref{3.104}, or
\be
\bpi=\mathbf{p}-e\mathbf{A},\label{a17}
\ee
and the magnetic field strength is determined, analogously to (\ref{a12}), by
\be
e\mathbf{B}=\frac1\rmi\bpi\times\bpi=(\bnabla\times e\mathbf{A})
-\rmi e\mathbf{A}\times e\mathbf{A}=-\mathbf{S\cdot\hat r}\frac{\mathbf{\hat r}}
{r^2}.
\ee
This reduces to the Abelian field strength (\ref{a1}) in an eigenstate
of $\mathbf{S\cdot\hat r}$,
\be
(\mathbf{S\cdot\hat r})'=-eg,
\ee
which is a possible state, since $\mathbf{S\cdot\hat r}$ is a constant of the
motion,
\be
[\mathbf{S\cdot\hat r},\bpi]=0.
\ee
The Abelian description is recovered from this one by means of the
unitary transformation (\ref{4.28})
\be
U=\exp(-\rmi \phi S_3)\exp(\rmi \theta S_2)\exp(\rmi \phi S_3).
\ee
Under this transformation, the mechanical momentum, (\ref{a17}),
takes on the Abelian form,
\be
U\bpi U^{-1}=\mathbf{p}+\hat{\bphi}\frac{S_3}r\tan\frac\theta2,\label{a21}
\ee
where we see the appearance of the Abelian potential
\be
e\mathbf{A}=-S_3\frac{\hat{\bphi}}r\tan\frac\theta2,
\ee
corresponding to a string along the $-z$ axis.  In an eigenstate
of $S_3$,
\be
S_3'=(U\mathbf{S\cdot \hat r}U^{-1})'=-eg,
\ee
this is the Dirac vector potential \eref{3.6a}.  
To find the relation between this
vector potential and the field strength, we apply the unitary transformation
(\ref{a21}) to the operator 
\be
e\mathbf{B}=\bnabla \times e\mathbf{A}+e\mathbf{A}\times\bnabla-\rmi e
\mathbf{A}\times e\mathbf{A}
\ee
to obtain, using Stokes' theorem,
\be
Ue\mathbf{B}U^{-1}=(\bnabla\times e\mathbf{A})-\rmi U\bnabla\times
\bnabla U^{-1}=(\bnabla\times e\mathbf{A})-S_3\mathbf{f(r)},
\ee
where $\bf f$ is the particular string function
\be
\mathbf{f(r)}=-4\pi \mathbf{\hat k}\eta(-z)\delta(x)\delta(y),\label{partst}
\ee
$\eta$ being the unit step function.  In this way the result (\ref{a2})
is recovered.

\subsubsection{Commentary}

There is no classical Hamiltonian theory of magnetic charge, since, without
introducing an arbitrary unit of action 
\cite{yan67,Brandt:1976hk}, unphysical elements (strings)
are observable.  In the quantum theory, however, there is a unit of action,
$\hbar$, and since it is not the action $W$ which is observable, but
$\exp(\rmi W/\hbar)$, a well-defined theory exists provided charge
quantization conditions of the form (\ref{dyon}) or (\ref{quant}) are
satisfied.  The precise form of the quantization condition depends on the
nature of the strings, which define the vector potentials.  It may be worth
noting that the situation which first comes to mind, namely, a single
vector potential with a single string, implies Schwinger's symmetrical
formulation with integer quantization \cite{Schwinger:1975ww}.

We have seen in the nonrelativistic treatment of the two-dyon system
that the charge quantization condition is essential for all aspects of the
self-consistency of the theory.  
Amongst these we list the nonobservability of
the string, the single-valuedness and gauge-covariance of the wavefunctions, 
and the compatibility with the commutation relations of angular momentum.
In fact, all these properties become evident when it is recognized that
the theory may be derived from an angular momentum formulation
\cite{Brandt:1977ks,hurst,lipkin}.

\subsection{Nonrelativistic Hamiltonian}
\label{sec:nrelham}
We now must turn to explicit solutions of the
Schr\"odinger equation to obtain numerical results for cross sections.
For a system of two interacting dyons the Hamiltonian 
corresponding to symmetrical string along the entire $z$ axis is
\be
{\cal H}=-\frac{\hbar^2}{2\mu}\left(\nabla^2+\frac{2m'}{r^2}
\frac{\cos\theta}
{\sin^2\theta}\frac{1}{\rmi}\frac{\partial}{\partial \phi}
-\frac{m^{\prime2}}{r^2}\cot^2\theta\right)+\frac{q}{r},\label{hamdyons}
\ee
where the quantity $\kappa$ in (\ref{qandkappa}) is replaced by the
magnetic quantum number $m'$ defined in \eref{dyon}.
[This is \eref{nrham2pot} with $\boldsymbol{\mathcal{A}}=
\boldsymbol{\mathcal{A}}'$ given by the second form in
\eref{potential} with $\mathbf{\hat n=\hat z}$.]
Even though this is much more complicated than the Coulomb
Hamiltonian, the wavefunction still may be separated:
\be
\Psi({\bf r})=R(r)\Theta(\theta)\rme^{\rmi m\phi},\ee
where the radial and angular factors satisfy
\numparts
\begin{eqnarray}
\left(\frac{\rmd^2}{\rmd r^2}+\frac{2}{r}\frac{\rmd}{\rmd r}+k^2
-\frac{2\mu}{\hbar^2}\frac{q}{r}
-\frac{j(j+1)-m^{\prime2}}{r^2}\right)R=0,\label{3.129a}\\
-\left[\frac{1}{\sin\theta}\frac{\rmd}{\rmd\theta}
\left(\sin\theta\frac{\rmd}{\rmd\theta}\right)
-\frac{m^2-2mm'\cos\theta+m^{\prime2}}{\sin^2\theta}\right]\Theta
=j(j+1)\Theta.
\end{eqnarray}
\endnumparts

The solution to the $\theta$ equation is the rotation matrix element:
($x=\cos\theta$)
\begin{eqnarray}
\fl U_{m'm}^{(j)}(\theta)=\langle jm'|\rme^{\rmi J_2\theta/\hbar}|jm\rangle
\propto(1-x)^\frac{m'-m}{2}(1+x)^\frac{m'+m}{2}
P_{j-m}^{(m'-m,m'+m)}(x),
\end{eqnarray}
where $P_j^{(m,n)}$
 are the Jacobi polynomials, or ``multipole harmonics''
\cite{Yang:1978td}.
This forces $m'$ to be an integer.  The radial solutions are, as with
the usual Coulomb problem, confluent hypergeometric functions,
\numparts
\bea
R_{kj}(r)=\rme^{-\rmi kr}(kr)^LF(L+1-\rmi\eta,2L+2,2\rmi kr),\\
\eta=\frac{\mu q}{\hbar^2k},\qquad k=\frac{\sqrt{2\mu E}}{\hbar},\qquad
L+\frac{1}{2}=\sqrt{\left(j+\frac{1}{2}\right)^2-m^{\prime2}}.
\eea
\endnumparts
Note that in general $L$ is not an integer.

We solve the Schr\"odinger equation such that a distorted incoming
plane wave is incident,
\be
\Psi_{\rm in}=\exp\left\{\rmi\left[{\bf k\cdot r}+\eta\ln(kr-{\bf k\cdot r})
\right]\right\}.
\ee
Then the outgoing wave has the form \eref{3.13}, 
(here $\theta$ is the scattering angle)
\be
\Psi_{\rm out}\sim\frac{1}{r}\rme^{\rmi(kr-\eta\ln2kr)}
f(\theta),
\ee
where the scattering amplitude is given by \eref{3.49}, or
\be
2\rmi kf(\theta)=\sum_{j=|m'|}^\infty (2j+1)U^{(j)}_{m'm'}(\pi-\theta)
\rme^{-\rmi(\pi L-2\delta _L)}
\ee
in terms of the Coulomb phase shift \eref{coulombph},
\be
\delta_L=\arg\Gamma(L+1+\rmi\eta).\ee
Note that the integer quantization of $m'$ results from the use of an
infinite (``symmetric'') string; an unsymmetric string allows
$m'=\mbox{integer}+\frac{1}{2}$.

We reiterate that
we have shown that reorienting the string direction gives rise to an
unobservable phase.  Note that this result is completely general:  the
incident wave makes an arbitrary angle with respect to the string 
direction. {\em Rotation of the string direction is a gauge 
transformation.}

\begin{figure}
\centering
\includegraphics*[height=10cm, scale=1.8, angle=-01]{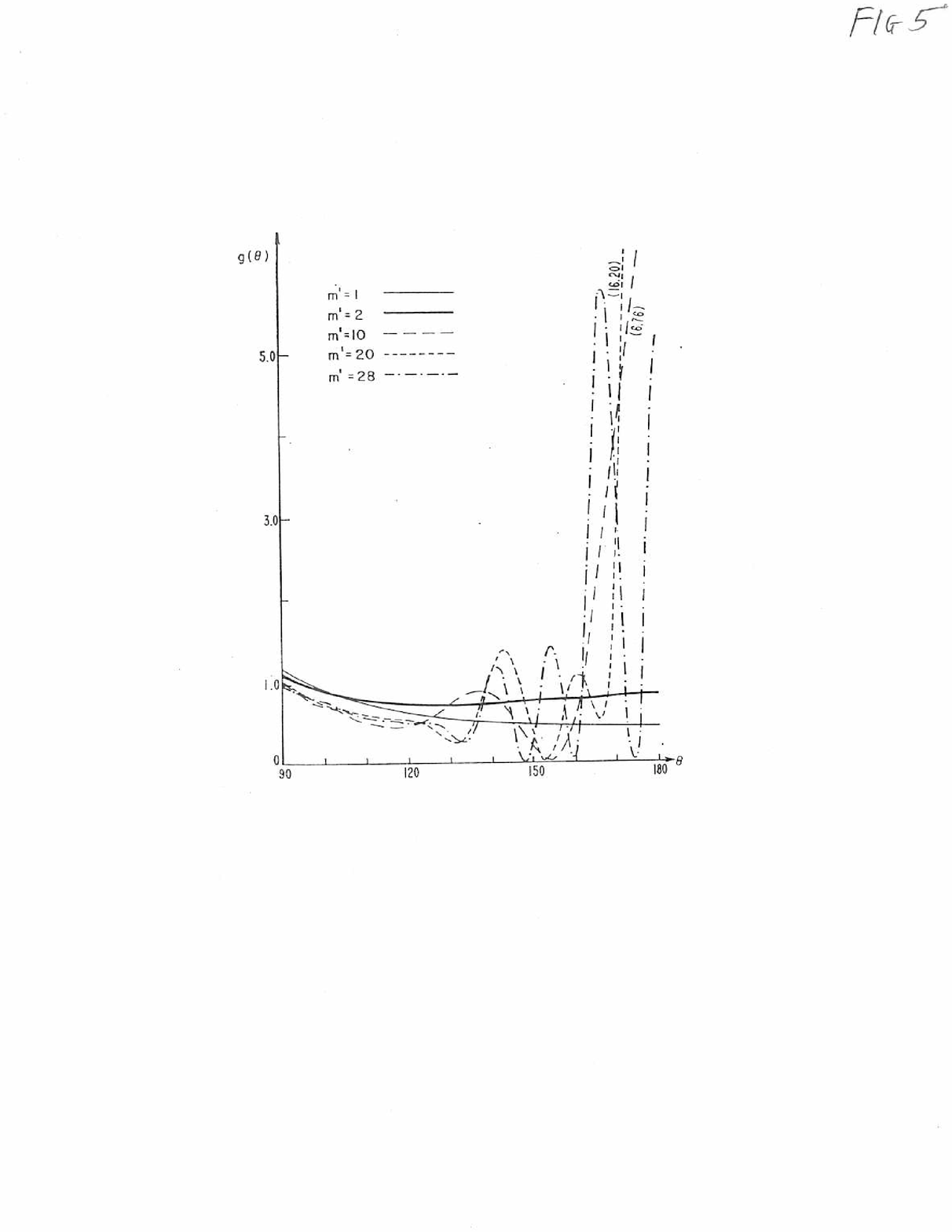}
\caption{\label{fig6}Quantum electron-monopole scattering.}
\end{figure}

\begin{figure}
\centering
\includegraphics*[height=10cm, scale=1.7]{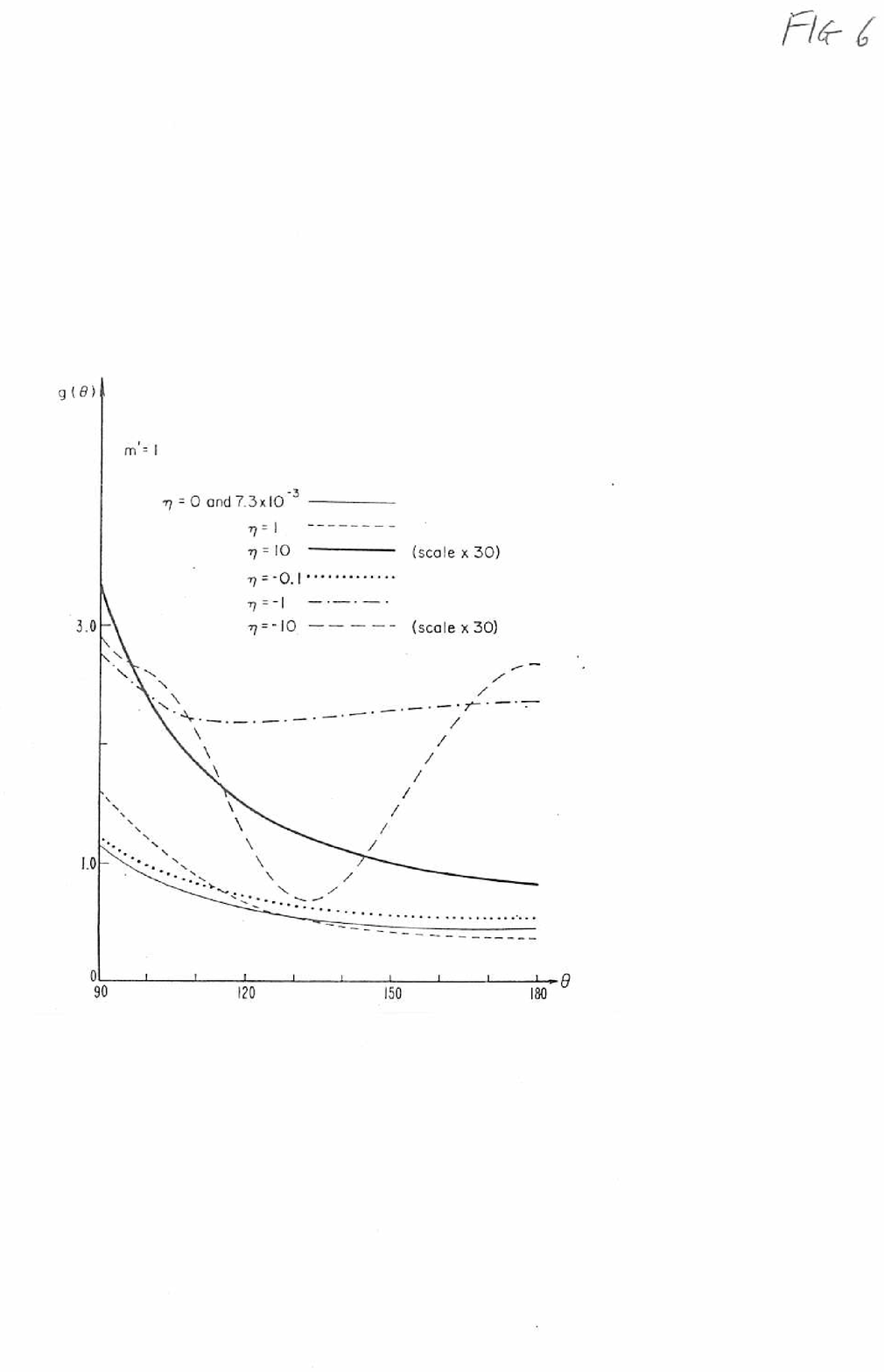}
\caption{\label{fig7}Quantum dyon-dyon scattering, $m'=1$.}
\end{figure}

\begin{figure}
\centering
\includegraphics*[height=10cm, angle=91]{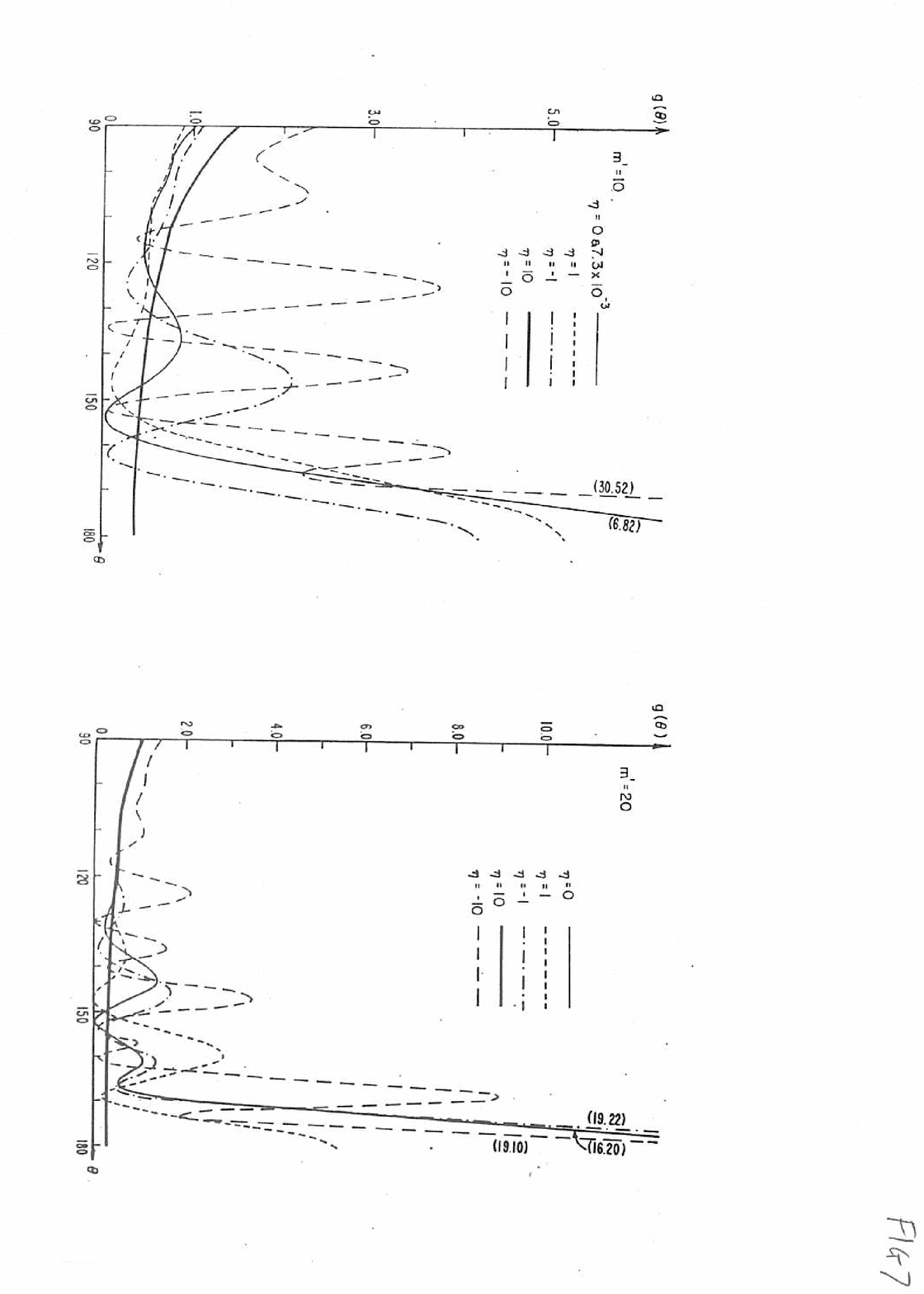}
\caption{\label{fig8}Quantum dyon-dyon scattering,  $m'=10, 20$.}
\end{figure}

By squaring the scattering amplitude, we can numerically extract
the scattering cross section.  Analytically, it is not hard to see
 that small angle scattering is still given by the Rutherford 
formula (\ref{ruth}):
\be
\frac{\rmd\sigma}{\rmd\Omega}\approx\left(\frac{m'}{2k}\right)^2\frac{1}
{\sin^4\theta/2},
\qquad\theta\ll1,\label{ruth2}
\ee
for electron-monopole scattering.  The classical result is good roughly up
to the first classical rainbow.  In general, one must proceed numerically.
In terms of
\be
g(\theta)=\frac{k^2}{m^{\prime2}}|f(\theta)|^2
\ee
we show various results in figures \ref{fig6}--\ref{fig8}.
Structures vaguely reminiscent of classical rainbows appear for large $m'$,
particularly for negative $\eta$, that is, with Coulomb attraction.

\subsubsection{Magnetic dipole interaction}
\label{sec:magdipole}
\begin{figure}
\centering
\includegraphics*[height=10cm, angle=90, width=5.0in]{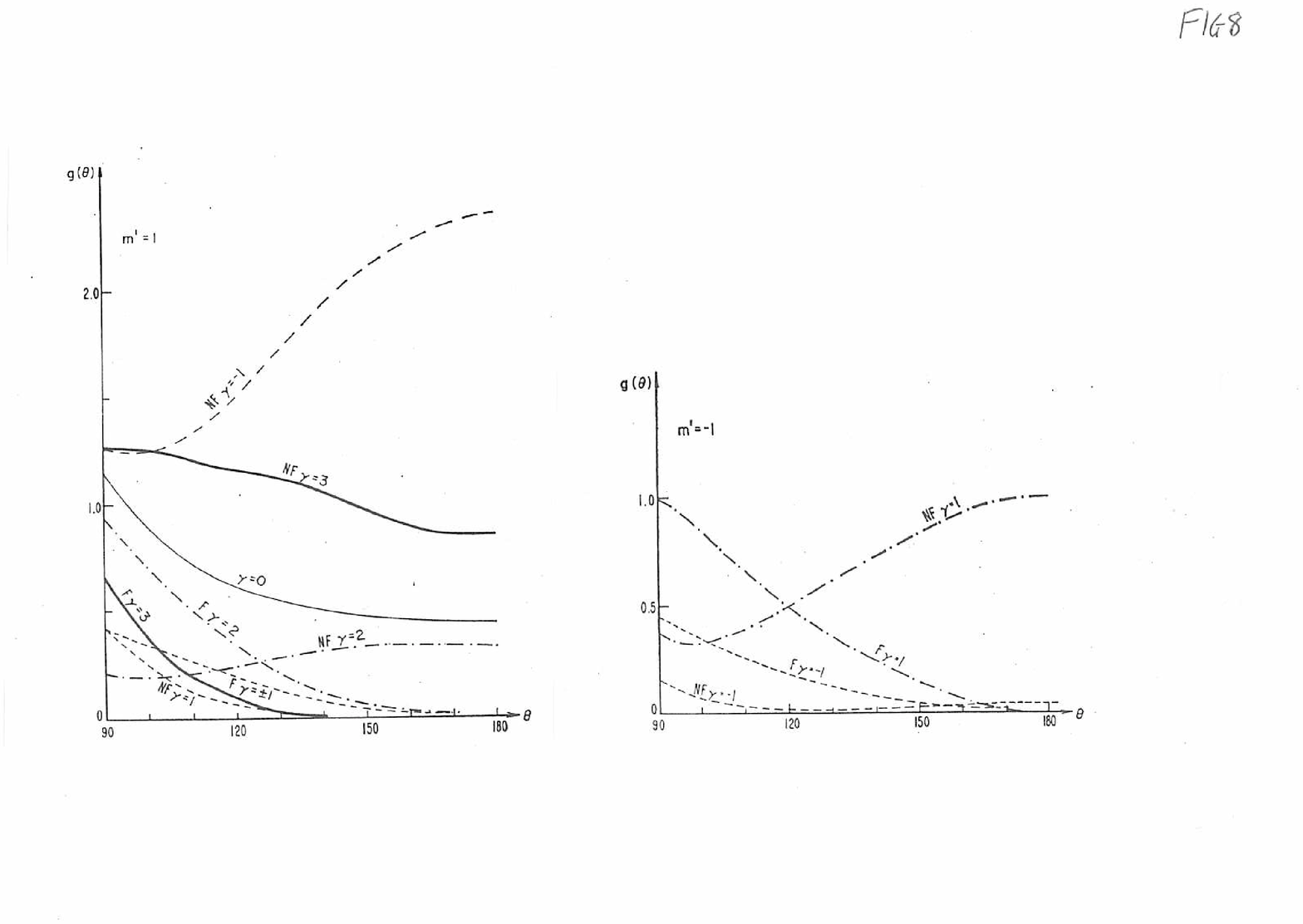}
\caption{\label{fig9}Spinflip (F) and nonflip (NF) cross sections for various
gyromagnetic ratios.  The first graph shows $m'=+1$, the second $m'=-1$.}
\end{figure}

\begin{figure}
\centering
\includegraphics*[height=10cm, angle=90, width=5.0in]{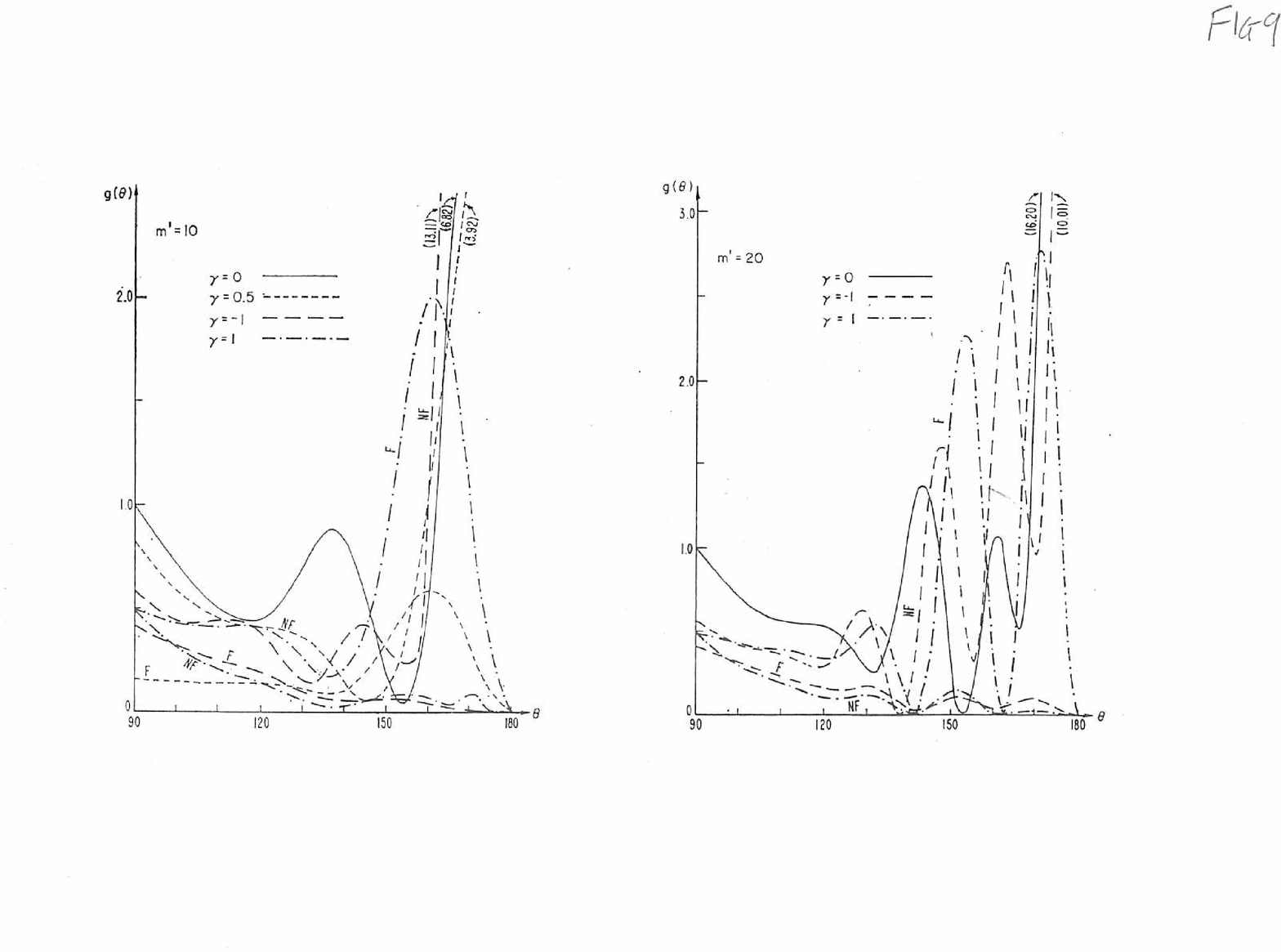}
\caption{\label{fig10}Spinflip and nonflip cross sections for $m'=10, 20$.}
\end{figure}

We can also include the effect of a magnetic dipole moment interaction, 
by adding a spin term to the Hamiltonian,
\be
\mathcal{H}_S
=-\frac{e\hbar}{2\mu c}\gamma\mbox{\boldmath{$\sigma$}}\cdot{\bf B},\qquad
{\bf B}=g\frac{{\bf r}}{r^3}.\label{magdipoleham}
\ee
For small scattering angles, the spin-flip and spin-nonflip cross sections
are for $\gamma=1$
 ($\theta\ll1$)
\begin{eqnarray}
\left.\frac{\rmd\sigma}{\rmd\Omega}\right|_{\rm F}
\approx\left(\frac{m'}{2k}\right)^2\frac{\sin^2
\theta/2}{\sin^4\theta/2},\qquad
\left.\frac{\rmd\sigma}{\rmd\Omega}\right|_{\rm NF}
\approx\left(\frac{m'}{2k}\right)^2\frac{\cos^2
\theta/2}{\sin^4\theta/2},
\end{eqnarray}
Numerical results are shown in \fref{fig9} and \fref{fig10}.
Note from the figures that {\em the spin flip amplitude always vanishes
in the backward direction;}
 the spin nonflip amplitude also vanishes there
for conditions almost pertaining to an electron:
$m'>0$, $\gamma=1$.

The calculations shown in figures \ref{fig6}--\ref{fig10}
were done many years ago \cite{Schwinger:1976fr},
which just goes to show that ``good work ages more slowly
than its creators.'' The history of the subject goes much further
back.  Tamm \cite{tamm} calculated the wavefunction for the electron-monopole
system immediately following Dirac's suggestion \cite{Dirac:1931kp};
while Banderet \cite{banderet}, following Fierz \cite{fierz}, 
was the first to suggest a partial-wave 
expansion of the scattering amplitude for the system.  
The first numerical work was carried out by Ford and Wheeler 
\cite{fordandwheeler}, while the comparison with the classical theory
can be found, for example, in \cite{lapidus,nadeau}.

\subsection{Relativistic calculation}
\label{sec:relscatt}
\begin{figure}
\centering
\includegraphics[height=10cm,angle=270]{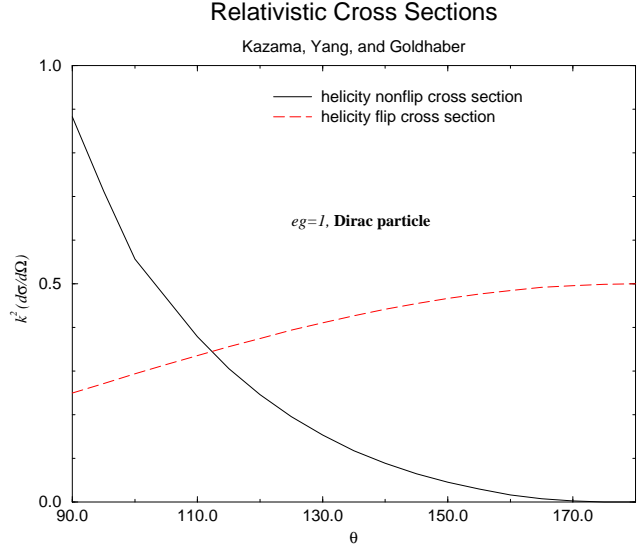}
\caption{\label{figrs}
Relativistic helicity-flip and helicity-nonflip cross sections.
Note for $\theta=\pi$, helicity nonflip corresponds to spin flip,
while helicity flip means spin nonflip.}
\end{figure}

A relativistic calculation of the scattering of a spin-1/2 Dirac particle
by a heavy monopole was given by
Kazama, Yang, and Goldhaber \cite{Kazama:1977fm}.  They used Yang's
formulation of the vector potential described above in \sref{yang}.
In order to arrive a result, they had to add an extra infinitesimal
magnetic moment term, in order to prevent the charged particle from passing 
through the monopole.  
The sign of this term would have measurable consequences
in polarization experiments.  It does not, however, appear in the
differential cross sections. It also does not affect the helicity flip and 
helicity nonflip cross sections which are shown in \fref{figrs}.
The vanishing of the helicity nonflip cross section in the backward direction
precisely corresponds to the vanishing of the nonrelativistic spinflip cross 
section there. The correspondence with the nonrelativistic calculation with
spin seems quite close.

\section{Non-Abelian monopoles}
\label{namono}
Although the rotationally symmetric, static solution of the Yang-Mills 
equations was found by Wu and Yang in 1969 \cite{wuyang}, 
it was only in 1974 when 
't Hooft and Polyakov included the Higgs field in the theory that
a stable monopole solution was found \cite{'tHooft:1974qc,Polyakov:1974ek}.
Dyonic configurations, that is, ones with both magnetic and arbitrary
electric charge, were found by Julia and Zee \cite{Julia:1975ff}.
Here we discuss the unit monopole solution first obtained by Prasad
and Sommerfield \cite{Prasad:1975kr}.  Referred to as BPS monopoles,
which saturate the Bogomolny energy bound \cite{Bogomolny:1975de},
they are static solutions of the SU(2) theory
\bea
\mathcal{L}=-\frac1{8\pi}\tr(F_{\mu\nu}F^{\mu\nu})+\tr(D_\mu H D^\mu
H)-\frac14\lambda\left(2\tr H^2-v^2\right)^2\nonumber\\
 =-\frac1{16\pi}F_a^{\mu\nu}F_{a\mu\nu}+\frac12(D_\mu H)^a(D^\mu H)^2
-\frac\lambda4(H^a H^a-v^2)^2,
\eea
where 
\be
D_\mu H=\partial_\mu H-e A_\mu\times H,
\ee
for an isotopic triplet Higgs field $H$.  (We denote the coupling
strength by $e$ to avoid confusion with the magnetic charge $g$.)
This is the Georgi-Glashow model \cite{Georgi:1972cj}, in which
the massive vector boson has mass
\numparts
\be
m_W=\sqrt{4\pi}ev,
\ee
while the Higgs  boson mass is
\be
m_H=\sqrt{\lambda}v.
\ee
\endnumparts

This model possesses  nontrivial topological sectors. 
The topological charge is
\be
k=-\frac{e}{16\pi}\int(\rmd \mathbf{x})\epsilon_{ijk}\tr (F_{jk}D_i H).
\ee
In the limit of zero Higgs coupling, $\lambda=0$,
where the Higgs boson mass vanishes,  Bogomolny showed that
the classical energy of the configuration was bounded by the charge,
\be
E\ge \sqrt{4\pi} k\frac{v}e.
\ee
The solution found by Prasad and Sommerfield achieves the lower bound on the 
energy, $E=\sqrt{4\pi} kv/e$, with unit charge, $k=1$, and has the
form
\numparts
\be
H=\frac{\mathbf{x}}r\cdot \bsigma h(r),
\qquad e\mathbf{A}=\bsigma\times\frac{\mathbf{x}}{r^2} f(r),
\label{psmono}
\ee
where with $\xi=\sqrt{4\pi}evr$
\be
h(r)=v\left(\coth\xi-\frac1{\xi}\right) ,\qquad f(r)=1-\frac\xi{\sinh \xi},
\ee
\endnumparts
If we regard $\sigma^a/2$ as the isotopic generator, the forms of the
isotopic components of the Higgs field and vector potential are
\be
H^a=2h(r)\frac{x^a}r,\qquad eA^a_i=2f(r)\epsilon_{aib}\frac{x^b}{r^2}.
\ee
These fields describe a monopole centered on the origin.  Far away from
that (arbitrary) point, the behavior of the fields is given by
\be
r\to\infty:\qquad h(r)\to v,\qquad f(r)\to 1,
\ee
so we see that the vector potential (\ref{psmono}) indeed describes a
monopole of twice the
Dirac charge, according to \eref{spinvp}, because the spin is $\mathbf{S}=
\frac12\bsigma$.  However, this monopole has structure, and is not
singular at the origin, because both $f$ and $h$ vanish there.
The energy of the configuration, the mass of the monopole, is finite,
as already noted, 
\be E=\int (\rmd \mathbf{r})\nabla^2\tr H^2=\frac12\int_{S_\infty}
\rmd \mathbf{S}\cdot\bnabla h^2(r)=\sqrt{4\pi}\frac{v^2}{ve},
\ee
which takes into account the relation between $r$ and $\xi$, as claimed.
In physical units this gives a mass of this monopole solution of 
\be
M=\frac{m_W}\alpha,
\ee
in terms of the ``fine structure'' constant of the gauge coupling,
$\alpha=e^2$ in Gaussian units.
If the electroweak phase transition produced monopoles,
then, we would expect them to have a mass of about 10 TeV
\cite{Preskill:1984gd,Kirkman:1981ck}.  There are serious
doubts about this possibility \cite{coleman,Cho:1996qd,Cho:1997jn,Cho:1997zc}
so it is far more likely to expect such objects at the GUT scale.
Magnetic monopole solutions for gauge theories with
arbitrary compact simple gauge groups, as well as noninteracting multimonopole 
solutions for those theories have been found \cite{Weinberg:1982ev}.

In general, the classical 't Hooft-Polyakov monopole mass in the Georgi-Glashow
model is with nonzero Higgs mass is
\be
M_{\rm cl}=\frac{m_W}\alpha \mu(z), \qquad z=\frac{m_H}{M_W}.
\ee
The function $\mu(0)=1$, and is less than 2 for large $z$.
Quantum corrections to the classical mass have been considered on the
lattice \cite{Rajantie:2005hi}.
 
We will not further discuss non-Abelian monopoles in this review,
because that is such a vast subject, and there are many excellent
reviews such as \cite{Sutcliffe:1997ec,Manton:2004tk}, as well
as textbook discussions \cite{Nair}.  
We merely note that asymptotically,
an isolated non-Abelian monopole looks just like a Dirac one, so that
most of the experimental limits apply equally well to either point-like
or solitonic monopoles.  Even the difficulties with the Dirac string,
which have not been resolved in the second quantized version,
to which we now turn our attention, persist with non-Abelian monopoles
\cite{Khvedelidze:2005rv}.
 
\section{Quantum field theory}
\label{sec:qft}
The quantum field theory of magnetic charge has been developed
by many people, notably Schwinger 
\cite{Schwinger:1966nj,schwingermc1,schwingermc2,Schwinger:1968rq,
Schwinger:1975ww} and Zwanziger 
\cite{Zwanziger:1968rs,Zwanziger:1970hk,Brandt:1977be,brandtandneri}.
We should cite the review article by Blagojevi\'c and Senjanovi\'c
\cite{Blagojevic:1985sh}, which cites earlier work by those authors.
A recent formulation suitable for eikonal calculations is given in
\cite{Gamberg:1999hq}, and will be described in \sref{ssqft} and following.

\subsection{Lorentz invariance}
Formal Lorentz invariance of the dual quantum electrodynamics system with
sources consisting of electric charges $\{e_a\}$ and magnetic charges
$\{g_a\}$ was demonstrated provided the quantization
condition holds:
\begin{eqnarray}
e_{a}g_{b}-e_{b}g_{a}=m'=
\Bigg\{
\begin{array}{l}
\frac{n}{2}\, , {\mbox{unsymmetric}}\\
n\, , \mbox{symmetric}
\end{array}
\Bigg\}, \qquad n \in\, Z.
\end{eqnarray}
``Symmetric'' and ``unsymmetric'' refer to the presence or
absence of dual symmetry in the solutions of Maxwell's equations,
reflecting the use of infinite or semi-infinite strings, respectively.

\subsection{Quantum action}
\label{sec:qa}
The electric and magnetic currents are the sources of the field
strength and its dual (here, for consistency, we denote by $j^\mu$,
$\jg^\mu$ what we earlier called $j^\mu_e$, $j^\mu_m$, respectively):
\begin{equation}
\partial^\nu F_{\mu\nu}=4\pi j_\mu \qquad {\mbox {\rm and}} \qquad
\partial^{\nu}\fag_{\mu\nu}=4\pi\jg_{\mu},\label{max}
\end{equation}
where
\begin{equation}
\fag_{\mu\nu}=\frac{1}{2}\epsilon_{\mu\nu\sigma\tau}F^{\sigma\tau},
\end{equation}
which imply the dual conservation of electric and magnetic
currents, $j_\mu$ and $\jg_\mu$, respectively,
\be
\partial_\mu j^{\mu}=0,\qquad \mbox{and}\qquad
\partial_\mu \jg^{\mu}=0.\label{con}
\ee
As we will detail below,
the relativistic interaction between an electric and a magnetic current
is
\begin{equation}
\fl W(j,\jg)=\int (\rmd x) (\rmd x') (\rmd x'')
\jg^\mu(x)\epsilon_{\mu\nu\sigma\tau}
\partial^{\nu} f^{\sigma}\left(x-x^{\prime}\right)
D_{+}\left(x^{\prime}-x^{\prime\prime}\right)j^{\tau}
\left(x^{\prime\prime}\right).\label{interact}
\end{equation}
Here the electric and magnetic currents are
\be
j_\mu=e\bar\psi\gamma_\mu\psi\qquad \mbox{and}
\qquad\jg_\mu=g\bar\chi\gamma_\mu\chi,
\ee
for example, for spin-1/2 particles.
The photon propagator is denoted by $D_+(x-x')$ and $f_\mu(x)$ is
the Dirac string function which satisfies the differential equation
\begin{eqnarray}
\partial_\mu f^{\mu}(x)=4\pi\delta(x),\label{dfstng}
\end{eqnarray}
the four-dimensional generalization of \eref{2.6}.
A formal solution of this equation is given by
\begin{eqnarray}
f^{\mu}(x)=4\pi n^\mu\left(n\cdot\partial\right)^{-1}\delta(x),\label{stngsl}
\end{eqnarray}
where $n^\mu$ is an arbitrary constant vector. [\Eref{partst} results if 
$\mathbf{\hat n=-\hat z}$, in which case $\mathbf{f(r},t)=\mathbf{f(r)}
\delta(t)$.]

\subsection{Field theory of magnetic charge}
\label{ssqft}

In order to facilitate the construction of the dual-QED
formalism we recognize that the well-known
continuous global ${\rm U}(1)$ {\it dual} symmetry \eref{dualtransf} 
\cite{Schwinger:1966nj,Schwinger:1968rq,Schwinger:1975ww}
implied by (\ref{max}), (\ref{con}), given by
\numparts
\bea
\left( \begin{array}{c}
j^{\prime}  \\  {\jg}^{\prime}
\end{array} \right)
=\left( \begin{array}{cc} \cos\theta &  \sin\theta \\
-\sin\theta & \cos\theta
\end{array} \right)
\left( \begin{array}{c} j  \\  {\jg} \end{array} \right),
\\
\left( \begin{array}{c}
F^{\prime} \\ {\fag}^{\prime}
\end{array} \right)
=\left( \begin{array}{cc} \cos\theta & \sin\theta  \\
-\sin\theta & \cos\theta
\end{array} \right)
\left( \begin{array}{c} F  \\  {\fag} \end{array} \right) ,
\eea
\endnumparts
suggests the introduction of an auxiliary vector potential $B_\mu(x)$
dual to $A_\mu(x)$.
In order to satisfy the Maxwell and charge conservation
equations, Dirac \cite{Dirac:1948um}  
modified the field strength tensor according to
\be
F_{\mu\nu}=\partial_{\mu}A_{\nu}-\partial_{\nu}A_{\mu}+\Gg_{\mu\nu},
\label{fs}
\ee
where now (\ref{max}) gives rise to the
consistency condition on $G_{\mu\nu}(x)=-G_{\nu\mu}(x)$
\be
\partial^{\nu}\ ^{\ast}F_{\mu\nu}=-\partial^{\nu}
G_{\mu\nu}=4\pi{\jg}_\mu .
\label{subc}
\ee
We then obtain the following inhomogeneous solution
to the dual Maxwell's equation (\ref{subc}) for the tensor
$G_{\mu\nu}(x)$ in terms of the string function $f_\mu$ and the magnetic
current $\jg_\nu$:
\bea
G_{\mu\nu}(x)=4\pi\left(n\cdot\partial\right)^{-1}
\left[n_{\mu}\jg_{\nu}(x)-n_{\nu}\jg_{\mu}(x)\right]
\nn
=\int \dy \left[f_{\mu}(x-y)\jg_{\nu}(y)
-f_{\nu}(x-y)\jg_{\mu}(y)\right],
\label{gten}
\eea
where use is made of (\ref{con}), (\ref{dfstng}), and (\ref{stngsl}).
A minimal generalization of the QED Lagrangian including
electron-monopole interactions reads
\bea
\mathcal{L}=-\frac{1}{16\pi}F_{\mu\nu}F^{\mu\nu}
+\bar\psi \left(\rmi\dslash+e\Aslash-m_{\psi}\right)\psi
+\bar\chi \left(i\dslash-m_{\chi}\right)\chi
\label{lag0} ,
\eea
where the coupling of the monopole field $\chi(x)$ to the electromagnetic
field occurs through the quadratic field strength term according to
(\ref{fs}).
We now rewrite the Lagrangian~(\ref{lag0})
to  display more clearly that interaction
by introducing the auxiliary potential $B_\mu(x)$.

Variation of (\ref{lag0}) with respect to the field variables,
$\psi$, $\chi$ and $A_\mu$, yields in addition to the Maxwell
equations for the field strength,
$F_{\mu\nu}$, (\ref{max}) where
$j^{\mu}(x)=e\bar{\psi}(x)\gamma^\mu\psi(x)$,
the equation of motion for the electron field
\be
\left(\rmi\dslash + e\Aslash(x) - m_{\psi}\right)\psi(x) =0,
\label{diracl}
\ee
and the nonlocal equation of motion for the monopole field,
\be
\left(\rmi\dslash - m_{\chi}\right)\chi(x)-\frac{1}{8\pi}
\int \left(\rmd y\right)
\fag^{\mu\nu}(y)\frac{\delta G_{\mu\nu}(y)}{\delta\bar\chi(x)} = 0\, .
\label{diracm}
\ee
[We regard $G_{\mu\nu}(x)$
as dependent on $\bar{\chi}$, $\chi$ but not $A_\mu$.  Thus,
the dual Maxwell equation is given by the subsidiary condition~(\ref{subc}).]
It is straightforward to see from the Dirac
equation for the monopole (\ref{diracm}) and the construction (\ref{gten})
that introducing the auxiliary dual field (which is a
functional  of $F_{\mu\nu}$ and depends on the string function $f_\mu$)
\be
B_{\mu}(x)=-\frac1{4\pi}\int \dy f^\nu\left(x-y\right)\fag_{\mu\nu}(y)\, ,
\label{gfb}
\ee
results in the following Dirac equation for the monopole field
\be
\left(\rmi\dslash +g\Bslash(x) - m_{\chi}\right)\chi(x)=0.
\label{diracm2}
\ee
Here we have chosen the string to satisfy the oddness condition
[this is the ``symmetric'' solution, generalizing \eref{strcon2}]
\be
f^\mu(x)=-f^\mu(-x),\label{symmstring}
\ee
which as we have seen is related to Schwinger's integer quantization condition 
\cite{Milton:1976jq,psf1}.
Now (\ref{diracl}) and (\ref{diracm2})
display the dual symmetry expressed in Maxwell's equations
(\ref{max}) and (\ref{con}).
Noting that  $B_\mu$   satisfies [like taking $\lambda_m=0$ in \eref{lambdam}]
\be
\int (\rmd x') f^{\mu}(x-x^{\prime})B_{\mu}(x^{\prime})=0,\label{lambdab}
\ee
we see that (\ref{gfb}) is a gauge-fixed vector field \cite{dir55,sis87}
defined in terms of the field strength through an {\em inversion}
formula (see \sref{ss:gauge}).
In terms of these fields the ``dual-potential''   action
can be re-expressed in terms of the vector potential $A_\mu$
and field strength tensor $F_{\mu\nu}$ (where $B_\mu$ is the functional
(\ref{gfb}) of $F_{\mu\nu}$) in first-order formalism as
\numparts
\bea
\fl W=\int\dx \bigg\{-\frac{1}{8\pi}F^{\mu\nu}(x)
\left(\partial_\mu A_\nu\left(x\right)-\partial_\nu A_\mu\left(x\right)\right)+
\frac{1}{16\pi}F_{\mu\nu}(x)F^{\mu\nu}(x)
\nn
\fl\mbox{}+\bar\psi(x)\left(\rmi\dslash+e{\Aslash}(x)
-m_{\psi}\right)\psi(x)
+\bar\chi(x)
\left(\rmi\dslash+g{\Bslash}(x)-m_{\chi}\right)\chi(x)\bigg\},
\label{act1}
\eea
or in terms of {\em dual\/} variables,
\bea
\fl W=\int \dx \bigg\{
-\frac{1}{8\pi}{\fag}^{\mu\nu}(x)\left(\partial_\mu B_\nu\left(x\right)
-\partial_\nu B_\mu\left(x\right)\right)
+\frac{1}{16\pi}{\fag}^{\mu\nu}(x){\fag}_{\mu\nu}(x)
\nonumber \\
\fl\mbox{}+\bar\psi(x)\left(\rmi\dslash+e{\Aslash}(x)-m_{\psi}\right)\psi(x)
+\bar\chi(x)\left(\rmi\dslash+g{\Bslash}(x)-m_{\chi}\right)\chi(x) \bigg\}.
\label{act2}
\eea
\endnumparts
In (\ref{act1}), $A_\mu(x)$ and $F_{\mu\nu}(x)$ are the independent
field variables 
and $B_\mu(x)$ is given
by (\ref{gfb}),  while in (\ref{act2}) the dual fields
are the independent variables, in which case,
\bea
\fl A_\mu(x)=-\frac1{4\pi}\int \dy f^\nu\left(x-y\right)\ F_{\mu\nu}(y)
={1\over8\pi}\epsilon_{\mu\nu\lambda\sigma}\int \dy f^\nu(x-y)\fag^{\lambda
\sigma}(y).
\label{gfa0}
\eea
[Note that (\ref{act2}) may be obtained from the form (\ref{act1}) by
inserting (\ref{gfa0}) into the former and then identifying $B_\mu$
according to the construction (\ref{gfb}).  In this way the sign of
${1\over16\pi}F_{\mu\nu}F^{\mu\nu}=-{1\over16\pi}{}^*F_{\mu\nu}{}^*F^{\mu\nu}$
is flipped.]
Consequently, the field equation relating $\fag^{\mu\nu}$ and $B^\mu$ is
\be
{\fag}_{\mu\nu}=\partial_{\mu}{B}_{\nu}-\partial_{\nu}{B}_{\mu}
-\int \dy \,{}^{\ast}\left(f_{\mu}(x-y)j_{\nu}(y)
-f_{\nu}(x-y)j_{\mu}(y)\right),
\label{dualfs}
\ee
which is simply obtained (\ref{fs}) by making the duality
transformation \eref{duality}.

\subsection{Quantization of dual QED: Schwinger-Dyson equations}
\label{ss:quant}
Although the various actions describing the interactions of
point electric and magnetic poles can be described in terms
of a set of Feynman rules which one conventionally uses
in perturbative calculations, the large value of
$\alpha_{g}$ or $eg$ renders them useless
for this purpose. In addition, calculations of physical processes using
the perturbative approach from string-dependent actions
such as (\ref{act1}) and (\ref{act2})
have led only to string dependent results \cite{Deans:1981qs}.
In conjunction with a nonperturbative
functional approach, however, the Feynman rules serve to elucidate
the electron-monopole interactions.  We express these interactions
in terms of the ``dual-potential'' formalism
as a quantum generalization of the relativistic
classical theory of section \ref{ssqft}.
We use the Schwinger action principle \cite{Schwinger:1951ex,Schwinger:1951hq}
to quantize the electron-monopole  system by
solving the corresponding Schwinger-Dyson equations for the
generating functional. Using a functional Fourier
transform of this generating functional in terms
of a path integral for the electron-monopole system,
we  rearrange the generating
functional into a form that is well-suited for the purpose of
nonperturbative calculations.

\subsubsection{Gauge symmetry}
\label{ss:gauge}

In order to construct the generating functional for Green's
functions in the electron-monopole system we must restrict
the gauge freedom resulting from the local gauge invariance
of the action (\ref{act1}).
The {\it inversion}
formulae for $A_\mu$ and  $B_\mu$, (\ref{gfa0}) and (\ref{gfb})
respectively, might suggest using the technique of gauge-fixed
fields \cite{dir55,Mandelstam:1962mi} as was adopted in \cite{Deans:1981qs}.
However, we use the technique of gauge
fixing according to methods outlined by Zumino~\cite{Zumino:1959wt,bb62}
and generalized by Zinn-Justin~\cite{zin86} in the language of
stochastic quantization.

The gauge fields are obtained in terms of the string and the gauge
invariant field strength, by contracting the field
strength (\ref{fs}), (\ref{gten}) with the Dirac string, $f^{\mu}(x)$,
in conjunction with (\ref{dfstng}), yielding
the following inversion formula for the equation of motion,
\be
A_\mu(x)=-\frac1{4\pi}\int \dxp f^{\nu}(x-x^{\prime})F_{\mu\nu}(x^{\prime})
+\partial_{\mu}\tilde{\Lambda}_{e}(x)
\label{sta},
\ee
where we use the suggestive notation, ${\tilde\Lambda}_{e}(x)$
\be
\tilde{\Lambda}_{e}(x)
=\frac1{4\pi}\int \dxp f^{\nu}(x-x^{\prime}) A_\nu\left(x^{\prime}\right).
\label{agf}
\ee
In a similar manner, given the dual field strength (\ref{dualfs})
the dual vector potential takes the following form [cf.~(\ref{gfb}), 
\eref{lambdab}]
\numparts
\be
B_{\mu}(x)
=-\frac1{4\pi}\int\dxp f^{\nu}(x-x^{\prime}){\fag}_{\mu\nu}(x^{\prime})+
\partial_{\mu}\tilde{\Lambda}_{g},
\label{gfb2}
\ee
where
\be
\tilde{\Lambda}_{g}(x)=\frac1{4\pi}\int\dxp f^{\mu}(x-\xp)B_{\mu}(x').
\label{bgf}
\ee
\endnumparts

It is evident that (\ref{sta}) transforms consistently under
a  gauge transformation
\be
A_\nu(x)\longrightarrow A_\nu(x)+\partial_{\nu}\Lambda_{e}(x),
\ee
while in addition we note that the Lagrangian  (\ref{act1})
is invariant under the gauge transformation,
\numparts
\be
\psi \rightarrow  {\exp}\left[\rmi e\Lambda_{e}\right]\psi,
\qquad
A_\mu \rightarrow  A_\mu + \partial_\mu\Lambda_{e} ,
\ee
as is the dual action (\ref{act2}) under
\be
\chi\rightarrow  {\rm exp}\left[\rmi g\Lambda_{g}\right]\chi,
\qquad
B_\mu \rightarrow  B_\mu + \partial_\mu\Lambda_{g}.
\ee\endnumparts
Assuming the freedom to  choose $\tilde{\Lambda}_e(x)=-\Lambda_{e}(x)$,
we bring the vector potential into gauge-fixed form,
coinciding with (\ref{gfa0}),
\be
A_{\mu}(x)=-\frac1{4\pi}\int \dy f^\nu\left(x-y\right)\ F_{\mu\nu}(y)
\label{gfa}
\ee 
where the gauge choice is equivalent
to a {\it string-gauge\/} condition
\be
\int\ \dxp f^{\mu}(x-x^{\prime})A_{\mu}(x^{\prime})=0.
\label{gfca}
\ee
[This is the analog of \eref{lambdab}, and is equivalent to the gauge
choice $\lambda_e=0$, see \eref{lambdae}, used in \sref{sec:strings}.
It is worth
noting the similarity of this condition to the Schwinger-Fock gauge
in ordinary QED, $x\cdot\A(x)=0$ which yields the gauge-fixed
photon field, $\A_{\mu}(x)=-x^\nu\int_0^1 ds\,s F_{\mu\nu}(xs)$.]
Taking the divergence of (\ref{gfa}) and using (\ref{max}),
the gauge-fixed condition (\ref{gfa}) can be written as
\be
\partial_\mu A^\mu = \int \dy f^\mu\left(x-y\right) j_{\mu}(y),
\label{stgc1}
\ee
which is nothing other than the gauge-fixed condition of Zwanziger
in the two-potential formalism \cite{Zwanziger:1970hk}.

More generally, the fact that a gauge function exists,
such that
${\Lambda}_e(x)=-\tilde\Lambda_e(x)$ [cf.~\eref{agf}], implying
that we have the freedom to consistently fix the gauge,
is in fact not a trivial claim. If this were not true, it
would certainly derail the consistency
of incorporating monopoles into QED while
utilizing the Dirac string  formalism.  On the contrary, the
{\it string-gauge condition} (\ref{gfca}) is in fact
a class of possible consistent gauge
conditions characterized by the symbolic operator function (\ref{stngsl})
depending on a unit vector $n^\mu$ (which may be either spacelike or 
timelike).

In order to quantize this system we must
divide out the equivalence class of field values
defined by a gauge trajectory in field space; in this
sense the gauge condition restricts the
vector potential to a hypersurface of field space
which is embodied in the generalization of (\ref{gfca})
\be
\frac1{4\pi}\int\dxp f^\mu(x-\xp)A_\mu(\xp)=\Lambda^f_{e}(x),
\label{stgc2}
\ee
where here $\Lambda^f_e$ is any function defining a unique gauge
fixing hypersurface in field space.

In a path integral formalism, we enforce the condition (\ref{stgc2})
by introducing a $\delta$ function, symbolically written as
\bea
\delta\left(\frac1{4\pi}f^{\mu}A_{\mu}-\Lambda^f_e\right)=\int
[\rmd\lambda_e]\exp\Bigg[\rmi\int\dx\lambda_e(x)\nn
\times\left(\frac1{4\pi}
\int\dxp f^{\mu}(x-\xp)A_{\mu}(\xp)-\Lambda^f_e(x)\right)\Bigg],
\label{delta}
\eea
or by introducing a Gaussian functional integral
\bea
\Phi\left(\frac1{4\pi}f^{\mu}A_{\mu}-\Lambda^f_e\right)=\int
[\rmd\lambda_e]\exp\Bigg[-\frac{\rmi}{2}\int\dx\dxp
\lambda_e(x)M(x,\xp)\lambda_e(x')
\nn
\mbox{}+\rmi\int\dx\lambda_e(x)\left(\frac1{4\pi}
\int\dxp f^{\mu}(x-\xp)A_{\mu}(\xp)-\Lambda^f_e(x)\right)\Bigg],
\label{gfcq}
\eea
where the symmetric matrix $M(x,\xp)=\kappa^{-1}\delta(x-\xp)$
describes the spread of the
integral $\int\dxp f^{\mu}(x-\xp)A_{\mu}(\xp)$ about the gauge
function, $\Lambda^f_e(x)$. That is, we  enforce the gauge fixing
condition (\ref{stgc2}) by adding the quadratic form appearing here
to the action (\ref{act1}) and in turn eliminating
$\lambda_{e}$ by its ``equation of motion''
\be
\lambda_{e}(x)=\kappa\left(\frac1{4\pi}\int (\rmd y)\,f^\mu(x-y)
A_\mu(y)-\Lambda^f_e(x)\right).
\label{agfl}
\ee
Now the equations of motion (\ref{max}) take the form,
\numparts
\bea
\partial^\nu F_{\mu\nu}(x)-\int(\rmd x') \lambda_e(\xp)f_\mu(\xp-x)=4\pi
j_\mu(x),
\label{gfmax1}
\\
\partial^\nu \fag_{\mu\nu}(x)-\int(\rmd x') \lambda_g(\xp)f_\mu(\xp-x)=4\pi
\jg_\mu(x),
\label{gfmax2}
\eea
\endnumparts
where the second equation refers to a similar gauge fixing in the dual sector.
Taking the divergence of (\ref{gfmax1}) implies $\lambda_e=0$
from (\ref{dfstng}) and (\ref{con}), which
consistently yields the gauge condition (\ref{stgc2}).
Using our freedom to make a transformation to the gauge-fixed condition
(\ref{gfa}), $\Lambda^f_e=0$, the equation of motion (\ref{gfmax1})
for the potential becomes
\be
\fl \Bigg[-g_{\mu\nu}\partial^2 +
\partial_\mu\partial_\nu +4\pi\kappa n_{\mu}(n\cdot\partial)^{-2}n_{\nu}
\Bigg]A^\nu(x)
= 4\pi j_\mu(x) +\epsilon_{\mu\nu\sigma\tau}
\frac{4\pi n^\nu}{\left(n\cdot\partial\right)}\partial^{\sigma}
\jg^\tau(x),
\label{difa2}
\ee
where we now have used the symbolic form of the string function~(\ref{stngsl}).
Even though \eref{agfl} now implies $n^\mu A_\mu=0$,
we have retained the term proportional to $n_\mu n_\nu$ in
 the  kernel, scaled by the arbitrary parameter $\kappa$,
\be
K_{\mu\nu}=
\Bigg[-g_{\mu\nu}\partial^2 +
\partial_\mu\partial_\nu+4\pi \kappa\, n_{\mu}(n\cdot\partial)^{-2}n_{\nu}
\Bigg],
\label{ker}
\ee
so that $K_{\mu\nu}$ possesses an inverse
\be
\fl D_{\mu\nu}(x)=\Bigg[g_{\mu\nu}
-\frac{n_{\mu}\partial_{\nu}+n_{\nu}\partial_{\mu}}{(n\cdot\partial)}
+n^{2}\left(1-\frac{1}{4\pi\kappa}\frac{(n\cdot\partial)^{2}\partial^{2}}
{n^{2}}\right)
\frac{\partial_{\mu}\partial_{\nu}}{(n\cdot\partial)^{2}}\Bigg]D_+(x),
\ee
that is, $\int\dxp K_{\mu\alpha}(x-\xp)D^{\alpha\nu}(\xp-\xpp)=
g_\mu^\nu\delta(x-x'')$,
where $D_+(x)$ is the massless scalar propagator,
\be
D_+(x)=\frac{1}{-\partial^{2}-\rmi \epsilon}\delta(x).
\ee
This in turn enables us to rewrite (\ref{difa2}) as
an integral equation, expressing the vector potential in terms of the
electron and monopole currents,
\bea
A_{\mu}(x)=4\pi\int\dxp D_{\mu\nu}(x-\xp)j^{\nu}(\xp)
\nn
\mbox{}+\epsilon^{\nu\lambda\sigma\tau}\int\dxp\dxpp
D_{\mu\nu}(x-\xp)f_{\lambda}(\xp-\xpp)\partial''_{\sigma}\jg_{\tau}(\xpp).
\label{inta}
\eea
The steps for $B_\mu(x)$ are analogous.

\subsubsection{Vacuum persistence amplitude and the path integral}
\label{ss:vpa}

Given the gauge-fixed but string-dependent action, we
are prepared to quantize this theory of dual QED.
Quantization using a path integral formulation of such
a string-dependent action is by no means straightforward;
therefore we will develop the generating functional making use of
a functional approach.
Using the quantum action principle
(cf.~\cite{Schwinger:1951ex,Schwinger:1951hq}) 
we write the generating functional 
for Green's functions (or the vacuum persistence
amplitude)   in the presence of external sources~$\K$,
\be
Z({\cal K})=
\langle 0_+\left|\right. 0_-\rangle^{\cal K},
\ee
for the electron-monopole system. Schwinger's action principle
states that under an arbitrary variation
\be
\delta\langle 0_+\left|\right. 0_-\rangle^{\K}=
\rmi\langle 0_+\left| \delta W(\K)\right| 0_-\rangle^{\K}\, ,
\ee
where $W(\K)$ is the action given in (\ref{act1}) externally
driven by the sources, $\K$, which for the present case are
given by the set $\{J,\Jg,\bar{\eta},\eta,\bar{\xi},\xi\}$:
\be
W({\cal K})=W+\int (\rmd x)\left\{J^\mu A_\mu+{}^* J^\mu B_\mu+\bar\eta\psi
+\bar\psi\eta+\bar\xi\chi+\bar\chi\xi\right\},
\ee
$\eta$ ($\bar\eta$), $\xi$ ($\bar\xi$) being the sources for electrons
(positrons) and monopoles (antimonopoles), respectively.
The one-point functions are then given by
\bea
\fl \frac{\delta}{\rmi\delta J^\mu(x)}
\log Z(\K)=
\frac{\langle 0_+|A_\mu(x)| 0_-\rangle^\K}
{\langle 0_+\left|\right. 0_-\rangle^{\cal K}}, \qquad
\frac{\delta}{\rmi\delta\Jg^\mu(x)}
\log Z(\K)=
\frac{\langle 0_+|B_\mu(x)| 0_-\rangle^\K}
{\langle 0_+\left|\right. 0_-\rangle^{\cal K}},
\nn
\fl\frac{\delta}{\rmi\delta\bar{\eta}(x)}
\log Z(\K)=
\frac{\langle 0_+| \psi(x)| 0_-\rangle^\K}
{\langle 0_+\left|\right. 0_-\rangle^{\cal K}}, \qquad
\frac{\delta}{\rmi\delta\bar{\xi}(x)}
\log Z(\K)=
\frac{\langle 0_+|\chi(x)| 0_-\rangle^\K}
{\langle 0_+\left|\right. 0_-\rangle^{\cal K}}.
\label{onept}
\eea
Using (\ref{onept}) we can write down
derivatives with respect to the charges (here
we redefine the electric and magnetic currents
$j\rightarrow ej$ and $\jg\rightarrow g\jg$) 
in terms of functional derivatives~\cite{Schwinger:1960qe,som63,jon65}
with respect to the external sources;
\bea
\fl\frac{\partial}{\partial e}
\langle 0_+| 0_-\rangle^{\K}
=\rmi\langle 0_+\big|\int\dx j^\mu(x)A_\mu(x)\big| 0_-\rangle^{\K}
=-\rmi\int\dx\left(\frac{\delta}{\delta\tA_\mu(x)}
\frac{\delta}{\delta J^\mu(x)}\right)
\langle 0_+\big| 0_-\rangle^{\K},
\nn
\fl\frac{\partial}{\partial g}\langle 0_+| 0_-\rangle^{\K}=
\rmi\langle 0_+\big|\int\dx \jg^\mu(x)B_\mu(x)\big| 0_-\rangle^{\K}
=-\rmi\int\dx\left(\frac{\delta}{\delta\tB_\mu(x)}
\frac{\delta}{\delta \Jg^\mu(x)}\right)
\langle 0_+\big| 0_-\rangle^{\K}.\nn
\eea
Here we have introduced an effective source to bring down the electron
and monopole currents,
\be
{\delta\over\delta \tA_\mu}\equiv{1\over \rmi}{\delta\over\delta \eta}\gamma^\mu
{\delta\over\delta \bar \eta},\qquad
{\delta\over\delta \tB_\mu}\equiv{1\over \rmi}{\delta\over\delta \xi}\gamma^\mu
{\delta\over\delta \bar \xi}.
\label{effcurrsource}
\ee
These first order differential equations can be integrated with the result
\bea
\langle 0_+\left|\right. 0_-\rangle^{\K}=
\exp\Bigg[-\rmi g\int\dx\left(\frac{\delta}{\delta\tB_\nu(x)}
\frac{\delta}{\delta\Jg^\nu(x)}\right)\nn
-\rmi e\int\dx\left(\frac{\delta}{\delta\tA_\mu(x)}
\frac{\delta}{\delta J^\mu(x)}\right)\Bigg]
\langle 0_+\left|\right. 0_-\rangle^{\K}_0,
\label{vacint}
\eea
where $\langle 0_+\left|\right. 0_-\rangle^{\K}_0$ is the vacuum
amplitude in the absence of interactions.
By construction, the vacuum amplitude and Green's functions
for the coupled problem are determined by functional derivatives
with respect to the external sources $\K$ of the uncoupled vacuum amplitude,
where $\langle 0_+\left|\right. 0_-\rangle^{\K}_0$ is the product
of the separate amplitudes for the quantized electromagnetic
and Dirac fields since they constitute completely
independent systems in the absence of coupling, that is,
\be
\langle 0_+\left|\right. 0_-\rangle^{\K}_0=
\langle 0_+\left|\right. 0_-\rangle^{(\bar\eta,\eta,\bar\xi,\xi)}_{0}
\langle 0_+\left|\right. 0_-\rangle^{(J,\Jg)}_{0}.
\label{fulfregen}
\ee
First we consider $\langle 0_+\left|\right. 0_-\rangle^{\K}_0$
as a function of $J$ and $\Jg$
\be
\frac{\delta}{\rmi\delta J^\mu(x)}\langle 0_+\left|\right. 0_-\rangle^\K_{0}=
\langle 0_+| A_\mu(x)| 0_-\rangle^\K_{0}.
\ee
Taking the matrix element of the integral
equation (\ref{inta}) but now with external sources rather than
dynamical currents we find
\bea
\fl\langle 0_+| A_\mu(x)| 0_-\rangle^\K_0=
\int\dxp D_{\mu\nu}(x-\xp)\left(4\pi J^{\nu}(\xp)
+\epsilon^{\nu\lambda\sigma\tau}\int\dxpp
f_{\lambda}(\xp-\xpp)\partial''_{\sigma}
\Jg_{\tau}(\xpp)\right)\nn
\times\langle 0_+|0_-\rangle_0^{\cal K}.
\label{qinta}
\eea
Using (\ref{difa2}) we arrive at the equivalent
gauge-fixed functional equation,
\bea
\Bigg[-g_{\mu\nu}\partial^2
+\partial_{\mu}\partial_{\nu}
+4\pi \kappa n_{\mu}(n\cdot\partial)^{-2}n_{\nu}
\Bigg]\frac{\delta}{\rmi\delta J^{\nu}(x)}
\langle 0_+\left|\right. 0_-\rangle^{\K}_0
\nn
=\left(4\pi J_{\mu}(x)+\epsilon_{\mu\nu\sigma\tau}\int\dxp
f^{\nu}(x-\xp)\partial^{\prime\sigma}\Jg^{\tau}(\xp)\right)
\langle 0_+\left|\right. 0_-\rangle^{\K}_0,
\label{qeqaf}
\eea
which is subject to the gauge condition
\numparts
\be
n^\nu{\delta\over\delta J^\nu}\langle 0_+|0_-\rangle_0^{\cal K}=0,
\ee
or
\be
\int (\rmd x')f^\nu(x-x'){\delta\over\delta J^\nu(x')}
\langle 0_+|0_-\rangle_0^{\cal K}=0.
\label{gcone}
\ee
\endnumparts
In turn, from (\ref{vacint})  we obtain the full functional equation for
$\langle 0_+\left|\right. 0_-\rangle^{\K}$:
\bea
\Bigg[-g_{\mu\nu}\partial^2
+\partial_{\mu}\partial_{\nu}
+4\pi \kappa n_{\mu}(n\cdot\partial)^{-2}n_{\nu}
\Bigg]\frac{\delta}{\rmi\delta J^\nu(x)}\langle 0_+\left|\right. 
0_-\rangle^{\K}
\nn
=\exp\left[-\rmi g\int\dy
\left(\frac{\delta}{\delta\tB_\alpha(y)}
\frac{\delta}{\delta \Jg^\alpha(y)}\right)
-\rmi e\int\dy
\left(\frac{\delta}{\delta\tA_\alpha(y)}
\frac{\delta}{\delta J^\alpha(y)}\right)\right]
\nn
\times\left(4\pi
J_{\mu}(x) +\epsilon_{\mu\nu\sigma\tau}\int\dxp
f^{\nu}(x-\xp)\partial^{\prime\sigma}\Jg^{\tau}(\xp)\right)
\langle 0_+\left|\right. 0_-\rangle^{\K}_0.
\label{fncteq1}
\eea
Commuting the external currents to the left of the exponential on
the right side of (\ref{fncteq1})
and using (\ref{onept}),
we are led to the Schwinger-Dyson equation for the vacuum amplitude,
where we have restored the meaning of the functional derivatives with
respect to $\tilde A$, $\tilde B$ given in (\ref{effcurrsource}),
\bea
\fl\Bigg\{\bigg[-g_{\mu\nu}\partial^2
+\partial_{\mu}\partial_{\nu}
+4\pi \kappa n_\mu(n\cdot\partial)^{-2}n_\nu
\bigg]\frac{\delta}{\rmi\delta J_\nu(x)}
\nn
\fl\mbox{}- 4\pi e\frac{\delta}{\rmi\delta\eta(x)}\gamma_\mu
\frac{\delta}{\rmi\delta\bar\eta(x)}
-\epsilon_{\mu\nu\sigma\tau}\int \dxp
f^{\nu}(x-\xp)\partial^{\prime\sigma}g
\frac{\delta}{\rmi\delta\xi(x')}\gamma^{\tau}
\frac{\delta}{\rmi\delta\bar\xi(x')}
\Bigg\}\langle 0_+\left|\right. 0_-\rangle^{\K}\nn
=\left(4\pi 
J_\mu(x)+\epsilon_{\mu\nu\sigma\tau}\int(\rmd x')f^\nu(x-x')\partial^{
\prime\sigma}\Jg^\tau(x')\right)\langle 0_+\left|\right. 0_-\rangle^{\K}
\label{dsi1}.
\eea
In an analogous manner, using
\be
\frac{\delta}{\rmi\delta \Jg^\mu(x)}\langle 0_+\left|\right. 0_-\rangle^\K_{0}=
\langle 0_+| B_\mu(x)| 0_-\rangle^\K_{0},
\ee
we obtain the functional  equation (which is consistent with duality)
\bea
\fl\Bigg\{\bigg[-g_{\mu\nu}\partial^2
+\partial_{\mu}\partial_{\nu}
+4\pi \kappa n_{\mu}(n\cdot\partial)^{-2}n_{\nu}
\bigg]\frac{\delta}{\rmi\delta \Jg_\nu(x)}
\nn
\fl\mbox{}- 4\pi g\frac{\delta}{\rmi\delta\xi(x)}\gamma_\mu
\frac{\delta}{\rmi\delta\bar\xi(x)}
+\epsilon_{\mu\nu\sigma\tau}\int \dxp
f^{\nu}(x-\xp)\partial^{\prime\sigma}
e\frac{\delta}{\rmi\delta\eta(x')}\gamma^{\tau}
\frac{\delta}{\rmi\delta\bar\eta(x')}
\Bigg\}\langle 0_+\left|\right. 0_-\rangle^{\K}
\nn
=\left(4\pi
\Jg_\mu(x)-\epsilon_{\mu\nu\sigma\tau}\int(\rmd x')f^\nu(x-x')\partial^{\prime
\sigma} J^\tau(x')\right)\langle 0_+\left|\right. 0_-\rangle^{\K},
\label{dsi2}
\eea
which is subject to the gauge condition
\be
\int(\rmd x')f^\mu(x-x')
{\delta\over\delta \Jg^\mu(x')}\langle 0_+|0_-\rangle^{
\cal K}=0.
\label{gctwo}
\ee
\numparts
In a straightforward manner we obtain the functional Dirac equations
\bea
\left\{\rmi\dslash+e\gamma^\mu\frac{\delta}{\rmi\delta J^\mu(x)}
-m_\psi\right\}
\frac{\delta}{\rmi\delta\bar{\eta}(x)}\langle 0_+\left|\right. 0_-\rangle^{\K}
=-\eta(x)\langle 0_+\left|\right. 0_-\rangle^{\K}
\label{dsi3},\\
\left\{\rmi\dslash+g\gamma^\mu\frac{\delta}{\rmi\delta\Jg^\mu(x)}
-m_\chi\right\}
\frac{\delta}{\rmi\delta\bar{\xi}(x)}
\langle 0_+\left|\right. 0_-\rangle^{\K}=-\xi(x)
\langle 0_+\left|\right. 0_-\rangle^{\K}
\label{dsi4}.
\eea
\endnumparts

In order to obtain a generating functional for Green's functions
we must solve the set of
equations~(\ref{dsi1}), (\ref{dsi2}), (\ref{dsi3}), (\ref{dsi4})
subject to (\ref{gcone}) and (\ref{gctwo}) for
$\langle 0_+\left|\right. 0_-\rangle^{\J}$.
In the absence of interactions, we can immediately
integrate the  Schwinger-Dyson equations;
in particular, (\ref{dsi2}) then integrates to
\bea
\langle 0_+\left|\right. 0_-\rangle^{J,\Jg}_{0}
=\N(J)\exp\bigg\{2\pi
\rmi\int(\rmd x)(\rmd x')\Jg_\mu\left(x\right)D^{\mu\nu}
(x-\xp)\Jg_\nu(\xp)
\nonumber \\
\mbox{}+ \rmi\epsilon_{\mu\nu\sigma\tau}
\int (\rmd x)(\rmd x')(\rmd x'')
\Jg_\beta(x)D^{\beta\mu}(x-\xp)\partial^{\prime\nu}
f^\sigma(\xp-\xpp) J^\tau(\xpp)\bigg\}.
\label{gen1}
\eea
We determine $\N$, which depends only on $J$, by inserting
(\ref{gen1}) into (\ref{dsi1}) or (\ref{qeqaf}):
\be
\ln\N(J)=2\pi\rmi\int\dx\dxp J_{\mu}(x)D^{\mu\nu}(x-\xp)J_{\nu}(\xp),
\ee
resulting in the generating functional for the photonic sector
\bea
\fl\langle 0_+\left|\right. 0_-\rangle^{(J,\Jg)}_{0}
=\exp \bigg\{2\pi\rmi
\int (\rmd x)(\rmd x') J_\mu(x)D^{\mu\nu}(x-x^\prime)J_\nu(x^\prime)\nonumber\\
\mbox{}+2\pi\rmi\int(\rmd x)(\rmd x'){\Jg}_{\mu}(x)
D^{\mu\nu}(x-x^\prime){\Jg}_{\nu}(x^\prime)\nn
\mbox{}-4\pi\rmi\int(\rmd x)(\rmd x')J_\mu(x)\tD^{\mu\nu}(x-\xp){\Jg}_\nu(\xpp)
\bigg\},
\label{phot}
\eea
where we use the shorthand notation for the ``dual
propagator'' that couples magnetic to electric charge
\be
\tD_{\mu\nu}\left(x-x'\right)=\frac1{4\pi}\epsilon_{\mu\nu\sigma\tau}
\int (dx'')
D_+\left(x-x''\right)
\partial^{\prime\prime\sigma} f^\tau\left(x''-x'\right).
\label{dualp}
\ee
The term coupling electric and magnetic sources has the same form as in
(\ref{interact}); here,  we have replaced
$D^{\kappa\mu}\to g^{\kappa\mu}D_+$, 
because of the appearance of the Levi-Civit\`a symbol in (\ref{dualp}).
[Of course, we may replace $D^{\mu\nu}\to g^{\mu\nu} D_+$ throughout
\eref{phot}, because the external sources are conserved, $\partial_\mu J^\mu
=\partial_\mu\Jg^\mu=0$.]
In an even more straightforward manner (\ref{dsi3}), (\ref{dsi4})
integrate to
\bea
\fl\langle 0_+\left|\right. 0_-\rangle^{(\bar{\eta},\eta,\bar{\xi},\xi)}_0=
\exp \left\{\rmi\int (\rmd x)(\rmd x')
\left[\bar\eta(x) G_{\psi}(x -x')\eta(x') + \bar\xi(x)
G_{\chi}(x -x')\xi(x')\right]\right\},\nn
\label{fermi}
\eea
where $G_{\psi}$ and $G_{\chi}$ are the free propagators for the
electrically and magnetically charged fermions, respectively,
\be
G_{\psi}(x)=\frac{1}{-\rmi\dslash +m_\psi}\delta(x),\qquad
G_{\chi}(x)=\frac{1}{-\rmi\dslash +m_\chi}\delta(x).
\ee
In the presence of interactions the coupled
equations~(\ref{dsi1}), (\ref{dsi2}), (\ref{dsi3}), (\ref{dsi4}) are
solved by substituting (\ref{phot}) and (\ref{fermi})
into (\ref{vacint}).
The resulting generating function is
\bea
Z(\K)
=\exp\left(-\rmi e\int (\rmd x)\frac{\delta}
{\delta\eta(x)}\gamma^{\mu}\frac{\delta}{\rmi\delta J^\mu(x)}
\frac{\delta}{\delta\bar\eta(x)}\right)\nn
\times\exp\left(-\rmi g\int(\rmd y)\frac{\delta}{\delta\xi(y)}\gamma^{\nu}
\frac{\delta}{\rmi\delta\Jg^\nu(y)}\frac{\delta}{\delta\bar\xi(y)}\right)
Z_0(\K).
\label{genfunct}
\eea

\subsubsection{Nonperturbative generating functional}
\label{ss:np}

Due to the fact that any expansion in $\alpha_g$ or $eg$
is not practically useful we recast the generating
functional (\ref{genfunct}) into a functional
form better suited for a nonperturbative calculation of the four-point
Green's function.

First we utilize the well-known Gaussian combinatoric
relation \cite{sym54,fri90};  moving
the exponentials containing the interaction vertices in
terms of functional derivatives with respect to fermion sources
past the free fermion propagators, we obtain (coordinate labels are now
suppressed)
\bea
\fl Z(\K)=
\exp\left\{\rmi\int\bar\eta\left(G_{\psi}
\left[1-e\gamma\cdot\frac{\delta}{\rmi\delta J}G_{\psi}\right]^{-1}\right)\eta
+\Tr\ln\left(1-e\gamma\cdot\frac{\delta}{\rmi\delta J}G_{\psi}\right)\right\}
\nonumber \\
\fl\times\exp\left\{\rmi\int\bar\xi\left(G_{\chi}
\left[1-g\gamma\cdot\frac{\delta}{\rmi\delta \Jg}G_{\chi}\right]^{-1}\right)\xi
+\Tr\ln\left(1-g\gamma\cdot\frac{\delta}{\rmi\delta \Jg}G_{\chi}\right)\right\}
Z_0(J,\Jg).
\label{functfull}
\eea
Now, we re-express (\ref{phot}), the noninteracting
part of the generating functional of the photonic action, $Z_0(J,\Jg)$,
using a functional Fourier transform,
\numparts
\be
Z_{0}(J,\Jg)=\int
\left[\rmd A\right]\left[\rmd B\right]\tilde{Z_{0}}\left(A,B\right)
\exp\left[\rmi\int\left(J\cdot A +
\Jg\cdot B\right)\right],
\label{piphot1}
\ee
or
\be
Z_0(J,\Jg)
=\int\, \left[\rmd A\right]\left[\rmd B\right]
\exp\left(\rmi\Gamma_0[A,B,J,\Jg]\right),
\ee
\endnumparts
where (using a matrix notation for integration over coordinates)
\be
\fl\Gamma_0[A,B,J,\Jg]=
\int\left(J\cdot A +\Jg \cdot B\right)
-\frac{1}{8\pi}\int A^{\mu} K_{\mu\nu} A^{\nu}
+\frac{1}{8\pi}\int
 B^{\prime\mu}\tilde{\Delta}^{-1}_{\mu\nu}
B^{\prime\nu},
\label{onepi}
\ee
with the abbreviation
\be
B^{\prime}_\mu(x)=B_\mu(x)-\frac1{4\pi}
\epsilon_{\mu\nu\sigma\tau}\int\dxp\partial^{\nu}f^{\sigma}(x-\xp) A^\tau(\xp)
\label{bprime}
\ee
and the string-dependent ``correlator''
\be
\fl\tilde{\Delta}_{\mu\nu}(x-\xp)=\frac1{(4\pi)^2}\int \dxpp
\left\{f^{\sigma}(x-\xpp)f_{\sigma}\left(\xpp-\xp\right)g_{\mu\nu}
-f_{\mu}\left(x-\xpp\right)f_{\nu}\left(\xpp-\xp\right)\right\}.
\label{strprp}
\ee
Using (\ref{onepi}) we recast (\ref{functfull}) as
\be
Z(\K)=
\int\, \left[\rmd A\right]\left[\rmd B\right]
 F_{1}(A)F_{2}(B)\exp\left(\rmi\Gamma_0[A,B,J,\Jg]\right).
\ee
Here the fermion functionals $F_1$ and $F_2$ are obtained by the replacements
${\delta\over\rmi \delta J}\to A$, ${\delta\over\rmi\delta\Jg}\to B$:
\numparts
\bea
\fl F_{1}(A)=\exp\left\{\Tr\ln\left(1-e\gamma\cdot A G_{\psi}\right)+
\rmi\int\bar\eta\left(G_{\psi}
\left[1-e\gamma\cdot A G_{\psi}\right]^{-1}\right)\eta\right\},
\\
\fl F_{2}(B)=\exp\left\{\Tr\ln\left(1-g\gamma\cdot B G_{\chi}\right) +
\rmi\int\bar\xi\left(G_{\chi}
\left[1-g\gamma\cdot B G_{\chi}\right]^{-1}\right)\xi\right\}.
\label{loops}
\eea\endnumparts
We perform a change of variables by shifting about the stationary
configuration of the effective action, $\Gamma_0[A,B,J,\Jg]$:
\be
A_\mu(x)=\bar{A}_\mu(x)+\phi_\mu(x),
\qquad
B'_\mu(x)=\bar{B}'_\mu(x)+\phi^\prime_\mu(x)
\label{exp}
\ee
where $\bar{A}$ and $\bar{B}$ are given by the solutions to
\be
\frac{\delta\Gamma_0\left(A,B,J,\Jg\right)}{\delta A^{\tau}}=0,
\qquad
\frac{\delta\Gamma_0\left(A,B,J,\Jg\right)}{\delta B^{\tau}}=0,
\ee
namely (most easily seen by regarding $A$ and $B'$ as independent variables),
\numparts
\bea
\fl\bar{A}_\mu(x)=\int\dxp D_{\mu\kappa}(x-\xp)\bigg(4\pi J^\kappa(\xp)
-\epsilon^{\kappa\nu\sigma\tau}\int\dxpp\partial^{\prime}_\nu
f_\sigma(\xp-\xpp)\Jg_\tau(\xpp)\bigg),\label{bgsol0}
\\
\fl\bar{B}_\mu(x)=\int\dxp D_{\mu\kappa}(x-\xp)\bigg(4\pi \Jg^\kappa(\xp)
+\epsilon^{\kappa\nu\sigma\tau}\int\dxpp\partial^{\prime}_\nu
f_\sigma(\xp-\xpp)J_\tau(\xpp)\bigg),
\label{bgsol}
\eea\endnumparts
reflecting the form of (\ref{inta}) and its dual.
Note that the solutions \eref{bgsol0},
(\ref{bgsol}) respect the dual symmetry, which is not
however manifest in the form of the effective action (\ref{onepi}).
Using the properties of Volterra expansions for functionals
and performing the resulting quadratic integration over
$\phi(x)$ and $\phi^\prime(x)$ 
we obtain a rearrangement of the generating functional
for the monopole-electron system that is well suited for
nonperturbative calculations:
\bea
\fl\frac{Z(\K)}{Z_0(J,\Jg)}=
\exp\Bigg\{2\pi\rmi\int(\rmd x)(\rmd x')
\Bigg(\frac{\delta}{\delta\bar{A}_{\mu}(x)}
 D_{\mu\nu}(x-x')\frac{\delta}{\delta\bar{A}_{\nu}(x')}\nn
\fl\mbox{}+\frac{\delta}{\delta\bar{B}_{\mu}(x)}
 D_{\mu\nu}(x-x')\frac{\delta}{\delta\bar{B}_{\nu}(x')}\Bigg)
-4\pi\rmi\int(\rmd x)(\rmd x')\frac{\delta}{\delta\bar{A}_{\mu}(x)}
\tD_{\mu\nu}\left(x-\xp\right)
\frac{\delta}{\delta\bar{B}_{\nu}(\xp)}\Bigg\}
\nn
\fl\times\exp\bigg\{ \rmi\int(\rmd x)(\rmd x')\bar{\eta}\left(x\right)
G(x,\xp|\bar{A})\eta\left(\xp\right)
+\rmi\int(\rmd x)(\rmd x')\bar{\xi}\left(x\right)
G(x,\xp|\bar{B})\xi\left(\xp\right)\bigg\}
\nn
\times \exp\bigg\{-\int_{0}^{e} \rmd e^{\prime} \,
\Tr\gamma\bar{A} G(x,x|\bar{A})
-\int_{0}^{g} \rmd g^{\prime}\, \Tr\gamma\bar{B}
G(x,x|\bar{B})\bigg\}.
\label{master}
\eea
Here the two-point fermion Green's functions $G(x_1,y_1|\bar{A})$, and
$G(x_2,y_2|\bar{B})$ in the background of the stationary
photon field $\bar{A},\bar{B}$ are given by
\numparts
\bea
G(x,x'|\bar{A})=\langle x|
(\gamma p +m_\psi-e\, \bAslash)^{-1}| x'\rangle \, ,
\\
G(x,x'|\bar{B})=\langle x|
(\gamma p +m_\chi-g\, \bBslash)^{-1}|x'\rangle\, ,
\eea
\endnumparts
where the trace includes integration over spacetime.
This result is equivalent to the functional
Fourier transform given in (\ref{piphot1}) including the fermionic
monopole-electron system:
\bea
Z({\cal K})=\int
\left[\rmd A\right]\left[\rmd B\right]
\det\left(-\rmi\gamma D_A+m_\psi\right) \det\left(-\rmi\gamma D_B+m_\chi\right)
\nn
\times
\exp\Bigg\{ \rmi\int(\rmd x)(\rmd x')\bigg(\bar{\eta}\left(x\right)
G(x,\xp|A)\eta\left(\xp\right) + \bar{\xi}\left(x\right)
G(x,\xp|B)\xi\left(\xp\right)\bigg)\Bigg\}
\nn
\times
\exp\left\{-\frac{\rmi}{8\pi}\int\left( A^{\mu} K_{\mu\nu} A^{\nu}-
 B^{\prime\mu}\tilde{\Delta}^{-1}_{\mu\nu}B^{\prime\nu}\right)
+ \rmi\int\left(J\cdot A +\Jg \cdot B\right)\right\},
\label{world}
\eea
where we have integrated over the fermion degrees of freedom.

Finally, from our knowledge of the manner in which electric
and magnetic charge couple to photons through Maxwell's equations
we can immediately write the generalization of (\ref{master})
for dyons, the different species of which are labeled by the index $a$:
\bea
Z(\K)=\exp\Bigg\{2\pi\rmi
\int(\rmd x)(\rmd x')\J^\mu(x)\D_{\mu\nu}\left(x-x^\prime\right)
\J^\nu\left(x^\prime\right)\Bigg\}
\nn
\times
\exp\Bigg\{2\pi\rmi\int(\rmd x)(\rmd x')
\frac{\delta}{\delta\bar{\A}_{\mu}(x)}
\D_{\mu\nu}(x-x')\frac{\delta}{\delta\bar{\A}_{\nu}(x')}\Bigg\}
\nn
\times
\exp\bigg\{ \rmi\sum_a\int(\rmd x)(\rmd x')\bar{\zeta}_a\left(x\right)
G_a(x,\xp|\bar{\A_a})\zeta_a\left(\xp\right)\bigg\}
\nn
\times
\exp\bigg\{-\sum_a\int_{0}^{1} \rmd q \,
\Tr\gamma\bar{\A_a} G_a(x,x|q\bar{\A_a})\bigg\}.
\label{masterd}
\eea
where $\A_a=e_aA+g_aB$, $\zeta_a$ is the source for the dyon of species $a$,
and a matrix notation is adopted,
\numparts
\be
\J^{\mu}(x)=
\left(
\begin{array}{c}
J(x)\\ \Jg(x)
\end{array}
\right),
\qquad
\frac{\delta}{\delta\bar{\A}_{\mu}(x)}=
\left(
\begin{array}{c}
\delta/\delta\bar{A}_{\mu}(x) \\
\delta/\delta\bar{B}_{\mu}(x)
\end{array}
\right),
\ee
\endnumparts
and
\numparts
\bea
\D_{\mu\nu}\left(x-x^\prime\right)=
\left(
\begin{array}{cc}
D_{\mu\nu}\left(x-x^\prime\right) & -\tD_{\mu\nu}\left(x-x^\prime\right) \\
\tD_{\mu\nu}\left(x-x^\prime\right) & D_{\mu\nu}\left(x-x^\prime\right)
\end{array}
\right).
\eea
\endnumparts

\subsubsection{High energy scattering cross section}
\label{s:ea}
In this subsection we provide evidence for  the string independence
of the dyon-dyon and charge-monopole (the latter being a
special case of the former) scattering cross section.  We will use
the generating functional (\ref{masterd}) developed in the last
subsection to calculate the  scattering cross section
nonperturbatively.  We are not able in general to demonstrate
the phenomenological string invariance of the scattering cross section.
However, it appears that in much the same manner as the Coulomb phase
arises as a soft effect in high energy charge scattering,
the string dependence arises from the exchange of soft
photons, and so in an appropriate eikonal approximation, the string-dependence
appears only as an unobservable phase.

To calculate the dyon-dyon scattering cross section
we obtain the four-point Green's
function for this process from (\ref{masterd})
\be
G(x_1,y_1;x_2,y_2)=
\frac{\delta}{\rmi\delta\bar{\zeta}_1(x_1)}
\frac{\delta}{\rmi\delta{\zeta}_1(y_1)}
\frac{\delta}{\rmi\delta\bar{\zeta}_2(x_2)}
\frac{\delta}{\rmi\delta{\zeta}_2(y_2)}
Z(\K)\bigg|_{{\cal K}=0}.
\label{gnfnct3}
\ee
The subscripts on the sources refer to the two different dyons.

\begin{figure}
\centerline{
\epsfig{figure=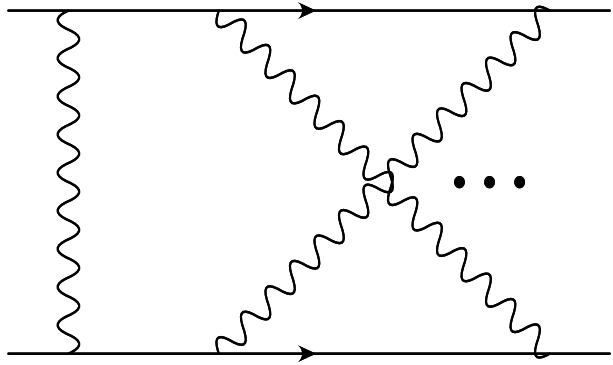,height=5cm,width=8cm}}
\caption{Dyon-dyon scattering amplitudes  in the quenched approximation.}
\label{figlink}
\end{figure}

Here we confront our calculational limits; these are not too
dissimilar from those encountered in diffractive scattering  or
in the strong-coupling regime
of QCD~\cite{Nachtmann:1991ua,Korchemsky:1993hr,Korchemskaya:1994qp,%
Gellas:1998sh,Karanikas:1998tn}.
As a first step in analyzing the string dependence
of the scattering amplitudes,
we study high-energy forward scattering
processes where {\em soft\/} photon contributions dominate.
In diagrammatic language, in this kinematic regime
it is customary to restrict attention to
that subclass in which there are no closed fermion loops
and the photons are exchanged
between fermions~\cite{Nachtmann:1991ua}.  In the context of Schwinger-Dyson
equations this amounts to quenched or ladder
approximation (see \fref{figlink}).
In this approximation
the linkage operators, $\L$,
connect two fermion propagators
via photon exchange, as we read off from (\ref{masterd}):
\be
\e^{\L_{12}} =\exp\Bigg\{4\pi\rmi
\int \dx\dxp\frac{\delta}{\delta\bar{\A}^{\mu}_1(x)}
\D^{\mu\nu}\left(x-\xp\right)
\frac{\delta}{\delta \bar{\A}^{\nu}_2(\xp)}\Bigg\}.
\label{link}
\ee
In this approximation (\ref{gnfnct3}) takes the form
\be
G\left(x_1,y_1;x_2,y_2\right)=
-\e^{\L_{12}}
G_1(x_1,y_1|\bar\A_1)G_2(x_2,y_2|\bar\A_2)\Big|_{\bar{A}=\bar{B}=0},
\label{scatt}
\ee
where we express the two-point function using the proper-time
parameter representation of an ordered exponential
\be
\fl G_a(x,y|\bar{\A}_a)=\rmi\int_{0}^{\infty}\rmd\xi\,
\rme^{-\rmi\xi\left(m_{a}-\rmi\gamma\partial\right)}
\exp\left\{{\rmi\int_{0}^{\xi}\rmd\xi^{\prime}{\rme}^{\xi^{\prime}
\gamma\partial}\gamma\bar\A_a{\rme}^{-\xi^{\prime}
\gamma\partial}}\right\}_{+}\delta(x-y),
\label{gextful}
\ee
where ``$+$'' denotes path ordering in $\xi^{\prime}$.
The 12 subscripts in $\L_{12}$ emphasize that only photon lines that
link the two fermion lines are being considered.

Adapting techniques outlined in \cite{Quiros:1976fk,Fried:1996uv} we consider
the connected form of (\ref{scatt}).
We use the connected two-point function and the identities
\be
\e^{\L}=1+\int_0^1 \rmd a\, \e^{a\L}\L
\label{par}
\ee
and
\be
\frac{\delta}{\delta\bA_{\mu}(x)} G(y,z|\bar{\A})
=e G(y,x|\bar{\A})\gamma^\mu
G(x,z|\bar{\A}).
\label{gderiv}
\ee
Using (\ref{scatt}) and (\ref{gextful})
one straightforwardly  is led to the
following representation of the four-point
Green function,
\bea
\fl G(x_{1},y_{1};x_{2},y_{2})=
-4\pi\rmi\int_0^1 \rmd a\int \dzi\dzii\left(\cq\,
D_{\mu\nu}(z_1-z_2)
-\wq\tD_{\mu\nu}(z_1-z_2)\right)
\nn
\fl\times\e^{a\L_{12}}
G_1(x_{1},z_1|\bar{\A}_1)\gamma^\mu
G_1(z_1,y_{1}|\bar{\A}_1)
G_2(x_{2},z_2|\bar{\A}_2)\gamma^\nu
G_2(z_2,y_{2}|\bar{\A}_2)\Bigg|_{\bar A=\bar B=0},
\label{mxptl}
\eea
where the charge combinations  invariant under duality transformations are
\bea
\fl\cq=e_1e_2+g_1g_2=q,\qquad
\wq=e_1g_2-g_1e_2=-m'\hbar c=-\kappa c.
\label{chginv}
\eea

In order to account for  the soft nonperturbative effects of
the interaction between electric and magnetic charges
we consider the limit in which the momentum exchanged by the photons
is small compared to the mass of the fermions.
This affords a substantial simplification in evaluating
the path-ordered exponential in (\ref{gextful});
in conjunction with the assumption of small
momentum transfer compared to the incident and outgoing
momenta, $q/p_{(1,2)}\ll\, 1$, this amounts to
the Bloch-Nordsieck~\cite{Bloch:1937pw} or {\it eikonal approximation}
(see~\cite{fri65,Fried:1971tz,Levy:1969cr,Levy:1970yn,Abarbanel:1969ek,%
Dittrich:1970vv,Dittrich:1972ya}; for more modern
applications in diffractive and strong coupling QCD processes see
\cite{Nachtmann:1991ua,Korchemsky:1993hr,Korchemskaya:1994qp,%
Gellas:1998sh,Karanikas:1998tn}).
In this approximation (\ref{gextful}) becomes
\be
\fl G_a(x,y|\bar{\A})\approx
\rmi\int_{0}^{\infty}\rmd\xi\,
\e^{-\rmi\xi m}\delta\left(x-y-\xi\frac{p}{m}\right)
\exp\left\{\rmi\int_{0}^{\xi}\rmd\xi^{\prime}
\frac{p}{m}\cdot\bar{\A}\left(x-\xi'\frac{p}{m}\right)
\right\}.
\label{gext}
\ee
With this simplification each propagator
in (\ref{scatt}) can be written
as an exponential of a linear function of the gauge field.
Performing mass shell amputation on each external coordinate
and taking the Fourier transform of (\ref{mxptl}) we obtain
the scattering
amplitude,  $T(p_{1},p^{\prime}_{1};p_{2},p^{\prime}_{2})$:
\bea
\fl\frac{ T(p_1,\pp_1 ; p_2,\pp_2)}{-4\pi\rmi}
=\int_0^1\! \rmd a\,
\rme^{a\L_{12}}\!\!
\int \dzi\dzii\left(\cq\,
D_{\mu\nu}(z_1-z_2)
-\wq\tD_{\mu\nu}(z_1-z_2)\right)
\nn
\times
\int\dxi\e^{-\rmi p_1x_1}
\bar{u}(p_1)\left(m_1+v_1\cdot p_1\right)
G_1(x_1,z_1|\bar{\A}_1)\gamma^\mu\nn
\times\int\dyi\e^{\rmi\ppi y_1}
G_1(z_1,y_1|\bar{\A}_1)
\left(m_1+v_1^\prime\cdot\ppi\right)u(\ppi)
\nn
\times
\int\dxii\e^{-\rmi x_2 p_2}
\bar{u}(p_2)\left(m_2+v_2\cdot p_2\right)
G_2(x_2,z_2|\bar{\A}_2)\gamma^\nu\nn
\times\int\dyii\e^{\rmi\ppii y_2}
G_2(z_2,y_2|\bar{\A}_2)
\left(m_2+v_2^\prime\cdot\ppii\right)u(\ppii).
\label{4.10}
\eea
Substituting (\ref{gext}) into (\ref{4.10}), we simplify this to
\bea
\fl \frac{T(p_{1},p^{\prime}_{1};p_{2},p^{\prime}_{2})}{-4\pi \rmi}\approx
\int_0^1 \rmd a\,\int\dzi\dzii
\rme^{-\rmi z_{1}\left(p_{1}-p^{\prime}_{1}\right)}
\rme^{-\rmi z_{2}\left(p_{2}-p^{\prime}_{2}\right)}
\bar{u}(p^{\prime}_{1})\gamma^{\mu}u(p_{1})
\bar{u}(p^{\prime}_{2})\gamma^{\nu}u(p_{2})
\nn
\times\left(\cq\,
D_{\mu\nu}(z_{1}-z_{2})-\wq\tD_{\mu\nu}(z_{1}-z_{2})\right)
\e^{a\L_{12}}
\nn
\times
\exp\left[\rmi\int_{0}^{\infty}\rmd\alpha_{1}\left\{
p_{1}\cdot\bar{\A}_1\left(z_{1}+\alpha_{1} p_{1}\right)
+p^{\prime}_{1}\cdot
\bar{\A}_1\left(z_{1}-\alpha_{1} p^{\prime}_{1}\right)
\right\}\right]\nn
\times\exp\left[
\rmi\int_{0}^{\infty}\rmd\alpha_{2}
\left\{
p_{2}\cdot\bar{\A}_2\left(z_{2}+\alpha_{2} p_{2}\right)
+p^{\prime}_{2}\cdot
\bar{\A}_2\left(z_{2}-\alpha_{2} p^{\prime}_{2}\right)
\right\}\right].
\label{smp}
\eea

Choosing the incoming momenta to be in the $z$
direction, in the center of momentum frame,
 $p_1^\mu=(E_1,0,0,p)$, $p_2^\mu=(E_2,0,0,-p)$,
invoking the approximation of small recoil
and passing  the linkage operator through the exponentials containing
the photon field, we find from (\ref{smp})
\bea
\fl \frac{T(p_{1},p^{\prime}_{1};p_{2},p^{\prime}_{2})}{-4\pi\rmi}\approx
\int_0^1 \rmd a\int\dzi\dzii
\rme^{-\rmi z_{1}\left(p_{1}-p^{\prime}_{1}\right)}
\rme^{-\rmi z_{2}\left(p_{2}-p^{\prime}_{2}\right)}
\bar{u}(p^{\prime}_{1})\gamma_{\mu}u(p_{1})
\bar{\upsilon}(p^{\prime}_{2})\gamma_{\nu}\upsilon(p_{2})
\nn
\times\left(\cq\,
D^{\mu\nu}\left(z_{1}-z_{2}\right)
-\wq\tD^{\mu\nu}\left(z_{1}-z_{2}\right)\right)
\rme^{\rmi a\Phi\left(p_1,p_2;z_1-z_2\right)},
\label{smp2}
\eea
where the ``eikonal phase'' integral is
\bea
\fl\Phi(p_1,p_2;z_1-z_2)\nn
\fl=4\pi p_{1}^{\kappa}p_{2}^{\lambda}
\int_{-\infty}^{\infty}\rmd\alpha_{1}\, \rmd\alpha_2
\left(\cq\,
D_{\kappa\lambda}
 -\wq\tilde D_{\kappa\lambda}\right)
 \left(z_{1}-z_{2}+\alpha_1 p_{1}-\alpha_2 p_{2}\right).
\label{phase1}
\eea
We transform to the center of momentum coordinates, by decomposing the
relative coordinate accordingly,
\be
\left(z_1-z_2\right)^\mu
=x_{\bot}^\mu-\tau_1 p_1^\mu+\tau_2 p_2^\mu,
\ee
where the Jacobian of the transformation is
\be
J=p\sqrt{s}\label{jacobian}
\ee
and $s=-(p_1+p_2)^2$ is the square of the center of mass energy.
Here we use the {\em symmetric} 
infinite string function, as discussed in \sref{sec:qt},
which has the momentum-space form,
\be
f^\mu(k)=4\pi\frac{n^\mu}{2\rmi}\left(\frac{1}{n\cdot k-\rmi\epsilon}+
\frac{1}{n\cdot k+\rmi\epsilon}\right).
\label{infstring}
\ee
Inserting the momentum-space representation of the propagator, 
and recalling \eref{dualp}, we cast
(\ref{phase1}) into the form
\bea
\Phi(p_1,p_2;x)\approx
4\pi p_{1}^{\kappa}p_{2}^{\lambda}
\int_{-\infty}^{\infty}\rmd\alpha_1\, \rmd\alpha_2
 \int\frac{(\rmd k)}{\left(2\pi\right)^{4}}
\frac{\e^{\rmi\, k\cdot\left(x+\alpha_1 p_{1}-\alpha_2 p_{2}\right)}}
{ k^{2}+\mu^2}
\nn
\times\left[\cq\, g_{\kappa\lambda} -
\wq\epsilon_{\kappa\lambda\sigma\tau}k^{\sigma}
\frac{n^{\tau}}{2}
\left(\frac{1}{n\cdot k-\rmi\epsilon}+\frac{1}{n\cdot k+\rmi\epsilon}
\right)\right],
\label{phase2}
\eea
where we have introduced the standard infrared photon-mass regulator,
$\mu^2$.
The delta functions that result from performing the integrations
over the parameters $\alpha_1$ and $\alpha_2$ in (\ref{phase2})
in the eikonal phase suggests the momentum decomposition
\be
\fl k^\mu = k_{\perp}^\mu + \lambda_1 e_1^\mu + \lambda_2 e_2^\mu,\qquad
\mbox{where}\qquad
\lambda_1 = p_2\cdot k,\qquad \mbox{and}\qquad
\lambda_2 = p_1\cdot k ,
\ee
and the four-vector basis is given by
\be
e_1^\mu=\frac{-1}{\sqrt{s}}\left(1,0,0,\frac{p_1^0}{p}\right)\qquad
\mbox{and}\qquad
e_2^\mu=\frac{-1}{\sqrt{s}}\left(1,0,0,-\frac{p_2^0}{p}\right),
\ee
which have the following properties, in terms of the masses $m_1$ and
$m_2$ of the two dyons,
\be
e_1\cdot e_1=\frac{1}{s}\frac{m_1^2}{ p^2}, \qquad
e_2\cdot e_2=\frac{1}{s}\frac{m_2^2}{ p^2},
\quad\mbox{and}\qquad
e_1\cdot e_2 =\frac{1}{s}\frac{p_1\cdot p_2}{ p^2}.
\ee
The corresponding measure is
\be
\left(\rmd k\right)=J^{-1}\rmd^2\mathbf{k}_\perp \rmd\lambda_1
\rmd\lambda_2,
\ee
in terms of the Jacobian in \eref{jacobian}. Using the
definition of the M{\o}ller amplitude, $M(s,t)$, given by removing the
momentum-conserving delta function,
\be
T(p_{1},p^{\prime}_{1};p_{2},p^{\prime}_{2})
=(2\pi)^4\delta^{(4)}(P-P^{\prime})M(s,t),
\ee
we put (\ref{smp2}) into the form
\be
\fl M(s,t)\approx
-\rmi\int_0^1 \rmd a\int  \rmd^2\mathbf{x}_{\bot}
\rme^{-\rmi\mathbf{q}_{\bot}\cdot\mathbf{x}_{\bot}}
\bar{u}(p^{\prime}_{1})\gamma^{\mu}u(p_{1})
\bar{u}(p^{\prime}_{2})\gamma^{\nu}u(p_{2}) I_{\mu\nu}
\rme^{\rmi a\Phi(p_1,p_2 ; x)},
\label{smp3}
\ee
where
\bea
I_{\mu\nu}=4\pi\int\frac{\rmd^2\mathbf{k}_{\bot}}{(2\pi)^2}
\frac{\rmd\lambda_{1}}{2\pi}\frac{\rmd\lambda_{2}}{2\pi}
\frac{\e^{\rmi\mathbf{k}_{\bot}\cdot\mathbf{x}_{\bot}}
2\pi\delta(\lambda_1)2\pi\delta(\lambda_2)}
{\left(\mathbf{k}_{\bot}^2+\mu^2+ \frac{1}{s\, p^2}
\left(\lambda_1^2 M_1^2+\lambda_2^2 M_2^2
+2\lambda_1\lambda_2 p_1\cdot p_2\right)\right)}
\nn
\times\left[\cq\, g_{\mu\nu}
-\wq\epsilon_{\mu\nu\sigma\tau}
k^{\sigma}\frac{n^{\tau}}{2}
{\left(\frac{1}{n\cdot k-\rmi\epsilon}+\frac{1}{n\cdot k+\rmi\epsilon}
\right)}\right].
\eea
Here $P=p_1+p_2$ and $P^\prime=p_1^\prime+p_2^\prime$,
and $q=p_1-p_1'$ is the momentum transfer. The factor
\be
\exp(\rmi\tau_1p_1\cdot q-\rmi\tau_2p_2\cdot q)=
\exp\left[\rmi{1\over2}q^2(\tau_1+\tau_2)\right]
\ee
has been omitted because it is unity in the eikonal limit, and correspondingly,
we have carried out the integrals on $\tau_1$ and $\tau_2$.
The eikonal phase (\ref{phase2}) now takes the very similar form
\be
\Phi(p_1,p_2;x)=
\frac{p_{1}^{\kappa} p_{2}^{\lambda}}{p\sqrt{s}}I_{\kappa\lambda}.
\label{phase4}
\ee
Choosing a spacelike
string in order to have a local interaction in momentum space,
$n^\mu=(0,\mathbf{\hat n})$, integrating over
the coordinates $\lambda_1 , \lambda_2$,
and introducing ``proper-time'' parameter representations
of the propagators, we reduce (\ref{phase4}) to
\bea
\Phi( p_1, p_2;x)={4\pi\over p\sqrt{s}}
\int \frac{\rmd^2\mathbf{k}}{(2\pi)^2}\rme^{\rmi\mathbf{k}\cdot\mathbf{x}}
\int_{0}^{\infty} \rmd\ps\,\rme^{-\ps\left(\mathbf{k}^2+\mu^2\right)}
\nn
\fl\times\Bigg\{\cq\,p_1\cdot p_2  -
\wq p_1^\mu p_2^\nu\epsilon_{\mu\nu\sigma\tau}\frac{n^\sigma}{2\rmi}
\frac{\partial}{\partial n_\tau}
\left(
\int_{0}^{\infty}\,\frac{d\pat}{\rmi\pat}\,
\e^{i\pat\left(\mathbf{n}\cdot\mathbf{k}+i\epsilon\right)}
-\int_{-\infty}^{0}\,\frac{d\pat}{i\pat}\,
\e^{i\pat\left(\mathbf{n}\cdot\mathbf{k}-i\epsilon\right)}\right)
\Bigg\}
\nn
\fl=2\cq\, \frac{p_1\cdot p_2}{p\sqrt{s}}
K_0\left(\mu\left|\mathbf{x}\right|\right)-
\wq\epsilon_{3jk} n^j\frac{\partial}{\partial n^k}
\int\frac{d\pat}{\pat}\,
K_0\left(\mu\left|\left(\mathbf{x}+\pat\mathbf{n}\right)\right|\right),
\label{phase5}
\eea
in terms of modified Bessel functions,
where we have dropped the subscript $\boldsymbol{\perp}$.

We perform the parameter integral over $\pat$
in the limit of small photon mass $\mu^2$:
\be
\fl-{1\over 2}\mathbf{\hat z\cdot(\hat n\times x)}
\left[\int_0^\infty-\int_{-\infty}^0
\right]{\rmd\pat\,\rme^{-\epsilon|t|}\over(\pat+\mathbf{\hat n\cdot x})^2+x^2-(
\mathbf{\hat n\cdot
x})^2}=\arctan\left[{\mathbf{ \hat n\cdot x}\over
\mathbf{\hat z\cdot(\hat n\times x)}}\right],
\ee
so the phase is
\be
\fl\Phi( p_1, p_2;x)\approx2\left\{\cq
\ln\left(\tilde{\mu}\left|\mathbf{x}\right|\right)
-\wq\arctan\left[\frac{\mathbf{\hat n}\cdot\mathbf{x}}
{\mathbf{\hat z}\cdot\left(\mathbf{\hat n}\times\mathbf{x}\right)}\right]
\right\}.\label{5.110}
\ee
In this limit we have used the asymptotic behavior of
the modified Bessel function
\be
K_0(x)\sim -\ln\left(\frac{\e^\gamma x}{2}\right),\qquad x\to0,
\ee
where $\gamma=0.577\dots$ is Euler's constant
and we have defined
$\tmu=\rme^\gamma\mu /\,2$.
Similarly,  (\ref{smp3}) becomes
\bea
\fl M(s,t)\approx
-2\rmi\int_0^1 \rmd a\int \rmd^2\mathbf{x}\,\rme^{-\rmi\mathbf{q}\cdot
\mathbf{x}}
\bar{u}(p^{\prime}_{1})\gamma^{\mu}u(p_{1})
\bar{u}(p^{\prime}_{2})\gamma^{\nu}u(p_{2})
\nn
\fl\times\Bigg\{
g_{\mu\nu}\cq K_0\left(\mu\left|\mathbf{x}\right|\right)
-\epsilon_{\mu\nu\sigma\tau}\wq
n^\tau\frac{\partial}{\partial n_\sigma}
{1\over2}\int\,\frac{d\pat}{\pat}\,
K_0
\left(\mu\left|\left(\mathbf{x}+\pat\mathbf{\hat n}\right)\right|\right)\Bigg\}
\rme^{\rmi a\Phi(p_1,p_2 ; x)}.
\label{smp5}
\eea

Although in the eikonal limit, no spin-flip processes occur, it is,
as always, easier to calculate the helicity amplitudes, of which there
is only one in this case.  In the high-energy limit, $p^0\gg m$, the Dirac
spinor in the helicity basis is
\be
u^\sigma(p)=\sqrt{p^0\over 2m}(1+\rmi\gamma_5\sigma)v_\sigma,
\ee
where the $v_\sigma$ may be thought of as two-component spinors satisfying
$\gamma^0v_\sigma=v_\sigma$.  They are further eigenstates of the helicity
operator $\mathbf{\bsigma\cdot\hat p}$ with eigenvalue $\sigma$:
\numparts
\bea
v_+^\dagger(\boldsymbol{\hat{p}}^\prime)=\left(
\cos\frac{\theta}{2}, \sin\frac{\theta}{2}\right),
\qquad
v_-^\dagger(\boldsymbol{\hat{p}}^\prime)=
\left(-\sin\frac{\theta}{2},\cos\frac{\theta}{2}
\right),
\\
v_+(\boldsymbol{\hat{p}})=\left(
\begin{array}{c} 1  \\  0
\end{array} \right),
\qquad
v_-(\boldsymbol{\hat{p}})=\left(
\begin{array}{c} 0  \\  1
\end{array} \right).
\eea
\endnumparts
We employ the definition
\be
\gamma_5=\gamma^0\gamma^1\gamma^2\gamma^3
\ee
and consequently $\gamma^0\boldsymbol{\gamma}
=\rmi\gamma^5\boldsymbol{\sigma}$, 
where $\sigma_{ij}=\epsilon_{ijk}\sigma^k$.
We then
easily find upon integrating over the parameter $a$ that the spin nonflip
part of (\ref{smp5}) becomes ($\theta\to0$)
\be
M(s,t)=\frac{s}{2m_1m_2}
\Bigg\{\int \rmd^2\mathbf{x}\, \e^{-\rmi\mathbf{q}\cdot\mathbf{x}}
\rme^{\rmi\Phi(p_1,p_2 ; x)}
-\left(2\pi\right)^2\delta^2(\mathbf{q})
\Bigg\}.
\label{scat9}
\ee

Now notice that the arctangent function in \eref{5.110} is discontinuous
when the $xy$ component of $\mathbf{\hat n}$
 and $\mathbf{x}$ lie in the same direction.  We require that the eikonal phase
factor $e^{i\Phi}$
be continuous, which leads to the Schwinger quantization condition
 (\ref{dyon}):
 \be
\wq= -m',
\ee
where $m'$ is an integer.
Now using the integral form for
the Bessel function of order $\nu$
\be
\rmi^\nu\/J_\nu(t)=\int_0^{2\pi}\frac{\rmd\phi}{2\pi}
\rme^{\rmi\left(t\cos\phi-\nu\phi\right)}\, ,
\ee
we find the dyon-dyon  scattering amplitude (\ref{scat9}) to be
[see also \eref{5.14} below]
\be
M(s,t)=\frac{\pi s}{m_1m_2}\rme^{-\rmi 2m'\psi}
\int_0^\infty \rmd x\,x \,J_{2m'}(qx)
\rme^{\rmi 2\tilde{\alpha}\ln\left(\tmu x\right)}\, ,
\label{scatbes}
\ee
where $\tilde{\alpha}=\cq$, and $\psi$ is the angle between
$\mathbf{q}_\perp$ and $\mathbf{\hat n}_\perp$.
The integral over $x$ is just a ratio of gamma functions,
\be
\frac1{\tilde\mu}\int_0^\infty \rmd x\,\left(\tmu x\right)^{1+2\rmi\talpha}
 J_{2m'}(qx)
=\frac{1}{2\tmu^2}
\left(\frac{4\tmu^2}{q^2}\right)^{\rmi\talpha+1}
\frac{\Gamma\left(1+m'+\rmi\talpha\right)}
{\Gamma\left(m'-\rmi\talpha\right)}.
\label{infr}
\ee
Then (\ref{scatbes}) becomes
\be
M(s,t)\approx\frac{ s}{m_1m_2}{2\pi\over q^2}(m'-\rmi\talpha)
\rme^{-\rmi 2m'\psi}
\left(\frac{4\tmu^2}{q^2}\right)^{\rmi\talpha}
\frac{\Gamma\left(1+m'+\rmi\talpha\right)}
{\Gamma\left(1+m'-\rmi\talpha\right)}.
\ee
This result is almost identical in structure to the nonrelativistic
form of the scattering amplitude for the Coulomb potential, which result
is recovered by setting $m'=0$.  (See, for example, \cite{gottfried}.)
Following the standard convention \cite{Itzykson:1980rh} we calculate
the spin-averaged cross section for dyon-dyon scattering
in the high energy limit,
\be
\frac{\rmd\sigma}{\rmd t}=4\pi
\frac{\left(\cq\right)^2+\left(\wq\right)^2}{ t^2}.
\label{dyoncross}
\ee
While the Lagrangian is string-dependent, because of
the charge quantization condition, the cross section, (\ref{dyoncross}),
is string independent.  Not surprisingly, this coincides with the
Rutherford formula \eref{ruth}, \eref{ruth2}.

For the case of charge-monopole scattering $e_1=g_2=0$,
this result, of course, coincides with that
found by Urrutia \cite{Urrutia:1978kq},
which  is  also  string independent as a consequence
of (\ref{quant}). We should also mention the slightly
earlier work of Ore, demonstrating the Lorentz invariance
of charge-monopole scattering \cite{Ore:1975my}.
 This is to be contrasted with {\it ad hoc\/} prescriptions
that average over string directions  or eliminate its
dependence by simply dropping string-dependent
terms because they cannot contribute to any gauge invariant quantities
(cf.~\cite{Deans:1981qs}).

\subsection{Conclusion}

In this section we have responded to the challenge of Schwinger 
\cite{Schwinger:1975ww},
to construct a realistic theory of relativistic magnetic charges.  He
sketched such a development in source theory language, but restricted his
consideration to classical point particles, explicitly leaving the details
to the reader.  Urrutia applied this skeletal formulation in the eikonal
limit \cite{Urrutia:1978kq}, as already suggested by Schwinger.

We believe that we have given a complete formulation, in modern quantum
field theoretic language, of an interacting electron-monopole  or
dyon-dyon system.
The resulting Schwinger-Dyson equations, although to some extent implicit
in the work of Schwinger and others, were given in \cite{Gamberg:1999hq} 
for the first time.

The challenge remaining is
 to apply these equations to the calculation of monopole and dyon
processes.  Perturbation theory is useless, not only because of the strength
of the coupling, but more essentially because the graphs are fatally string-
or gauge-dependent.  The most obvious nonperturbative technique for
transcending these limitations in scattering processes lies in the
high energy regime where the
eikonal approximation is applicable; in that limit, our formalism
generalizes  the lowest-order result of Urrutia and
charts the way to include systematic corrections.
More problematic is the treatment of monopole production
processes---we must defer that discussion to  subsequent publications.
In addition we have also detailed
how the Dirac string dependence disappears from physical quantities.
It is by no means a result of string
averaging or a result of dropping
string-dependent terms as in \cite{Deans:1981qs}.
In fact, it is a result of summing the soft contributions to
the dyon-dyon or charge-monopole process.  There are good reasons
to believe that inclusion of hard scattering contributions will
not spoil this consistency. At the level of the eikonal
approximation and its corrections
one might suspect the occurrence of a
factorization of hard string-independent
and soft string-dependent contributions in a manner similar to that
argued in strong-coupling QCD.

It is also of interest to
investigate other nonperturbative methods of calculation in order
to demonstrate gauge covariance of Green's functions and scattering
amplitudes in both electron-monopole and dyon-dyon scattering
and in Drell-Yan production processes. In addition there is a
formalism employed in \cite{Fried:1995cm,Fried:1995zr,Fried:1995zx} 
based on Fradkin's \cite{fra66} Green's function representation,
which includes approximate vertex and self-energy polarization
corrections using nonperturbative techniques, which we
are adapting to the magnetic charge domain.  (For a first
pedagogical example of this formalism see \cite{Shajesh:2005dy}.) 
We hope in the future
to apply the techniques and results found here to the Drell-Yan
production mechanism, for example, and obtain phenomenologically
relevant estimates for the laboratory production of monopole-antimonopole
pairs.

\section{Renormalization}
As discussed in \sref{sec:qft}, Lorentz invariance (rotational invariance
for the nonrelativistic theory) is satisfied by the dual electrodynamics
of electric and magnetic charges interacting provided the quantization 
condition is obeyed.  But is the theory renormalizable?  
This question was addressed by Schwinger in 1966 
\cite{schwingermc1,schwingermc2}.  His view at the time was that 
renormalization described the connection between the particle and field level
description of reality.  At both of these levels consistency demanded that
the quantization condition must hold, so that if integer quantization is
appropriate,
\be
\frac{e_0g_0}{\hbar c}=n_0,\qquad\frac{eg}{\hbar c}=n,\label{renquant}
\ee
but that the integers $n$ and $n_0$ need not be the same.  The question is
whether electric and magnetic charges are renormalized by the same, or 
different factors.  He argued that the former was the case, because
charge renormalization refers to the electromagnetic field, not its sources.
That is
\be
\frac{e}{e_0}=\frac{g}{g_0}=C<1,
\ee
so in view of the charge quantization condition \eref{renquant} the quantum
numbers $n_0$ and $n$ are not the same:
\be
C^2=\frac{n}{n_0}.
\ee
The discreteness of renormalization of the dual theory is thus manifest from 
this point of view.

This is at odds with the modern understanding of renormalization as a 
continuous evolution of parameters, such as the charge, with change of
energy scale.  It would seem that this view of the renormalization group
may be difficult to maintain without a perturbative framework: That is, at
any energy scale $Q$, we might expect
\be
e(Q)g(Q)=n.\label{fixn}
\ee
For this reason Laperashvili and Nielsen \cite{Laperashvili:1999pu,%
Laperashvili:2000np,Laperashvili:2002xv,Laperashvili:2002jc,Das:2005iv},
following Zwanziger \cite{Zwanziger:1970hk,brandtandneri}
argue that \eref{fixn} holds at all scales, or in terms of the bare and
renormalized quantization numbers, $n=n_0$.  That is, the electric and
magnetic charges are renormalized by exactly inverse factors.  In terms of
the fine structure constants, for the minimal Dirac pole strength, $m'=1/2$,
this says
\be
{}^*\alpha(Q)\alpha(Q)=\frac14.
\ee Laperashvili, Nielsen, and collaborators have exploited the small
window which this seems to permit for perturbative calculations, where
neither $\alpha$ nor ${}^*\alpha$ are bigger than unity.

However, at best there is room for serious doubt about the essential
validity of this procedure.  In ordinary quantum electrodynamics 
charge renormalization can be regarded as arising entirely from vacuum
polarization.  Presumably, this is still the case in dual QED.  Using
lowest order perturbative graphs to describe vacuum polarization does
violence to the charge quantization condition; moreover, higher order
graphs involving both electrically and magnetically charged particles
necessarily bring in the Dirac string, which as we have repeatedly
emphasized can only disappear in a nonperturbative treatment.  Such is
as yet lacking in our analysis of renormalization in dual QED.

\section{Eikonal approximation}

It is envisaged that if monopoles are sufficiently light, they would be
produced by a Drell-Yan type of process occurring in $p\overline p$ collisions
at the Tevatron.  Two photon production channels may also be important.
The difficulty is to make a believable estimate of the
elementary process $q\overline q\to\gamma^*\to M\overline M$, where $q$
stands for quark and $M$ for magnetic monopole.   It is not known how
to calculate such a process using perturbation theory; indeed, perturbation
theory is inapplicable to monopole processes because of the quantization
condition (\ref{quant}).  It is only because of that consistency condition
that the Dirac string, for example, disappears from the result.

Only formally has it been shown that the quantum field theory of electric
and magnetic charges is independent of the string orientation, or, more
generally, is gauge and Lorentz invariant 
\cite{Schwinger:1966nj,schwingermc1,schwingermc2,Schwinger:1968rq,
Schwinger:1975ww,%
Zwanziger:1968rs,Zwanziger:1970hk,Brandt:1977be,brandtandneri}.
It has not yet proved possible to develop generally consistent schemes for
calculating processes involving real or virtual magnetically charged particles.
Partly this is because a sufficiently general field theoretic formulation
has not yet been given; a small step in remedying this defect 
was given in \cite{Gamberg:1999hq}, reviewed in \sref{sec:qft}.
However, the nonrelativistic scattering of magnetically charged particles
is well understood, as described in \sref{sec:qt}.
Thus it should not be surprising that
an eikonal approximation gives a string-independent result for 
electron-monopole
scattering provided the condition (\ref{quant}) is satisfied.  
In  \sref{sec:qft} we described the eikonal approximation in terms of the
full field-theoretic formulation.  Since that formalism is rather
elaborate, we give here a simplified pedagogical treatment, as 
described in \cite{binding}, 
and first worked out by Urrutia \cite{Urrutia:1978kq}.

The interaction between electric ($J^\mu$) and magnetic (${}^*J^\mu$)
currents is given by \eref{interact}, or
\begin{equation}
\fl W^{(eg)}=
-\epsilon_{\mu\nu\sigma\tau}\int(\rmd x)(\rmd x')(\rmd x'')
J^\mu(x)\partial^\sigma D_+(x-x')f^\tau(x'-x'')\Jg^\nu(x'').
\label{weg}
\end{equation}
Here $D_+$ is the usual photon propagator, and
the arbitrary ``string'' function $f_\mu(x-x')$ satisfies \eref{dfstng}, or
\begin{equation}
\partial_\mu f^\mu(x-x')=4\pi\delta(x-x').
\end{equation}
It turns out to be convenient for this
calculation to choose a symmetrical string, which
satisfies \eref{symmstring}, or 
\begin{equation}
f^\mu(x)=-f^\mu(-x).
\end{equation}
In the following we choose a string lying along the straight line $n^\mu$, in
which case the function may be written as a Fourier transform \eref{infstring},
or
\begin{equation}
f_\mu(x)=4\pi{n_\mu\over 2\rmi}\int{(\rmd k)\over(2\pi)^4}\rme^{\rmi kx}
\left({1\over n\cdot k
-\rmi\epsilon}+{1\over n\cdot k+\rmi\epsilon}\right).
\label{string}
\end{equation}

In the high-energy, low-momentum-transfer regime, the scattering amplitude
between electron and monopole is obtained from (\ref{weg}) by inserting
the classical currents,
\numparts
\begin{eqnarray}
J^\mu(x)=e\int_{-\infty}^\infty \rmd\lambda \,{p_2^\mu\over m}\,
\delta\left(x-{p_2\over m}\lambda\right),\\
\Jg^\mu(x)=g\int_{-\infty}^\infty \rmd\lambda'\,
{p_1^{\mu}\over M}\,\delta\left(x+b-{p'_2\over M}\lambda'\right),
\end{eqnarray}
\endnumparts
where $m$ and $M$ are the masses of the electron and monopole, respectively.
Let us choose a coordinate system such that
the incident ultrarelativistic
momenta of the two particles have spatial components
along the $z$-axis:
\numparts
\begin{equation}
p_1=(p,0,0,p),\quad p_2=(p,0,0,-p),
\end{equation}
and the impact parameter lies in the $xy$ plane:
\begin{equation}
b=(0,{\bf b},0).
\end{equation}
\endnumparts
Apart from kinematical factors, the scattering amplitude is simply the
transverse Fourier transform of the eikonal phase, which is the content
of \eref{scat9},
\begin{equation}
I({\bf q})=\int \rmd^2\mathbf{b}\, 
\rme^{-\rmi{\bf b\cdot q}}\left(\rme^{\rmi\chi}
-1\right),
\label{eikonal}
\end{equation}
where $\chi$ is simply $W^{(eg)}$ with the classical currents substituted,
and $\bf q$ is the momentum transfer.

First we calculate $\chi$; it is immediately seen to be, if $n^\mu$ has
no time component,
\begin{equation}
\chi=2\pi eg\int{\rmd^2\mathbf{k}_\perp
\over(2\pi)^2}{{\bf \hat z\cdot(\hat n\times k_\perp})\over k_\perp^2
-\rmi\epsilon}
\rme^{\rmi{\bf k_\perp\cdot b}}\left({1\over {\bf \hat n\cdot k_\perp}
-\rmi\epsilon}+{1\over {\bf \hat n\cdot k_\perp}+i\epsilon}\right),
\label{chi}
\end{equation}
where $\bf k_\perp$ is the component of the photon momentum perpendicular
to the $z$ axis.
From this expression we see that the result is independent of the angle $\bf 
\hat n$
makes with the $z$ axis.  We next use proper-time representations for the
denominators in (\ref{chi}),
\numparts
\begin{eqnarray}
{1\over k_\perp^2}=\int_0^\infty \rmd s\, \rme^{-sk_\perp^2},\\
{1\over {\bf \hat n\cdot k_\perp}-\rmi\epsilon}+
{1\over {\bf \hat n\cdot k_\perp}+\rmi\epsilon}={1\over\rmi}\left[\int_0^\infty
\rmd\lambda-\int_{-\infty}^0 \rmd\lambda\right]\rme^{\rmi\lambda 
{\bf \hat n\cdot k_\perp}}
\rme^{-|\lambda|\epsilon}.
\end{eqnarray}
\endnumparts
We then complete the square in the exponential and perform the Gaussian
integration to obtain
\numparts
\begin{equation}
\chi=eg{\bf \hat z\cdot(\hat n\times b)}\int^\infty_{-\infty}\rmd\lambda{1\over
(\lambda+{\bf b\cdot \hat n})^2+b^2-({\bf b\cdot\hat n})^2},
\end{equation}
or
\begin{equation}
\chi=2eg\arctan\left({\bf \hat n\cdot b}\over{\bf \hat z\cdot(b
\times \hat n})\right),
\end{equation}
\endnumparts
which is contained in \eref{5.110}.
Because $\rme^{\rmi\chi}$ must 
be continuous when $\bf \hat n_\perp$ and $\bf b$ lie in the
same direction, we must have the Schwinger quantization condition for
an infinite string,
\begin{equation}
eg=m',
\label{quant2}
\end{equation}
where $m'$ is an integer.

To carry out the integration in (\ref{eikonal}), choose $\bf b$ to
make an angle $\phi$ with $\bf q_\perp$, 
and the projection of $\bf \hat n$ in the
$xy$ plane to make an angle $\psi$ with $\bf q_\perp$; then
\begin{equation}
\chi=2eg(\phi-\psi-\pi/2).
\end{equation}
To avoid the appearance of a Bessel function,
as occurs in \eref{scatbes}, we first integrate over
$b=|{\bf b}|$, and then over $\phi$:
\begin{eqnarray}
I({\bf q})=\int_0^{2\pi}\rmd\phi\int_0^\infty b\,\rmd b\,\rme^{-\rmi bq
(\cos\phi-\rmi\epsilon)}
\rme^{2\rmi m'(\phi-\psi-\pi/2)}\nonumber\\
={4\over \rmi}{\rme^{-2\rmi m'(\psi+\pi/2)}\over q^2}
\oint_C{\rmd z\,z^{2m'-1}\over(z+1/z-\rmi\epsilon)^2}
=-{4\pi m'\over q^2}\rme^{-2\rmi m'\psi},\label{5.14}
\end{eqnarray}
where $C$ is a unit circle about the origin, and
where again the quantization condition (\ref{quant2}) has been used.
Squaring this and putting in the kinematical factors we obtain Urrutia's
result \cite{Urrutia:1978kq} [cf.~\eref{dyoncross}],
\begin{equation}
{\rmd\sigma\over \rmd t}=4\pi(eg)^2{1\over t^2}, \qquad t=q^2,
\end{equation}
which is exactly the same as the nonrelativistic, small angle result
found, for example, in (\ref{ruth}).

\section{Energy loss by magnetic monopoles traversing matter}
\label{sec:dedx}

The essence of the energy loss mechanism of charged particles traveling
through matter can be described by classical electrodynamics.  By a 
simple duality analysis, therefore, one should be able to describe the
rate at which a particle carrying magnetic charge loses energy when
it passes through matter.  The details of this argument can be found in
the last two chapters of \cite{CE}.  It is based on the fundamental
analyticity requirements of the electrical permittivity, demanded by
causality, the Kramers-Kronig relations.  In terms of positive
spectral functions $p(\omega)$ and $q(\omega)$, which satisfy
\be
\int_0^\infty \rmd \omega' p(\omega')=1,\qquad
\int_0^\infty \rmd \omega' q(\omega')=1,
\ee
the dielectric function obeys
\numparts
\bea
\varepsilon(\omega)=1+\omega_p^2\int_0^\infty \rmd \omega'\frac{p(\omega')}
{\omega^{\prime2}-(\omega+\rmi\epsilon)^2},\label{epsilonp}\\
\frac1{\varepsilon(\omega)}=1-
\omega_p^2\int_0^\infty \rmd \omega'\frac{q(\omega')}
{\omega^{\prime2}-(\omega+\rmi\epsilon)^2}.\label{epsilonq}
\eea
\endnumparts
Here $\omega_p$ is the plasma frequency, 
\be
\omega_p=\frac{4\pi n e^2}m,
\ee
in terms of the electron mass $m$, and density of free electrons $n$.

Using \eref{epsilonq}, in Chapter 52 of \cite{CE} we derive the following
formula for the energy loss $-\rmd E$ when a charged particle (charge
$Ze$) having velocity $v$ travels a distance $\rmd z$:
\begin{equation}
-{\rmd E\over \rmd z}={1\over2}{\omega_p^2(Ze)^2\over v^2}\left[
\ln {K^2 v^2\over\omega_{e}^2\left(1-{v^2\over c^2}\right)}-{v^2\over c^2}
 -\int_{1/\epsilon(0)}^{v^2/c^2}\rmd\left({v^{\prime 2}\over c^2}\right)
{\nu_{v'}^2\over \omega_p^2} \right],
 \label{48.36}
\end{equation}
where the last integral should be omitted if $v/c<1/\sqrt{\varepsilon(0)}$.
Here \be
\int_0^\infty \rmd \omega \,q(\omega)\ln\omega^2=\ln\omega_e^2,
\ee
and $\nu_v$ is given by the root of $1-v^2\varepsilon(\rmi \nu)/c^2$, that
is,
\be
\omega_p^2\int_0^\infty\rmd \omega'\frac{q(\omega')}{\omega^{\prime2}+\nu_v^2}
=1-\frac{v^2}{c^2}.
\ee
Further, $K$ is boundary between low momentum transfer events and high
momentum ones, such as $\delta$-rays.  For a more complete theory, and
extensive comparison with experiment,
the reader is referred to \cite{PDBook}, Passage of particles through matter.

In parallel with the above derivation, we can use \eref{epsilonp}
to derive the corresponding formula for the energy loss rate by
a magnetically charged particle:
\begin{eqnarray}
-{\rmd E\over \rmd z} ={1\over2}{ \omega_p^2 g^2\over c^2}\left[\ln{ K^2v^2
\over \omega_m^2\left(1-{v^2\over c^2}\right)} -1-\int_{c^2/v^2}^{\epsilon(0)}
\rmd\left({c^2\over v^{\prime 2}}\right){\nu_{v'}^2\over\omega_p^2}\right],
\end{eqnarray}
where the latter integral only appears if
${v\over c} > {1 \over\sqrt{\epsilon(0)}}$.
Here 
\be\int_0^\infty \rmd \omega\, p(\omega)\ln\omega^2\equiv
\ln\omega_m^2.
\ee
We note that the predominant change from the energy loss for
electrically charged particles lies in the replacement
\be
\frac{Ze}v\to\frac{g}c, 
\ee as earlier claimed in \eref{velsup}, provided $\omega_e^2\approx
\omega_m^2$.

\begin{figure}
\centering
\includegraphics[height=10cm]{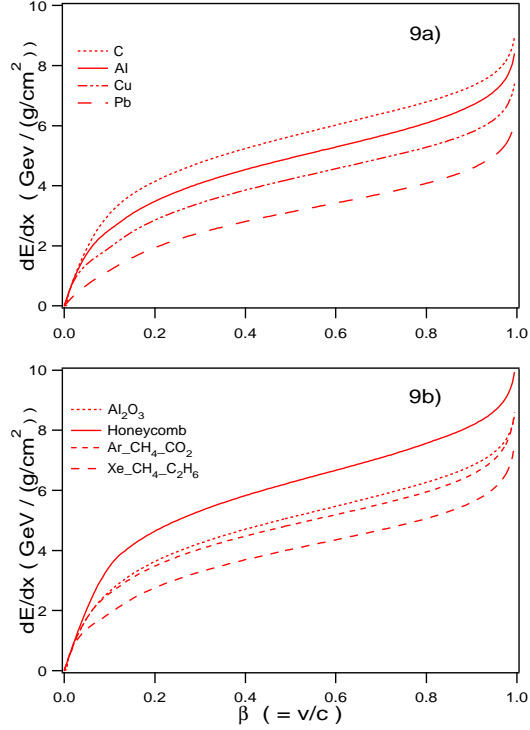}
\caption{\label{figel} Energy loss of a magnetic monopole in various 
materials. These $\rmd E/\rmd x$ curves are for a
magnetic charge value of $2|m'|=n=1$; apart from the correction terms $K(|n|)$
and $B(|n|)$, we multiply by $n^2$ for larger magnetic charge values.}
\end{figure}

This result should only be regarded as qualitative.  For our experimental
analysis, we will use the extensive results of Ahlen 
\cite{Ahlen:1982mx,Ahlen:1982rw}. As we discussed in \sref{sec:relscatt}
Kazama, Yang, and Goldhaber \cite{Kazama:1977fm} have obtained the relativistic
differential scattering cross section for an electron moving in the magnetic
field of a fixed magnetic pole. Ahlen
then used this cross section to
obtain the following expression for monopole stopping power:
\begin{eqnarray}
-\frac{\rm dE}{\rmd x}=\frac{4\pi}{c^2}
\frac{g^2e^2}{ m_e}N_e\bigg(\ln\frac{2m_ec^2\beta^2\gamma^2}
{I}+\frac12K(|n|)-\frac12\delta-\frac12-B(|n|)\bigg),
\label{e11}
\end{eqnarray}
where $N_e$ is the number density of electrons, $I$ is the
mean ionization energy,
$K(|n|)= 0.406$ (0.346) is the Kazama, Yang and Goldhaber correction
for magnetic charge  $2|m'|=n=1$ ($n\ge2$) respectively, $\delta$ is the usual 
density correction and  $B(|n|)= 0.248$ (0.672, 1.022, 1.685) is the Bloch
correction for $n=1$ ($n= 2, 3, 6$), respectively \cite{PDBook}.
(Of course, one must divide by the density to get $\rmd x$ in g/cm$^2$.)
 This formula is good only for velocities $\beta = v/c > 0.1$.
For velocities $\beta < 0.01$, we use (60) of \cite{Ahlen:1982mx}
as an approximation for all materials:
\begin{equation}
-\frac{\rmd E}{dx} = (45\mbox{ Gev/cm})  n^2 \beta,
\label{e13}
\end{equation}
which is linear in $\beta$ in this region.
The two $\rmd E/\rmd x$ velocity
regions are
joined by an empirically fitted polynomial in the region of 
$\beta = 0.01$--0.1 in order
to have a smooth function of $\beta$. For the elemental and composite materials
found in the D0 and CDF detectors, we show the resulting $\rmd E/\rmd x$ curves
we used in \fref{figel}.  (See \cite{Luo:2002tm}.)

\section{Binding}
\label{sec:binding}
To this point in this review, we have concentrated on the scattering of
monopoles with charged particles, or on dyon-dyon scattering.  Now we
turn to the question of the binding of these particles.  Our discussion
will be largely, although not exclusively, based on the nonrelativistic
description.

If $q=e_1e_2+g_1g_2<0$, and $m'=-(e_1g_2-e_2g_1)/\hbar c$, 
${\cal H}_{\rm NR}$ \eref{hamdyons} gives binding:
\be
E_{Nj}=-\frac{\mu}{2}q^2\left[N+\frac{1}{2}
+\left((j+1/2)^2-m^{\prime2}\right)^{1/2}\right]^{-2},
\ee where $N$ is a principal quantum number.
We will not further address the issue of dyons 
\cite{Osland:1984yu,Osland:1984ys,Osland:1985ma,Osland:1985kz,Osland:1985qw,%
Osland:1985va}, which for the correct sign
of the electric charge will always bind electrically to nuclei.
Monopoles will not bind this way; rather,
a magnetic moment coupling as in \eref{magdipoleham} is required;
for example, for spin-1/2,
\be
{\cal H}_S=-\frac{e\hbar}{2\mu c}\gamma\mbox{\boldmath{$\sigma$}}\cdot {\bf B},
\qquad \gamma=1+\kappa=\frac{g}{2}.
\ee
($\gamma=1$ or $g=2$ is the ``normal'' value.)
				   
Suppose monopoles are produced in a collision at the Tevatron, 
for example; they travel
through the detector, losing energy in a well-known manner (see, e.g.,
\cite{CE}, which results are summarized in \sref{sec:dedx}), 
presumably ranging out, and
eventually binding to matter in the detector (Be, Al, Pb,
for example).  The purpose of this section is to review the theory of the
binding of magnetic charges to matter.

We consider the binding of a monopole of magnetic charge $g$
to a nucleus of charge $Ze$, mass ${\cal M}=A m_p$, and magnetic moment
\begin{equation}
\mbox{\boldmath{$\mu$}}={e\over m_pc}\gamma{\bf S},
\end{equation}
$\bf S$ being the spin of the nucleus.  (We will assume here that the
monopole mass $\gg {\cal M}$, which restriction could be easily removed.)
The charge quantization condition is given by (\ref{quant}).
Because the nuclear charge is $Ze$, the relevant angular momentum
quantum number is ($m'$ being an integer or an integer plus 1/2)
\begin{equation}
l=|m'|Z.
\end{equation}

\subsection{Nonrelativistic binding for $S=1/2$}
In this subsection we follow the early work of Malkus \cite{Malkus:1950yc} and
the more recent papers of Bracci and Fiorentini \cite{Bracci:1983fe,%
Bracci:1983tq,Bracci:1984zx,Bracci:1984db}.
(There are also the results given in \cite{Sivers:1970zm}, but
this reference seems to contain errors.)

The neutron ($Z=0$) is a special case.  Binding will occur in the
lowest angular momentum state, $J=1/2$, if
\begin{equation}
|\gamma|>{3\over4|m'|}
\end{equation}
Since $\gamma_n=-1.91$, this condition is satisfied for all $m'$.

In general, it is convenient to define a reduced gyromagnetic ratio,
\begin{equation}
\hat\gamma={A\over Z}\gamma,\quad \hat\kappa=\hat\gamma-1.
\end{equation}
This expresses the magnetic moment in terms of the mass and charge of
the nucleus.
Binding will occur in the special lowest angular momentum state
$J=l-{1\over2}$ if
\begin{equation}
\hat \gamma>1+{1\over4l}.
\label{lowest}
\end{equation}
Thus binding can occur here only if the anomalous magnetic moment
 $\hat\kappa>1/4l$.  The proton, with $\kappa=1.79$, will bind.

Binding can occur in higher angular momentum states $J$ if and only if
\begin{equation}
|\hat\kappa|>\kappa_c={1\over l}\left|J^2+J-l^2\right|.
\label{kappac}
\end{equation}
For example, for $J=l+{1\over2}$, $\kappa_c=2+3/4l$, and for
$J=l+{3\over2}$, $\kappa_c=4+15/4l$.  Thus ${}^3_2$He, which is spin 1/2,
will bind in the first excited angular momentum state because $\hat\kappa
=-4.2$.

Unfortunately, to calculate the binding energy, one must regulate
the potential at $r=0$.  The results shown in \tref{tabb} assume a hard
core.

\subsection{Nonrelativistic binding for general $S$}
The reference here is \cite{Olaussen:1984xb}.  
The assumption made here is that
$l\ge S$.  (There are only 3 exceptions, apparently: ${}^2$H,
${}^8$Li, and ${}^{10}$B.)

Binding in the lowest angular momentum state $J=l-S$ is
given by the same criterion (\ref{lowest}) as in spin 1/2.
Binding in the next state, with $J=l-S+1$, occurs if
$\lambda_{\pm}>{1\over4}$ where
\begin{equation}
\fl \lambda_{\pm}=\left(S-{1\over2}\right){\hat\gamma\over S}l
-2l-1\pm\sqrt{(1+l)^2+(2S-1-l){\hat\gamma\over S}l+
{1\over4}l^2\left({\hat\gamma\over S}\right)^2}.
\end{equation}
 The previous result for $S=1/2$ is recovered, of course.
$S=1$ is a special case: Then $\lambda_-$ is always negative, while
$\lambda_+>{1\over4}$ if $\hat\gamma>\gamma_c$, where
\begin{equation}
\gamma_c={3\over4l}{(3+16l+16l^2)\over9+4l}.
\end{equation}

For higher spins, both $\lambda_\pm$ can exceed $1/4$:
\numparts
\begin{eqnarray}
\lambda_+>{1\over4} &\mbox{ for }& \hat\gamma>\gamma_{c-}\\
\lambda_->{1\over4} &\mbox{ for }& \hat\gamma>\gamma_{c+}
\end{eqnarray}
\endnumparts
where for $S={3\over2}$
\begin{equation}
(\gamma_c)_\mp={3\over4l}(6+4l\mp\sqrt{33+32l}).
\end{equation}
For ${}_4^9$Be, for which $\hat\gamma=-2.66$, we cannot have binding
because $3>\gamma_{c-}>1.557$, $3<\gamma_{c+}<8.943$, where the
ranges come from considering different values of $2|m'|$ from 1 to $\infty$.
For $S={5\over2}$,
\begin{equation}
(\gamma_{c})_\mp={36+28l\mp\sqrt{1161+1296l+64l^2}\over12l}.
\end{equation}  So ${}^{27}_{13}$Al will bind in either of
these states, or the lowest angular momentum state, because
$\hat\gamma=7.56$, and
$1.67>\gamma_{c-}>1.374$, $1.67<\gamma_{c+}<4.216$.

\subsection{Relativistic spin-1/2}
Kazama and Yang treated the Dirac
equation \cite{Kazama:1976sr}.  
See also \cite{Olaussen:1983qc,Olaussen:1983bm} 
and \cite{Osland:1984yu,Osland:1984ys,Osland:1985ma,Osland:1985kz,%
Osland:1985qw,Osland:1985va}.

In addition to the bound states found nonrelativistically, deeply
bound states, with $E_{\rm binding}=\mathcal{M}$ are found.
These states always exist for $J\ge l+1/2$.   For $J=l-1/2$, these
 (relativistic) $E=0$ bound
states exist only if $\kappa>0$.
 Thus (modulo the question of
form factors) Kazama and Yang \cite{Kazama:1976sr}
expect that electrons can bind to monopoles.
(We suspect that one must take the existence of these deeply bound
states with a fair degree of skepticism.  See also \cite{Walsh:1983bz}.)

As expected, for $J=l-1/2$
we have weakly bound states only for $\kappa>1/4l$, which is the same
as the nonrelativistic condition (\ref{lowest}), and for $J\ge l+1/2$, only if
$|\hat\kappa|>\kappa_c$, where $\kappa_c$ is given in (\ref{kappac}).

\subsection{Relativistic spin-1}
Olsen, Osland, and Wu considered this situation 
\cite{Olsen:1990jm,Olsen:1990jn}.

In this case, no bound states exist, unless an additional interaction
is introduced  (this is similar to what happens nonrelativistically,
because of the bad behavior of the Hamiltonian at the origin).
Bound states are found if an ``induced magnetization'' interaction
(quadratic in the magnetic field)
is introduced.  Binding is then found for the lowest angular momentum
state $J=l-1$ again if $\hat\kappa>1/4l$.  For the higher angular
momentum states, the situation is more complicated:
\begin{itemize}
\item for $J=l$: bound states require $l\ge 16$, and
\item for $J\ge l+1$: bound states require $J(J+1)-l^2\ge 25$.
\end{itemize}
But these results are probably highly dependent on the form of the
additional interaction.  The binding energies found are inversely
proportional to the strength $\lambda$ of this extra interaction.

\begin{table}[ht]
\caption{\label{tabb}Weakly bound states of nuclei to a magnetic monopole.
The angular momentum quantum number $J$ of the lowest bound state
is indicated.  In Notes, NR means nonrelativistic and R relativistic
calculations;
hc indicates an additional hard core interaction is assumed, while
FF signifies use of a form factor. IM represents induced magnetization, the
additional interaction employed for the relativistic spin-1 calculation.
We use  $|m'|=1/2$ except for the deuteron, where $|m'|=1$ is required for
binding.}
\begin{tabular}{@{}llllllll}\br
Nucleus&Spin&$\gamma$&$\hat\gamma$&$J$&$E_b$&Notes&Ref\\ \mr
$n$&${1\over2}$&$-1.91$&&${1\over2}$&350 keV&NR,hc&\cite{Sivers:1970zm}\\
${}_1^1$H&${1\over2}$&2.79&2.79&$l-{1\over2}=0$&15.1 keV
&NR,hc&\cite{Bracci:1983fe}\\
&&&&&320 keV
&NR,hc&\cite{Sivers:1970zm}\\
&&&&&50--1000 keV&NR,FF&
\cite{Olaussen:1984xb}\\
&&&&&263 keV&R&
\cite{Olaussen:1983qc,Olaussen:1983bm}\\
${}_1^2$H&$1$&$0.857$&1.71&$l-1=0$ ($|m'|=1$)&${130\over\lambda}$ keV&R,IM
&\cite{Olsen:1990jm,Olsen:1990jn}\\
${}_2^3$He&${1\over2}$&$-2.13$&$-3.20$&$l+{1\over2}={3\over2}$&
13.4 keV&NR,hc&\cite{Bracci:1983fe}\\
${}_{13}^{27}$Al&${5\over2}$&3.63&7.56&$l-{5\over2}=4$
&2.6 MeV&NR,FF&\cite{Olaussen:1983qc,Olaussen:1983bm}\\
${}_{13}^{27}$Al&${5\over2}$&3.63&7.56&$l-{5\over2}=4$
&560 keV&NR,hc&\cite{Goebel:1983xf}\\
${}_{48}^{113}$Cd&${1\over2}$&$-0.62$&$-1.46$&$l+{1\over2}={49\over2}$&
6.3 keV&NR,hc&\cite{Bracci:1983fe}\\
\br
\end{tabular}
\end{table}

\subsection{Remarks on binding}
Clearly, this summary indicates that the theory of monopole binding
to nuclear magnetic dipole moments is rather primitive.  The angular
momentum criteria for binding is straightforward; but in general
(except for relativistic spin 1/2) additional interactions have to
be inserted by hand to regulate the potential at $r=0$.  The results
for binding energies
clearly are very sensitive to the nature of that additional interaction.
It cannot even be certain that binding occurs in the allowed states.
In fact, however, it seems nearly certain that monopoles will bind
to all nuclei, even, for example, Be, because the magnetic field in
the vicinity of the monopole is so strong that the monopole will
disrupt the nucleus and will bind
to the nuclear, or even the subnuclear, constituents.

\subsection{Binding of monopole-nucleus complex to material lattice}
\label{bindingtolattice}
Now the question arises: Can the magnetic field in the detector
extract the monopole from the nucleus that binds it?  And if not,
is the bound complex of nucleus and monopole
rigidly attached to the crystalline lattice of the material?
To answer the former question we regard it as a simple tunneling situation.  
The decay rate is estimated by the WKB formula
\numparts
\begin{equation}
\Gamma\sim {1\over a}\exp\left[-\frac2\hbar
\int_a^b\rmd r\sqrt{2{\cal M}(V-E)}\right],
\end{equation}
where the potential is crudely that due to the dipole interaction and the
external magnetic field,
\begin{equation}
V=-{\mu g\over r^2}-gBr,
\end{equation}
\endnumparts
$\cal M$ is the nuclear mass $\ll$ monopole mass,
and the inner and outer turning points, $a$ and $b$ are the zeroes of
$E-V$.  Provided the following equality holds,
\begin{equation}
(-E)^{3}\gg g^3\mu B^2,
\end{equation}
which should be very well satisfied, since the right hand side is about
$10^{-19} |m'|^3$ MeV$^3$, for the CDF field of $B=1.5$ T,
we can write the decay rate as
\begin{equation}
\Gamma\sim |m'|^{-1/2} 10^{23} \mbox{s}^{-1}
 \exp\left[-{4\sqrt{2}\over 3 \cdot 137}
\left({-E\over m_e}\right)^{3/2}{B_0\over |m'|B}A^{1/2}\left(
{m_p\over m_e}\right)^{1/2}\right],
\end{equation}
where the characteristic field, defined by $eB_0=m^2_e$, is $4\times 10^9$ T.
If we put in $B=1.5$~T, and $A=27$, $-E=2.6$MeV, appropriate for
${}_{13}^{27}$Al, we have for the exponent, for $m'=1/2$, $-2\times 10^{11}$,
corresponding to a rather long time!  To get a 10 yr lifetime, the
binding energy would have to be only of the order of 1 eV.  Monopoles
bound with kilovolt or more energies will stay around forever.

Then the issue is whether  the entire Al atom-monopole
complex  can be extracted
with the 1.5 T magnetic field present in CDF.  The answer seems to be
unequivocally no.  The point is that the atoms are rigidly bound in a
lattice, with no nearby site into which they can jump.  A major disruption
of the lattice would be required to dislodge the atoms, which would
probably require kilovolts of energy.
Some such disruption was made by the monopole when it came to rest and
was bound in the material, but that disruption would be very unlikely to
be in the direction of the accelerating magnetic field.  Again, a simple
Boltzmann argument shows that any effective binding slightly bigger than 1 eV
will result in monopole trapping ``forever.''  This argument applies equally
well to binding of monopoles in ferromagnets.  If monopoles bind strongly
to nuclei there, they will not be extracted by 5 T fields, contrary to
the arguments of Goto \etal \cite{goto}.
The corresponding limits on monopoles from
ferromagnetic samples  of Carrigan \etal \cite{Carrigan:1978ku} are suspect.

\section{Searches for magnetic monopoles}
With the advent of ``more unified'' non-Abelian theories, classical composite
monopole solutions were discovered, as briefly discussed in \sref{namono}.
  The mass of these mono\-poles
would be of the order of the relevant gauge-symmetry breaking scale,
which for grand unified theories is of order $10^{16}$ GeV
or higher.  But there are models where the electroweak symmetry
breaking can give rise to monopoles of mass $\sim 10$ TeV
\cite{Preskill:1984gd,Kirkman:1981ck,coleman,Cho:1996qd}.
 Even the latter are not yet accessible to accelerator experiments,
 so limits on heavy monopoles depend either on cosmological
considerations (see for example \cite{turnerrev})
or detection of cosmologically produced (relic) monopoles
impinging upon the earth or
moon \cite{Eberhard:1971re,Ross:1973it,price84,cabrera,cabrera2,
Jeon:1995rf,jeonerr,ambrosio}. 
Since the revival of interest in monopoles in the 1970s, there have been
two well-known announcements of their discovery: that of Price \etal
\cite{Price:1975zt}, who found an cosmic ray track etched in a
plastic detector, and that of Cabrera \cite{cabrera}, 
who reported a single event in a induction loop.
The former interpretation was immediately refuted by Alvarez 
\cite{Alvarez:1975gm}, while the latter
has never been duplicated, so is presumed spurious.

However, {\it a priori}, there is no reason that Dirac/Schwinger monopoles
or dyons of arbitrary mass might not exist: {\em In this respect, it
is important to set limits below the 1 TeV scale.}

\subsection{Direct searches}
In this review we will concentrate on recently obtained limits,
since periodic reviews of the status of magnetic monopole searches have
been published \cite{gp,Giacomelli:2005xz}.  Before 2000,
the best previous direct
limit on magnetic monopoles was that obtained at Fermilab by Bertani \etal
\cite{Bertani:1990tq} who obtained cross section limits of $2\times 10^{-34}$
cm$^2$ for monopole masses below 850 GeV.  As we shall see below, the
Oklahoma experiment \cite{Kalbfleisch:2003yt}, while not extending to as
high masses, gives cross section limits some two orders of magnitude smaller.
The recent CDF experiment \cite{Abulencia:2005hb} sets a three order
of magnitude improvement
over \cite{Bertani:1990tq}.  (In contrast to \cite{gp,Giacomelli:2005xz},
we call all of these experiments ``direct,'' whether they are searching
for previously produced monopoles trapped in material or the ionization
and radiation produced by monopoles passing through a detector.)
As noted in \sref{bindingtolattice} there have been experiments to search 
for monopoles by extracting them from matter with strong magnetic fields
\cite{Carrigan:1978ku,Giacomelli:1984gq}; as remarked there, it is doubtful
such an experiment would succeed, since the binding energy of a monopole
to the lattice is probably at least in the keV range, while the energy 
acquired by a Dirac monopole in a 100 kG field over an atomic distance is only 
20 eV.

Cosmologically produced monopoles are commonly assumed to arise from a
GUT (Grand Unified Theory) where a grand unified group such as SU(5)
breaks down into the standard model group SU(3)$\times$SU(2)$\times$U(1).
Barring premature unification due to, say, large extra dimensions, the
mass of such a monopole is expected to be of the order $10^{16}$ GeV, 
so they are incapable of being produced in accelerators.  However,
since in the early universe at least one monopole should be produced per
causal domain, too many monopoles would have been produced 
\cite{Preskill:1979zi,Turner:1982ag}, and would come into conflict with
the Parker bound, which states that cosmic fields would be quenched if
the density of magnetic monopoles is too high \cite{Parker:1970xv}.
This is one of the problems solved by inflation.

Various experiments have been conducted to look for cosmic monopoles.
An interesting limit comes from the Rubakov-Callan mechanism for monopole
catalysis of proton decay \cite{Rubakov:1981rg,Callan:1982au},
\be
M+p\to M+e^++\pi^0,
\ee
where MACRO \cite{Ambrosio:2002qu}
found a limit on the flux of 3--8$\times 10^{-16}$ cm$^{-2}$
s$^{-1}$ sr$^{-1}$. However, MACRO's best limit
\cite{Ambrosio:2002qq}, based on scintillation
counters, limited streamer tubes, and nuclear track detectors, give
a much better limit of $1.4\times 10^{-16}$ cm$^{-2}$ s$^{-1}$ sr$^{-1}$
for monopole velocities in the range
 $4\times 10^{-5}<\beta<1$, roughly a factor of two improvement over 
previous limits, and well below the Parker bound of 
$\sim10^{-15}$ cm$^{-2}$ s$^{-1}$ sr$^{-1}$.  
Even smaller limits depend on the mechanism by which a monopole would
produce  a track in ancient mica \cite{Price:1988ge,Ghosh:1990ki}.
One should note that lower mass monopoles, with masses of order $10^{10}$ GeV,
arising from intermediate stages of symmetry breaking below the GUT
scale, would not catalyze proton decay \cite{Lazarides:1986rt,Kephart:2001ix},
but the more stringent MACRO limits still apply.

We will discuss the three recent direct search limits in sections 
\ref{sec:ok}--\ref{sec:cdf}.

\subsection{Indirect searches}
In the above, and in the following sections, we discussed direct searches,
where the monopoles are searched for as free particles.  The
{\it indirect\/} searches that have
been proposed and carried out rely on effects attributable to the
virtual existence of monopoles.  De
R\'ujula in 1995 \cite{DeRujula:1994nf} 
proposed looking at the three-photon decay of the $Z$ boson, 
where the process proceeds through a virtual monopole loop, as shown
in \fref{figlbl}. 
If we use his formula for the
branching ratio for the $Z\to3\gamma$
process, compared to the current
experimental upper limit \cite{acciarri}
for the branching ratio of $10^{-5}$, we can rule out
monopole masses lower than about 
400 GeV, rather than the 600 GeV quoted by De R\'ujula.
Similarly, Ginzburg and
Panfil in 1982 \cite{Ginzburg:1982fk}
and more recently Ginzburg and Schiller in 1999
\cite{Ginzburg:1998vb,Ginzburg:1999ej}
considered the production of two photons with
high transverse momenta by the collision of two photons produced either
from $e^+e^-$ or $q\overline{q}$  collisions. Again the final photons are
produced through a virtual monopole loop.  
Based on this theoretical scheme, an experimental limit was given by the 
D0 collaboration  \cite{Abbott:1998mw},
which sets the following bounds on the monopole mass $M$:
\be
\frac{M}{2|m'|}>\left\{\begin{array}{cc}
610 \mbox{ GeV}&\mbox{ for } S=0\\
870 \mbox{ GeV}&\mbox{ for } S=1/2\\
1580 \mbox{ GeV}&\mbox{ for } S=1 \end{array}\right.,
\ee
where $S$ is the spin of the monopole, and $m'=eg$ is the magnetic charge
quantization number.

It is worth noting that a lower mass limit of 120 GeV
 for a Dirac monopole has been
set by Graf, Sch\"afer, and
Greiner \cite{Graf:1991xe}, based on the monopole
contribution to the vacuum polarization correction to the muon anomalous
magnetic moment. (Actually, we believe that the correct limit, obtained
from the well-known textbook formula for the $g$-factor correction
due to a massive Dirac particle is 60 GeV.)

\subsubsection{Difficulty with indirect limits}
\begin{figure}[h]
\centerline{\psfig{figure=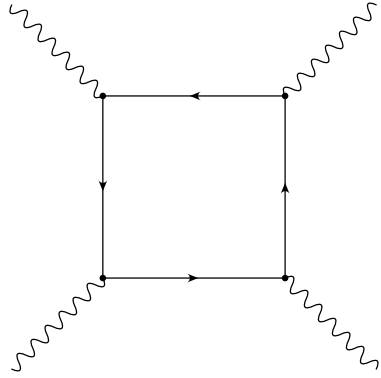,height=2in,width=2in,angle=270}}
\caption{The light-by-light scattering graph for either an electron
or a monopole loop.}
\label{figlbl}
\end{figure}

The indirect limits mentioned above rely up the
Feynman graph shown in \fref{figlbl}.
If the particle in the loop
is an ordinary electrically charged electron, this process is
well-known. If, further, the photons involved are of very low
momentum compared the the mass of the electron, then the result may be
simply derived from the well-known {\em Euler-Heisenberg Lagrangian}
\cite{Heisenberg:1935qt,weisskopf36,Schwinger:1951nm}
which for a spin-1/2 charged-particle loop in the presence of 
weak {\em homogeneous} electric and magnetic fields is
\begin{eqnarray}
{\cal L}&=&-\frac{1}{16\pi}F^2+\frac{\alpha^2}{360}\frac{1}{m^4}
\frac1{(4\pi)^2}[4(F^2)^2+7(F  {}^*F)^2],\label{weakeh}
\end{eqnarray}
where $m$ is the mass of the particle in the loop.
The Lagrangian for a spin-0 and spin-1 charged particle in the loop
is given by similar formulas, which are derived in 
\cite{Schwinger:1951nm,psf2}
and (implicitly) in Ref.~\cite{Dong:1992hg,Dong:1992fa,Jikia:1993tc},
 respectively.

Given this homogeneous-field effective Lagrangian, 
it is a simple matter to derive the cross section for the
$\gamma\gamma\to\gamma\gamma$ process in the
low energy limit. (These results can, of course, be directly calculated
from the corresponding one-loop Feynman graph with on-mass-shell photons.
See \cite{psf2,lp}.)
Explicit results for the differential cross section are given
in textbooks:
\be
\frac{\rmd\sigma}{\rmd\Omega}=
\frac{139}{32400\pi^2}\alpha^4\frac{\omega^6}{m^8}(3+\cos^2\theta)^2,
\ee
and the total cross section for a spin-1/2 charged particle in the loop
is
\be
\sigma=\frac{973}{10125\pi}\alpha^4\frac{\omega^6}{m^8},\qquad 
\omega/m\ll1,\qquad s=4\omega^2.\label{10.5}
\ee
The numerical coefficient in the total cross section
are $0.00187$, $0.0306$, and
$3.50$ for spin 0, spin 1/2, and spin 1 particles in the loop, respectively.

How is this applicable to photon scattering through a monopole loop?
This would seem impossible because of the existence of the string,
which renders perturbation theory meaningless.
Of course, no one has
attempted a calculation of the ``box'' diagram with the
monopole interaction.
Rather, De R\'ujula and Ginzburg (explicitly or implicitly) appeal to
{\em duality}, that is, the dual
symmetry \eref{duality} that the introduction of magnetic charge brings 
to Maxwell's equations:
\be
{\bf E}\to {\bf B}, \qquad {\bf B}\to -{\bf E},
\ee
and similarly for charges and currents.
 Thus the argument is that
for low energy photon processes
it suffices to compute the fermion loop graph in the presence of
zero-energy photons, that is, in the presence of static, constant fields.
Since the Euler-Heisenberg Lagrangian  is invariant
under the duality substitution on the fields alone,
this means we obtain the low energy
cross section 
$\sigma_{\gamma\gamma\to\gamma\gamma}$
 through the monopole
loop from the equation for the QED cross section by the substitution
{$e\to g$},  or
\be
\alpha\to\alpha_g=137m^{\prime2},\qquad 2|m'|=1,2,3,\dots.
\ee

It is critical to emphasize that the Euler-Heisen\-berg Lagrangian is
an effective
Lagrangian for calculations at the {\it one fermion loop level\/} for low
energy, i.e., $\omega/m\ll1$.
However, it becomes unreliable if radiative corrections are 
large. (The same has been noted in another context by
Bordag, Robaschik, Wieczorek, and Lindig 
\cite{Bordag:1998sw,Bordag:1983hk}.)
For example, the internal radiative correction to the box
diagram have been computed by
Ritus \cite{ritus} and by Reuter, Schmidt, and Schubert 
\cite{Reuter:1996zm,Fliegner:1997ra}
in QED.  In the $O(\alpha^2)$ term in the expansion of the EH Lagrangian
(\ref{weakeh}), the coefficients of the $(F^2)^2$ and the
$(F\tilde F)^2$ terms are multiplied by
\be
\left(1+\frac{40}{9}\frac{\alpha}{\pi}
+O(\alpha^2)\right)\quad\mbox{and}\quad
\left(1+\frac{1315}{252}\frac{\alpha}{\pi}+O(\alpha^2)
\right),
\ee
respectively.  The corrections become meaningless when we {\it replace\/}
$\alpha\to\alpha_g$.

\subsubsection{Unitarity bound}
 This would seem to be a
devastating objection to the results given by Ginzburg \etal 
\cite{Ginzburg:1998vb,Ginzburg:1999ej} and used
in the D0 analysis \cite{Abbott:1998mw}.  
But even if one closes one's eyes to higher order 
effects, it seems clear that the mass limits quoted are inconsistent.

If we take the cross section given by \eref{10.5} and make the duality
substitution, we obtain for the low energy light-by-light
scattering cross section in the presence of a monopole loop
($M$ is the monopole mass)
\be
\sigma_{\gamma\gamma\to\gamma\gamma}\approx \frac{973}{10125\pi}
\frac{m^{\prime8}}{\alpha^4}\frac{\omega^6}{M^8}
=1.08\times 10^7 \,m^{\prime8}\frac{1}{M^2}
\left(\frac{\omega}M\right)^6.
\ee
If the cross section were dominated by a single partial wave of angular
momentum $J$, the cross section would be bounded by
\be
\sigma\le\frac{\pi(2J+1)}{s}\sim \frac{3\pi}{s},\qquad J\sim1.
\ee
Comparing this with the cross section given above, we obtain the following
inequality for the cross section to be consistent with unitarity,
\be
\frac{M}{\omega}\ge 6|m'|.
\ee
But the limits quoted by D0 for the monopole mass are less than this:
\be
\frac{M}{2|m'|}>870 \mbox{ GeV}, \quad \mbox{spin } 1/2,
\ee
because, at best, a minimum estimate is $\langle\omega\rangle\sim 300$
GeV, so the theory cannot sensibly be applied below a monopole mass of 
about 1 TeV. (Note that changing the
value of $J$ in the unitarity limits has very little effect on the bound
 since an 8th root is taken: Replacing $J$ by 50 reduces
the limit only by 50\%.)

Similar remarks can be directed toward the De R\'ujula limits
\cite{DeRujula:1994nf}. That author, however, notes the
``perilous use of a perturbative expansion
in $g$.''  However, although he writes down the correct vertex,
 he does not, in fact, use it,
instead appealing to duality, and even so he admittedly omits
enormous radiative corrections of $O(\alpha_g)$ without any justification
other than what we believe is a specious reference to the use of effective
Lagrangian techniques for these processes.

As we will see, some of the same objections apply to the direct search limits.
The advantage, however, of the latter, is that the signal of a positive event
is more unambiguous, and in the Oklahoma and H1 experiments, a monopole, if
found, would be available for further study.

\section{Oklahoma experiment: Fermilab E882}
\label{sec:ok}
The best prior experimental limit on the direct accelerator production
of magnetic monopoles is that
of Bertani \etal in 1990  \cite{Bertani:1990tq} 
(see also Price \etal \cite{Price:1987py,Price:1990in}):
\be
\sigma\le2\times 10^{-34}\mbox{cm}^2 \quad\mbox{for a monopole mass}\quad
M\le850\,\mbox{GeV}.
\ee
The fundamental mechanism is supposed to be a Drell-Yan process,
\be
p+\bar p\to M+\bar M+X,
\ee
where the cross section is given by
\begin{eqnarray}\frac{\rmd\sigma}{\rmd \mathcal{M}}
=(68.5n^2)^2\beta^3\frac{8\pi\alpha^2}{9s}
\int \frac{\rmd x_1}{x_1}\sum_iQ_i^2q_i(x_1)\bar q_i
\left(\frac{\mathcal{M}^2}{sx_1}\right).
\end{eqnarray}
Here $\mathcal{M}$ 
is the invariant mass of the monopole-antimonopole pair, and we have
included a factor of $\beta^3$
to reflect (1) phase space and (2) the velocity
suppression of the magnetic coupling, as roughly implied by 
\eref{interact}--See also \eref{velsup}.
Note that we are unable to calculate the elementary process
$$q\bar q\to \gamma^*\to M\bar M$$ perturbatively, so we must use
nonperturbative estimates.

\begin{figure}
\centering
\includegraphics[height=7cm]{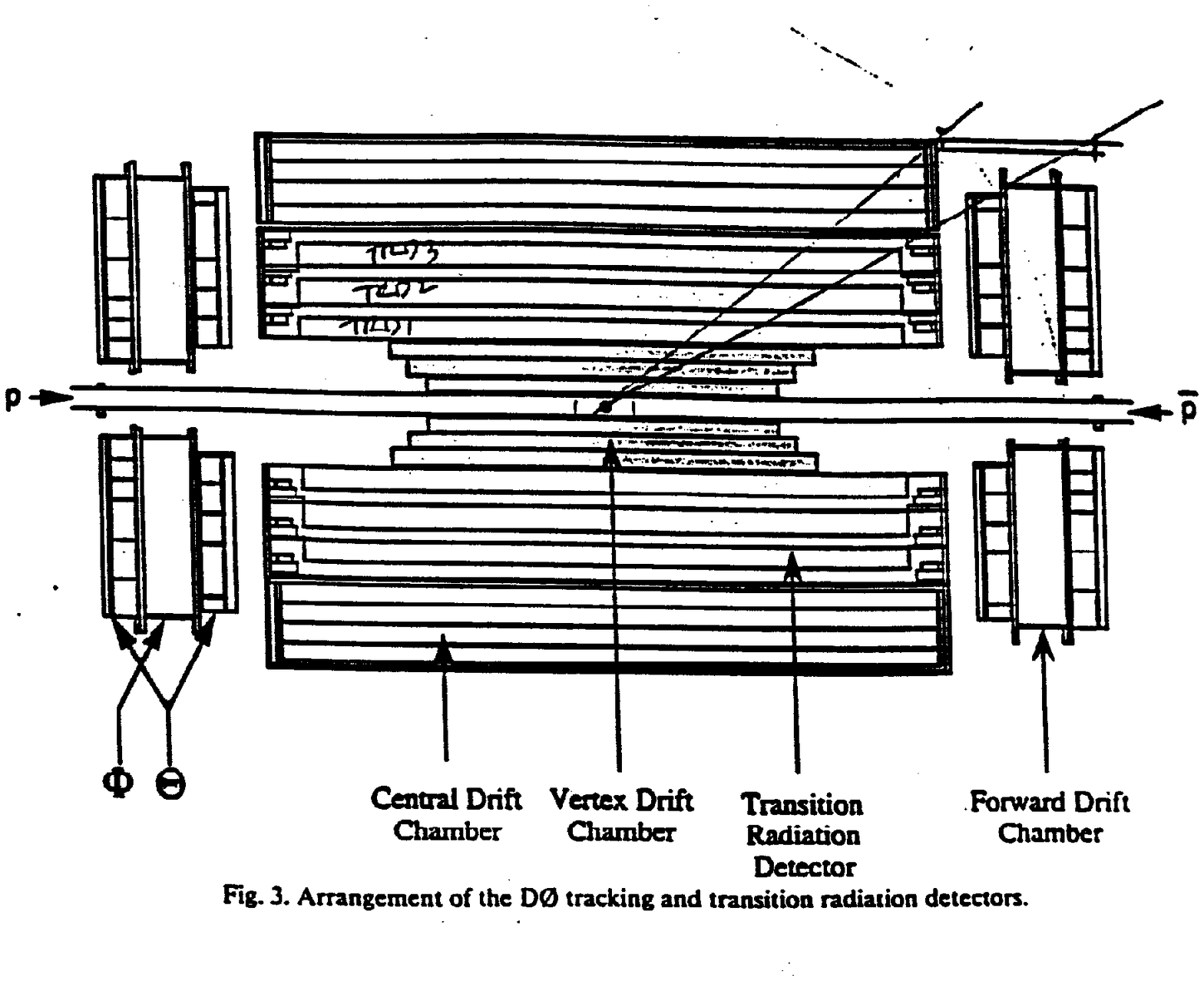}
\caption{\label{D0Detector} Arrangement of the D0 tracking and 
transition radiation detectors.}
\end{figure}

Any monopole produced at Fermilab is trapped in the detector elements with
100\% probability due to interaction with the magnetic moments of the
nuclei, based on the theory described in \sref{sec:binding}.
 The experiment consists of running samples obtained from the old
D0 and CDF detectors through a superconducting induction detector.
\Fref{D0Detector} is a sketch of the D0 detector.
We are able to set much better limits than Bertani \etal \cite{Bertani:1990tq}
because the integrated luminosity is $10^4$ times that of the previous 1990 
experiment:
\be
\int {\cal L}=172\pm8\,\mbox{pb}^{-1} \quad(\mbox{D0}).
\ee
We use energy loss formula of Ahlen
\cite{Ahlen:1982rw,Ahlen:1982mx,Ahlen:1978jy}, as described in \sref{sec:dedx}.
 The graph in \fref{figel} shows the energy loss 
$\rmd E/\rmd x$ for various materials.

\begin{figure}
\centering
\includegraphics[height=7cm]{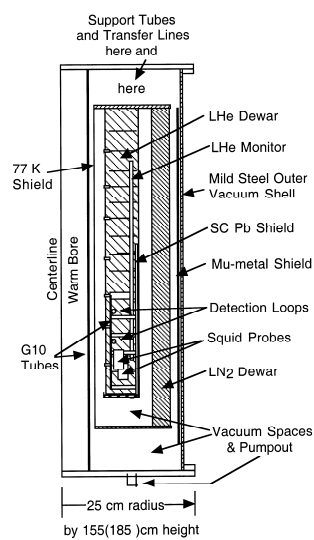}
\caption{\label{oudetector}
Sketch of the OU induction detector. Shown is a vertical cross
section; it should be imagined as rotated about the vertical axis labelled
``centerline.''}
\end{figure}

\Fref{oudetector} is a diagram of the OU magnetic monopole induction detector.
It is a cylindrical detector, with a warm bore of diameter 10 cm, surrounded
by a cylindrical liquid N$_2$ dewer, which insulated a liquid He dewer.  
The superconducting loop detectors were within the latter, concentric
with the warm bore.  Any current established in the loops was detected by a
SQUID.  The entire system was mechanically isolated from the
building, and magnetically isolated by $\mu$ metal and superconducting lead
shields.  The magnetic field within the bore was reduced with the help of
Helmholtz coils to about 1\% of the earth's field.  
Samples were pulled vertically through the warm bore with a
computer controlled stepper motor.  Each traversal took about 50 s; every
sample run consisted of some 20 up and down traversals.  Most samples
were run more than once, and more than 660 samples of Be, Pb, and Al from
both the old CDF and D0 detectors were analyzed over a period of 7 years.

\subsection{Monopole induced signal}

\begin{figure}
\centering
\includegraphics[height=10cm]{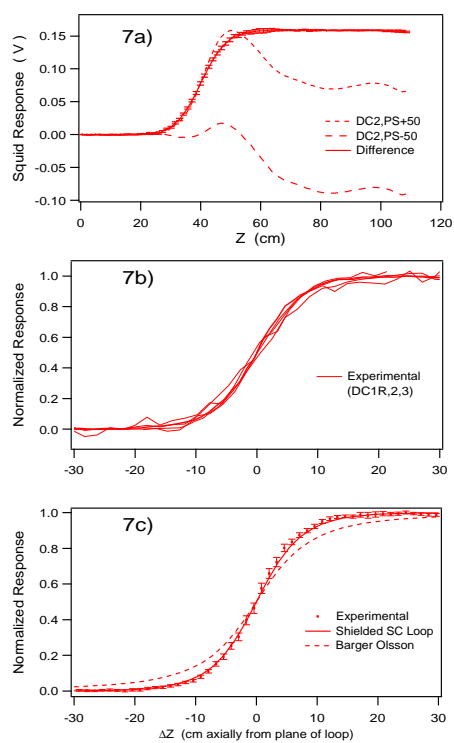}
\caption{\label{caldata}
Typical step plots: D0 aluminum, CDF lead, and CDF aluminum.
The experimental data was collected from pseudopole
simulations; the steps shown are for the difference between the results with
reversed polarizations of the pseudopole.  Data agrees well with the theory
which incorporates the effect of the shielded superconducting loops.
The theory without the shield, given by Barger and Ollson 
\cite{bargerandollson}, is also shown.}
\end{figure}

Note that if the shield were not present, the supercurrent induced 
by a monopole of strength $g$ passing through a
loop of radius $r$ and inductance $L$ would be given by
\be
I(t)=\frac{2\pi g}{Lc}\left(1-\frac{z(t)}{\sqrt{r^2+z(t)^2}}\right),
\ee
where $z(t)$ is the vertical position of the monopole relative to
the position of the center of the loop.  A more detailed theory is
described in the following. The theory can be verified
with a {\em pseudopole,} which is a long, $\sim 1$ m, electromagnetic solenoid,
which produces a field near one end very similar to that produced by a pure
magnetic pole.  The excellent agreement between theory and experiment is 
indicated in \fref{caldata}.

\subsubsection{Simplified theory of monopole detector}
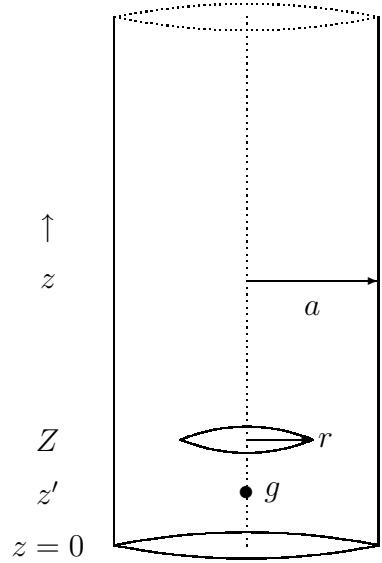
\begin{figure}
\centering
\begin{picture}(200,200)
\put(50,0){\line(0,1){200}}
\put(150,0){\line(0,1){200}}
\qbezier[80](100,0)(100,100)(100,200)
\qbezier(50,0)(100,10)(150,0)
\qbezier(50,0)(100,-10)(150,0)
\put(100,20){\makebox(0,0){$\bullet$}}
\put(110,20){\makebox(0,0){$g$}}
\put(25,20){\makebox(0,0){$z'$}}
\qbezier(75,40)(100,50)(125,40)
\qbezier(75,40)(100,30)(125,40)
\put(25,40){\makebox(0,0){$Z$}}
\put(100,40){\vector(1,0){25}}
\put(130,40){\makebox(0,0){$r$}}
\put(100,100){\vector(1,0){50}}
\put(125,90){\makebox(0,0){$a$}}
\put(25,100){\makebox(0,0){$z$}}
\put(25,120){\makebox(0,0){$\uparrow$}}
\put(25,0){\makebox(0,0){$z=0$}}
\qbezier[50](50,200)(100,210)(150,200)
\qbezier[50](50,200)(100,190)(150,200)
\end{picture}
\caption{Diagram of monopole detector.  The monopole $g$ is assumed to
be on the central axis at a height $z'$ above the bottom of the detector,
which we model as a cylindrical perfectly conducting can of radius $a$,
closed at the bottom.  The superconducting loop of radius $r$ is a height
$Z$ above the base.}
\label{figapp1}
\end{figure}

This subsection describes the basis of the functioning of our magnetic monopole
detector.  It works by detecting the magnetic flux intercepted by
a superconducting loop contained within a superconducting cylinder.
The detector is sketched in \fref{figapp1}.

In order to incorporate finite-size effects,
we consider first a perfectly conducting right circular cylinder of radius
$a$ of semi-infinite length, with axis along the $z$-axis, and with a
perfectly conducting circular bottom cap at $z=0$.  We use cylindrical
coordinates $\rho$, $\theta$, and $z$.

Because the boundaries are superconductors, the normal component of
$\bf B$ must vanish on the surfaces, that is,
\begin{equation}
B_\rho\bigg|_{{\rho=a}\atop{z>0}}=0,\quad B_z\bigg|_{z=0}=0.
\label{a0}
\end{equation}

Now suppose a magnetic pole of strength $g$ is placed on the
$z$ axis at $z=z'>0$.  This could either be a magnetic monopole (magnetic
charge) or one pole of a very long electromagnet (``pseudopole'').
Imagine a circular conducting loop of radius $r<a$ centered on the axis
of the cylinder and perpendicular to that axis, with center at $z=Z$.
Inside the cylinder and outside of the loop,  $\bf B$ is
derivable from a magnetic scalar potential,
\begin{equation}
{\bf B}=-\mbox{\boldmath{$\nabla$}}\phi_M,
\label{na1}
\end{equation}
since we may ignore the displacement current, because the
time variation is negligible.   $\phi_M$ satisfies
Poisson's equation, in cylindrical coordinates:
\begin{eqnarray}
\fl\mbox{\boldmath{$\nabla$}}\cdot\mathbf{B}=
-\left(\frac1\rho\frac\partial{\partial \rho}\rho\frac\partial{\partial \rho}
+\frac1{\rho^2}\frac{\partial^2}{\partial\theta^2}+\frac{\partial^2}
{\partial z^2}\right)\phi_M
=4\pi g\delta({\bf r-r'}),
\label{na2}
\end{eqnarray}
where $\mathbf{r}'$ is the position of the monopole, $\mathbf{r'}=(\rho',
\theta',z')$.
This is the equation for a Green's function, which we can express
in separated variables form.  That is, we write
\begin{eqnarray}
\phi_M=\frac2\pi\int_0^\infty \rmd k\cos kz\cos kz'
\sum_{m=-\infty}^\infty
\frac{1}{2\pi}\rme^{\rmi m(\theta-\theta')}g_m(\rho,\rho';k),
\label{na4}
\end{eqnarray}
where, in view of the first boundary condition in (\ref{a0}),
we may express the reduced Green's function in terms of modified Bessel
functions:
\begin{eqnarray}
g_m(\rho,\rho';k)=-4\pi gI_m(k\rho_<)
\left[K_m(k\rho_>)-I_m(k\rho_>)
\frac{K'_m(ka)}{I'_m(ka)}\right],
\label{na5}
\end{eqnarray}
where $\rho_<$ ($\rho_>$) is the lesser (greater) of $\rho$, $\rho'$.
If the monopole is confined to the $z$ axis, only the $m=0$ term survives:
\begin{eqnarray}
\phi_M=-\frac{4g}{\pi}\int_0^\infty \rmd k\cos kz\cos kz'
\left[K_0(k\rho)+I_0(k\rho)\frac{K_1(ka)}{I_1(ka)}\right],
\label{na6}
\end{eqnarray}
which uses
\begin{equation}
I_0'(x)=I_1(x),\qquad K_0'(x)=-K_1(x).
\label{na7}
\end{equation}

By integrating over the cross section of the loop using
\begin{eqnarray}
\int_0^x \rmd t\,t\,K_0(t)=-x\,K_1(x)+1,\qquad
\int_0^x \rmd t\,t \,I_0(t)=x\,I_1(x),
\label{na8}
\end{eqnarray}
we obtain the following formula for the magnetic flux subtended
by the loop,
\begin{equation}
\Phi=\int \rmd{\bf S\cdot B}=4\pi g\left[\eta(Z-z')-F(Z,z')
\right],\label{na9}
\end{equation}
where the step function is (\ref{stepfunction}),
and the response function is
\begin{eqnarray}
\fl F(z,z')=\frac2\pi\frac{r}{a}\int_0^\infty \rmd x\,\sin x\frac{z}{a}
\cos x\frac{z'}{a}
\left\{K_1(xr/a)-I_1(xr/a)\frac{K_1(x)}{I_1(x)}\right\}.
\label{na11a}
\end{eqnarray}

Now suppose that the pole is {\em slowly\/} moved from
a point far above the loop, $z'=+\infty$, to a point below the loop,
$z'=z_0$, $Z>z_0$.  Then from Maxwell's equation
\begin{equation}
\mbox{\boldmath{$\nabla$}}\times{\bf E}=-\frac1c\frac{\partial}{\partial t}
{\bf B}-\frac{4\pi}c{\bf J}_m,
\label{na11}
\end{equation}
where $\mathbf{J}_m$ is the magnetic current density,
the emf induced in the loop is
\begin{equation}
{\cal E}=\oint{\bf E}\cdot \rmd{\bf l},
=-\frac{\rmd\Phi}{c\rmd t}+\frac{4\pi}{c}g\delta(t),
\label{na13}
\end{equation}
if $t=0$ is the time at which the pole passes through the plane of the
loop.  The net change in emf gives rise to a persistent current $I$ in the
superconducting loop,
\begin{equation}
LI=\int_{-\infty}^\infty {\cal E} \rmd t=-\frac1c\Delta\Phi+\frac{4\pi}c g
=\frac{4\pi}c gF(Z,z_0),
\label{na14}
\end{equation}
where $L$ is the inductance of the loop, and the response function
$F$ is given in (\ref{na11a}).  
This is just a statement of the Meissner
effect, that the flux change caused by the moving monopole is cancelled by
that due to the current set up in the loop.

When the loop is very far from the bottom cap,
$Z\gg a$, only small $x$ contributes to the integral in Eq.~(\ref{na11a}),
and  it is easy to see that
\begin{equation}
\int_{-\infty}^\infty {\cal E}\rmd t= \frac{4\pi g}c\left(1-\frac{r^2}{a^2}\right),
\label{na15}
\end{equation}
so the signal is maximized by making the loop as small as possible, relative
to the radius of the cylinder.  We get the full flux of the monopole
only for a loop in empty space, $a/r\to\infty$.
This perhaps counterintuitive effect is
due to the fact that the superconducting walls confine
the magnetic flux to the interior of the cylinder.
Thus for the superconducting can, the induced current in the detection loop
caused by the passage of a monopole from $z'=\infty$ to $z'=0$ is
\begin{equation}
LI=\frac{4\pi g}c-\frac{\Delta\Phi}c=\frac{4\pi g}c-\frac{\Phi(z'=0)}c,
\label{na16}
\end{equation}
which yields the result (\ref{na15})
if one assumes that the magnetic field is uniform across the can's cross
section at the position of the loop when the pole is at the bottom, because
all the flux must pass up through the can.  If we consider, instead, an
infinite, open-ended, superconducting cylinder, with the monopole passing
from $z=+\infty$ to $z=-\infty$, at either extreme half the flux must
cross the plane of the loop, so with the uniformity assumption we get the
same result:
\begin{equation}
LI=\frac{4\pi g}c-\frac{\Delta\Phi}c
=\frac{4\pi g}{c}\left(1-\frac{r^2}{a^2}\right).
\label{na17}
\end{equation}
The simple assumption of a uniform magnetic field is apparently justified
by the exact result (\ref{na15}).

We conclude this discussion by noting how the exact calculation is modified
for an infinite superconducting cylinder.  In the magnetic scalar
potential, the integral over $k$ mode functions in (\ref{na4}) is replaced by
\begin{equation}
\int_{-\infty}^\infty\frac{\rmd k}{2\pi}\rme^{\rmi k(z-z')},
\label{na18}
\end{equation}
which has the effect of replacing the flux expression (\ref{na9})
by
\begin{equation}
\Phi=2\pi g[\epsilon(z-z')-F(Z-z',0)],
\label{na19}
\end{equation}
where $\epsilon(\xi)$ is given by \eref{epsilon}.
Then the induced current in the detection loop when the monopole passes
from a point above the loop  $z'=Z+\xi$ to a point, equidistant, below
the loop, $z'=Z-\xi$, is
\begin{equation}
LI=\frac{4\pi g}cF(\xi,0)\to \frac{4\pi g}c\left(1-\frac{r^2}{a^2}\right),
\label{na20}
\end{equation}
where the last limit applies if $\xi/a\gg1$.  This result coincides with
that in (\ref{na15}).
The function $R(\xi)=\frac12F(\xi,0)/(1-r^2/a^2)+\frac12$,
corresponding to a monopole starting from a point $z_1$ far above the loop,
$z_1-Z\gg a$, and ending at a point $z_0=Z-\xi$,
 is plotted as a function of $\xi$
for our parameter values in \fref{caldata}c, where it is shown to
agree well with experimental data.  This response
function  coincides
with the result obtained from (\ref{na14}),
because
\begin{eqnarray}
\fl F(Z,Z-\xi)=\frac12 F(2Z-\xi,0)+\frac12 F(\xi,0)
\approx\frac12\left(1-\frac{r^2}{a^2}\right)+\frac12F(\xi,0),
\end{eqnarray}
if $Z/a\gg1$. This shows that
the effect of the endcap (which of course is not present in actual detector)
is negligible,  demonstrating that the fact that the superconducting
shield is of finite length is of no significance.

\subsection{Background effects}
All nonmagnetic but conducting samples possess:
\begin{itemize}
\item Permanent magnetic dipole moments $\mu$, which give rise to signals
in free space of the form
\be
I(t)=-\frac{2\pi\mu_z}{Lc}\frac{r^2}{[r^2+z(t)^2]^{3/2}}
\ee
\item Induced magnetization:  Conducting samples passing though magnetic
gradients with speed $v$ produce time-varying magnetic fields which induce
signals in our detector,
\be
I(t)=\frac{v}{c^3}\frac{1}{L}\int(\rmd{\bf r})r^2\sigma(r^2)\frac{\partial
B_z}{\partial z'}(z')\frac{1}{r}H\left(\frac{z'}{r},\frac{a}{r}\right),
\ee
where $H$ is the response function, essentially that appearing in 
(\ref{na11a}),
\numparts
\begin{eqnarray}
H\left({z'\over r},{a\over r}\right)&=&\int_0^\infty \rmd 
y\,y \cos y{z'\over r}
\left[K_1(y)-I_1(y){K_1(ya/r)\over I_1(ya/r)}\right]\\
&\to&{\pi\over 2}{r^3\over(r^2+z^{\prime2})^{3/2}},\quad a/r\to\infty.
\end{eqnarray}
\endnumparts
\end{itemize}

\subsection{Calibration, real data, and limits}
The pseudopole data shown in \fref{caldata}
clearly shows that we could detect a Dirac pole.
We demonstrated that the detector (SQUID response)  was remarkably linear
over a range of 0.7--70 Dirac poles.

\begin{figure}
\centering
\includegraphics[height=10cm]{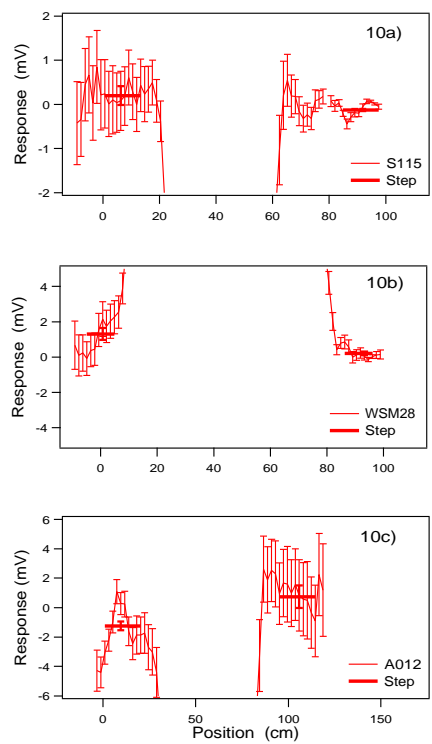}
\caption{\label{samples}Steps: D0 Al, CDF Pb, and CDF Al.}
\end{figure}

\begin{figure}
\centering
\includegraphics[height=6cm]{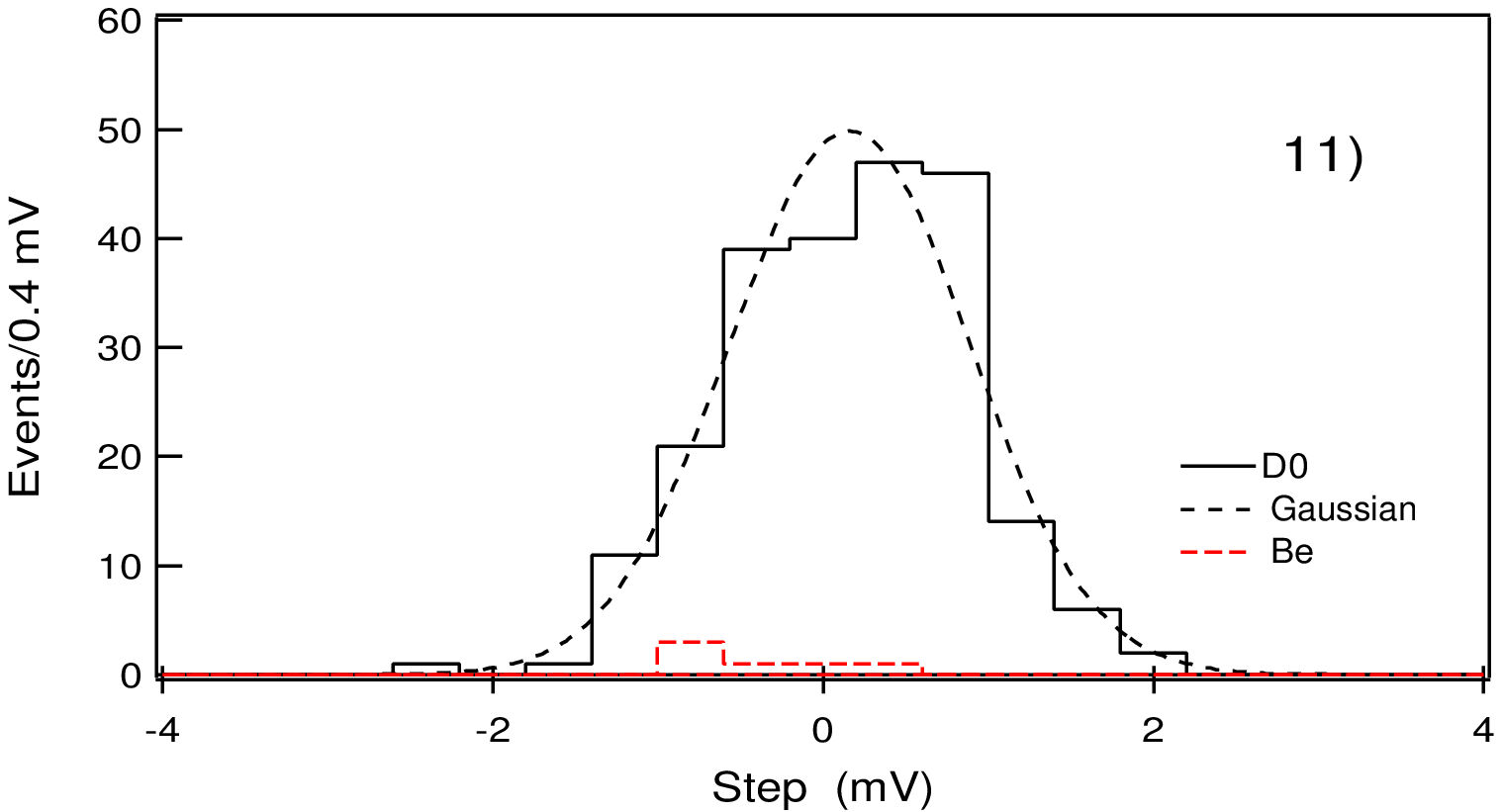}
\caption{\label{hist1}Steps from D0 samples.
A Dirac pole would appear as a step at 2.46 mV.}
\end{figure}

\begin{figure}
\centering
\includegraphics[height=10cm]{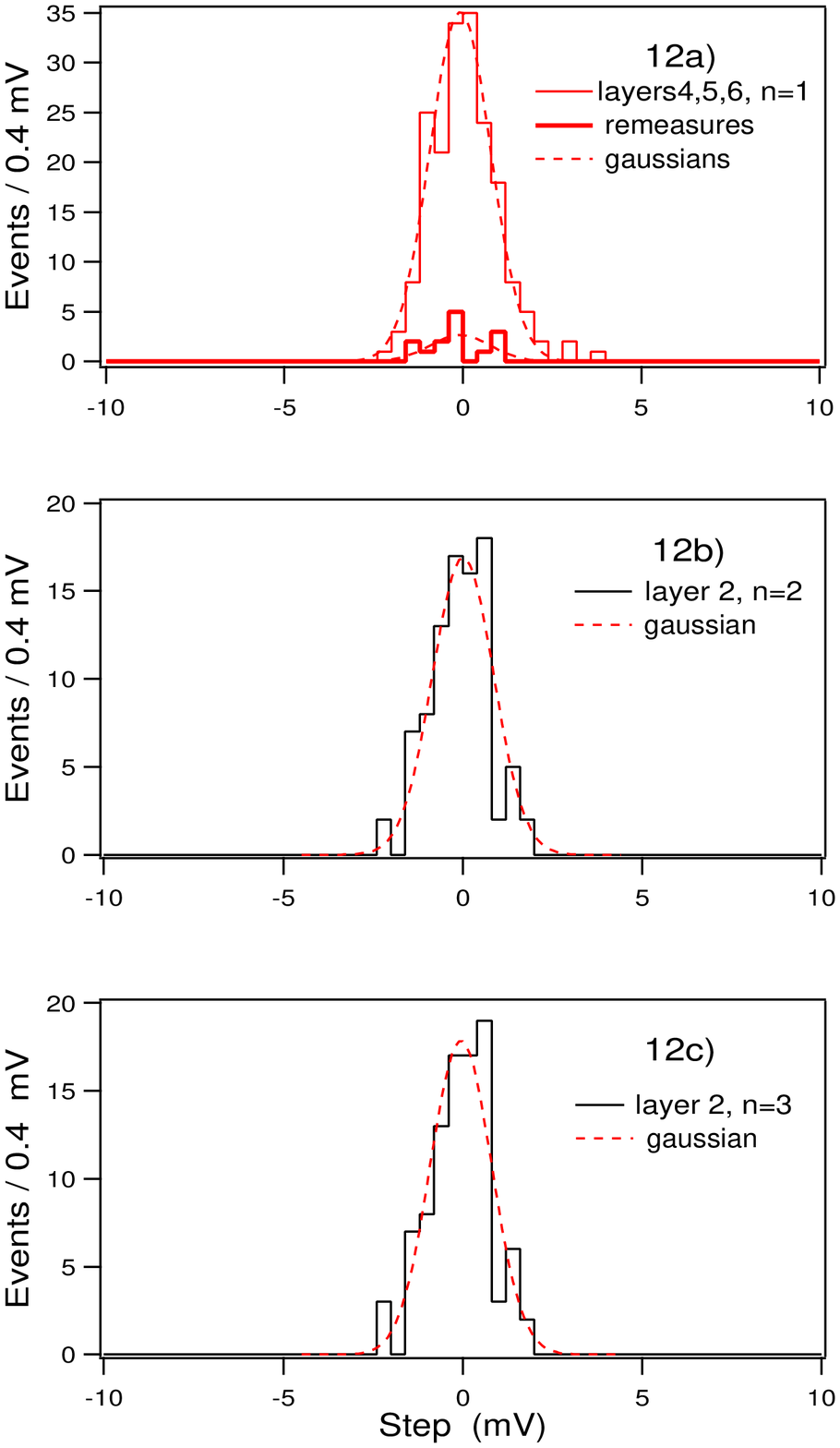}
\caption{\label{hist2}Steps from CDF Pb samples. A Dirac pole would appear
as a step at 2.46 mV.}
\end{figure}

\begin{figure}
\centering
\includegraphics[height=5cm]{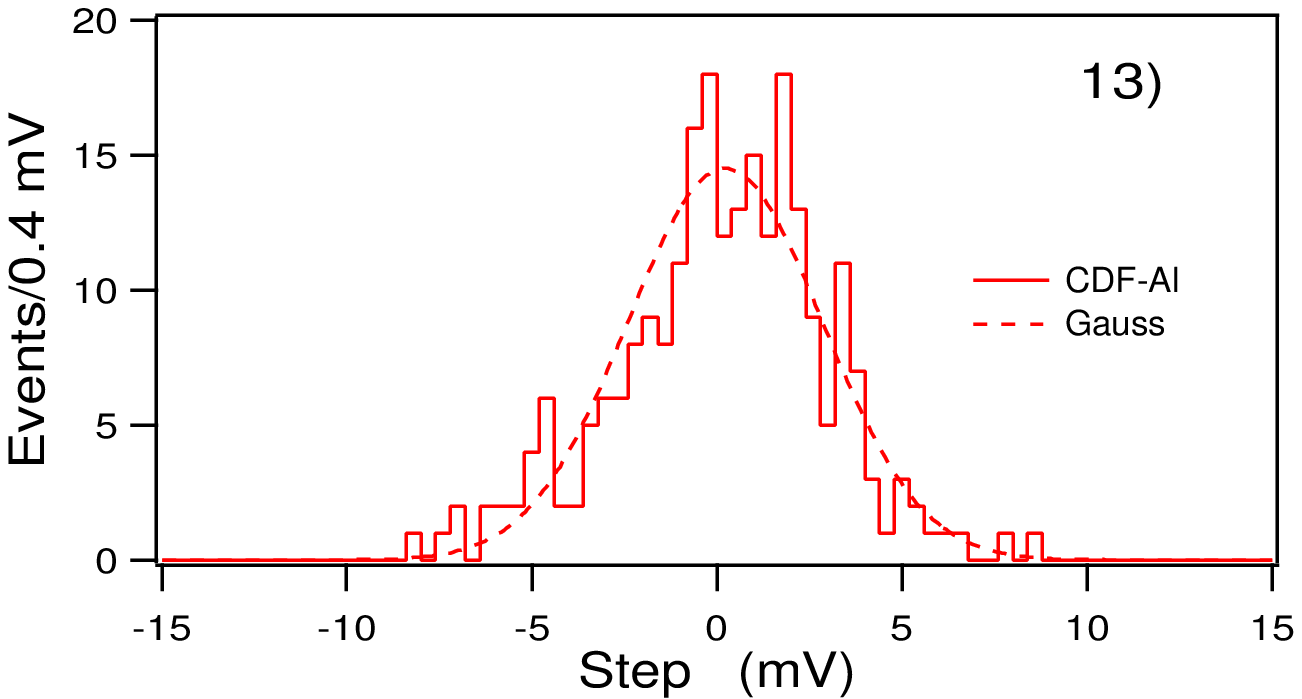}
\caption{\label{hist3}Steps from CDF Al samples. A Schwinger pole ($2g_D$)
would appear as a step at 10.64 mV.}
\end{figure}

As one sees from \fref{samples}, real samples have large dipole 
signals; what we are looking
for is an asymptotic step indicating the presence of a magnetic charge.
Steps seen are typically much smaller than that expected of a magnetic
pole of Dirac strength.  The histograms of steps are shown in 
figures \ref{hist1}--\ref{hist3}.

 For $m'=1/2$ the 90\% confidence upper limit is 4.2 signal events
for 8 events observed when 10 were expected \cite{cousins}. 
These 8 samples were remeasured and all fell within $\pm1.47$ mV of $m'=0$.
(More than 1.28$\sigma$ from $|m'|=1/2$.)
 For $m'=1$ the 90\% confidence upper limit is 2.4 signal events
for zero events observed and zero expected.

\begin{figure}
\centering
\includegraphics[height=10cm]{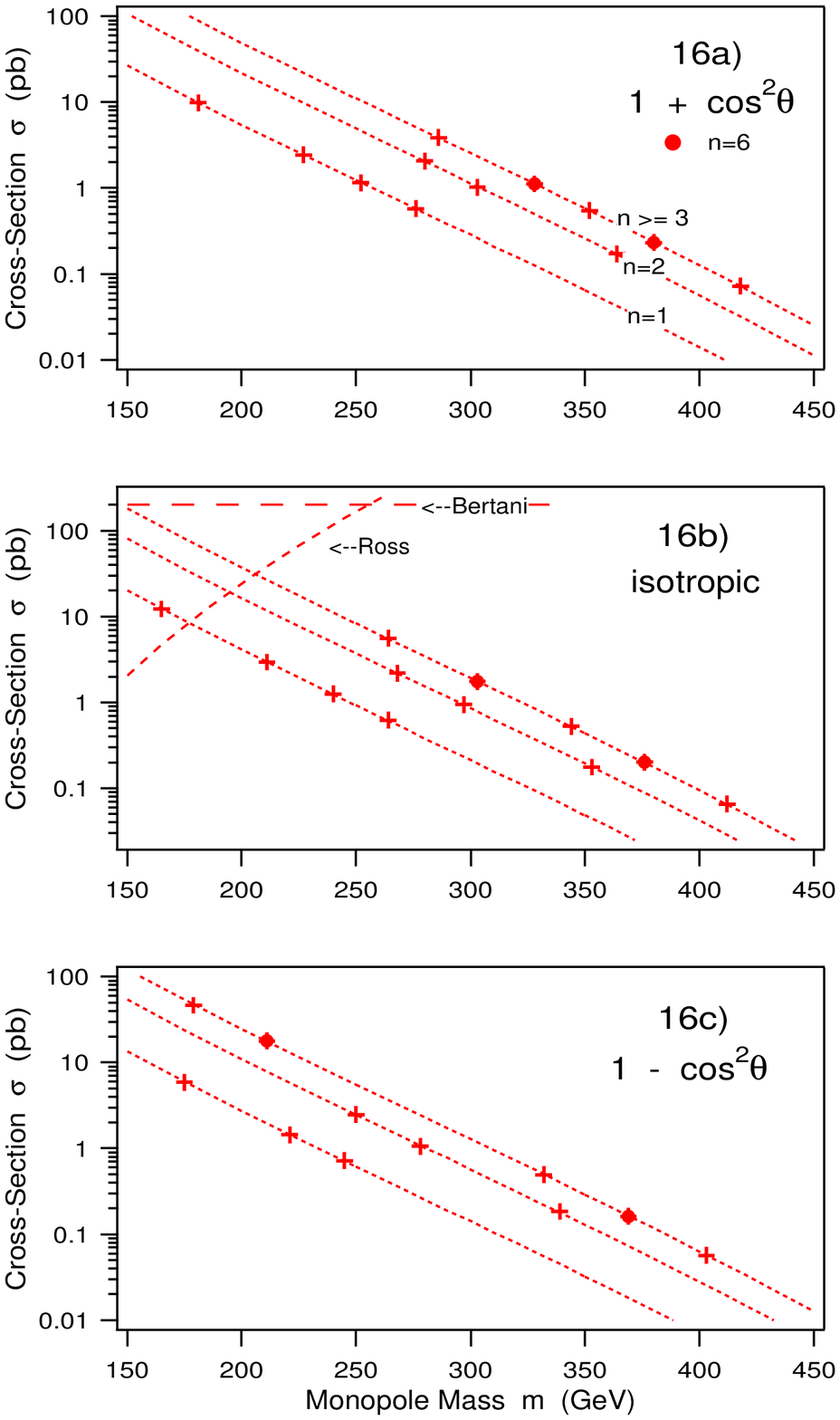}
\caption{\label{figmasslimits} Cross section vs.~mass limits.
The three graphs show three different assumptions about the angular
distribution, since even if we knew the spin of the monopole, we cannot
at present predict the differential cross section. Shown in the second
figure are the Bertani \cite{Bertani:1990tq} and lunar \cite{Ross:1973it} 
limits.}
\end{figure}

By putting in angular and mass acceptances we can get cross section
limits as shown in \tref{tablelimites}.
\begin{table}
\caption{\label{tablelimites} 
Alternative interpretations for different production angular
distributions of the monopoles, comparing $1$ and $1\pm\cos^2\theta$.
Here the cross section $\sigma_a$
corresponds to the distribution $1+a\cos^2\theta$, and similarly for the
mass limits (all at 90\% confidence level).}
\begin{indented}
\item[]\begin{tabular}{@{}llllllll}
\br
Set&$2m'$&$\sigma^{\rm ul}_{+1}$&$m^{\rm LL}_{+1}$&
$\sigma^{\rm ul}_{0}$&$m^{\rm LL}_{0}$&
$\sigma^{\rm ul}_{-1}$&$m^{\rm LL}_{-1}$\\
&&(pb)&(GeV/$c^2$)&(pb)&(GeV/$c^2$)&(pb)&(GeV/$c^2$)\\
\mr
1 Al&1&1.2&250&1.2&240&1.4&220\\
1 Al RM&1&0.6&275&0.6&265&0.7&245\\
2 Pb&1&9.9&180&12&165&23&135\\
2 Pb RM&1&2.4&225&2.9&210&5.9&175\\
1 Al&2&2.1&280&2.2&270&2.5&250\\
2 Pb&2&1.0&305&0.9&295&1.1&280\\
3 Al &2&0.2&365&0.2&355&0.2&340\\
1 Be&3&3.9&285&5.6&265&47&180\\
2 Pb&3&0.5&350&0.5&345&0.5&330\\
3 Al&3&0.07&420&0.07&410&0.06&405\\
1 Be&6&1.1&330&1.7&305&18&210\\
3 Al&6&0.2&380&0.2&375&0.2&370\\
\br
\end{tabular}
\end{indented}
\end{table}
These numbers reflect the new analysis, published in 2004 
\cite{Kalbfleisch:2003yt}, and so differ somewhat from our earlier
published results  \cite{Kalbfleisch:2000iz}.
To obtain the mass limits, we use the model cross sections given in 
\fref{figmasslimits}.

\begin{figure}[ht]
\centering
\includegraphics[width=3.4in]{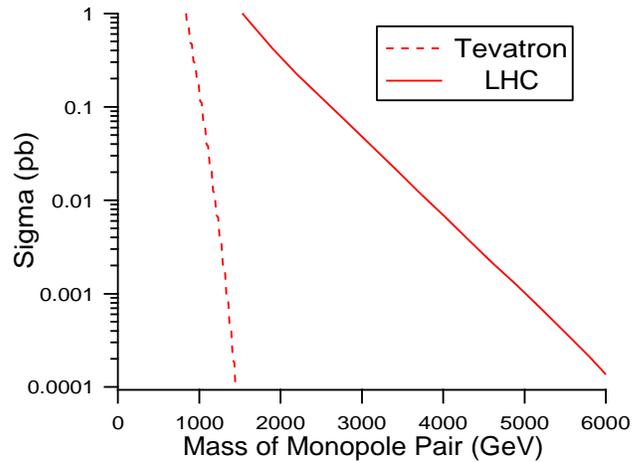}
\caption{\label{figlhc}Monopole pair masses as a function of the cross section
at the Tevatron ($p\bar p$ at 2 TeV) and at the LHC ($pp$ at 14 TeV).
Both include the $\beta^3$ correction, and are for a Dirac monopole,
$m'=1/2$.}
\end{figure}

 Finally, we show in \fref{figlhc} what might be achievable at the LHC,
using the same techniques applied here.

\section{H1 limits}
\label{sec:h1}
We now turn to the limits on monopole production 
obtained from $e^+p$ collisions at HERA
recently published by the H1 collaboration \cite{Aktas:2004qd}.  
This production
mechanism is intermediate between that of $p\bar p$ experiments such as
that given in \cite{Kalbfleisch:2003yt}, and that of the possible production
through $e^+e^-$ collisions \cite{acciarri}, and 
conceivably might yield a cleaner
interpretation if monopole condensates are responsible for the confinement of
quarks \cite{nambu,mandelstam:1976,Polyakov:1976fu,'tHooft:1981ht}.
Although the mass limits determined
are not as strong as in our experiment described in
the previous section, it is crucial that different physical domains be explored
carefully.

In their experiment, the aluminum beam pipe used at the H1 interaction point
at HERA during 1995--1997 was cut into 75 long and short
strips.  This beam pipe had been exposed to an integrated 
luminosity of $62\pm1$ pb$^{-1}$. These strips were then placed on
a conveyor belt and passed through a warm-bore magnetometer at Southhampton
Oceanographic Centre, UK.  If a monopole passed through the superconducting
coil, as in our experiment, it would establish a persistent current there, which
would be detected by a SQUID.  Again they calibrated their detector by a long,
thin solenoid, which at each end produced a pseudopole.  Calibration was within
some 10\% for pole strengths above $g_D=\hbar c/2e$.  Large dipole signals were
seen, but the signals always returned to the baseline unless a pseudopole was
present. A few persistent current events were seen, but they always disappeared
upon remeasurement.  Some of the runs exhibited large fluctuations of unknown
orgin, but none was consistent with a monopole event.

No monopole was detected in their experiment of strength greater than
$0.1 g_D$ for a sample consisting of $93\pm3\%$ of the beam pipe.  To interpret
this as a limit on the production cross section, models had to be adopted,
since perturbation theory was unreliable.  Two models were tried:
\numparts
\bea
e^+p\to e^+ M\bar M p,\qquad \mbox{spin 0 monopole},\label{modela}\\
e^+p\to e^+ M\bar M X,\qquad \mbox{spin 1/2 monopole},\label{modelb}
\eea
\endnumparts
where in both models the monopole pairs were produced by two-photon processes.
The effects on the produced monopoles by the H1 magnetic fields were included.
The stopping power was computed by using the classical results of Ahlen
\cite{Ahlen:1976jw,Ahlen:1978jy,Ahlen:1982mx}.  

The results of their analysis are expressed in plots of the upper limits on the
cross sections for a given monopole mass, up to a mass of 140 GeV, for
different models and magnetic charges.  
That is, cross sections above those limits
are excluded based on their experimental analysis.  
These limits are weakest for
the Dirac charge, $g_D$, and strongest for Schwinger quantization based on
quark charges, $6g_D$, of course.  For model (\ref{modela}) the cross section 
limits range from about 1 pb for several GeV to more that 100 pb for 
140 GeV for
$g_D$.  For higher charges the limits are relatively constant around 0.1 pb or
less.  For (\ref{modelb}) the limits are similar, except for $g_D$, where
the limit drops below 0.1 pb for 10 GeV masses or less.  
Thus their results our complementary to ours: Our mass limits are stronger,
but in some cases they reach smaller cross sections for lower masses.


\section{CDF limits}
\label{sec:cdf}
Quite recently, a new CDF experiment \cite{Abulencia:2005hb} has been
announced, which claims, on the basis of an integrated 
luminosity of 35.7 pb$^{-1}$,
a production cross section limit for spin-1/2 monopoles below 0.2 pb
for masses between 200 and 700 GeV, and hence, in a Drell-Yan model,
a lower mass limit of 360 GeV.  (These limits are quoted at the 95\%
confidence level.) This is based on quite different technology
than the Oklahoma or H1 experiments.  Rather, they looked at a sample
of $p\bar p$ events collected during 2003 by the CDF detector by a special 
trigger.  The signal for a monopole is the large ionization and heavy production
of delta rays by such a particle.  They use our crude model of replacing
$e$ by $g\beta$ in the Drell-Yan production mechanism 
\cite{Kalbfleisch:2000iz,Kalbfleisch:2003yt}, apart from this simply
replacing the lepton mass by the monopole mass.  Acceptance is affected by
production kinematics, which effect they estimate at 10\%.  Light monopoles
will be swept out of the detector by the magnetic field, while heavy monopoles
may reach the time-of-flight detector too late to cause a trigger.  Other
particles (``spoilers'') may cause a charge integration to start in the
detector before a monopole signal arrives; they estimate a few percent 
fraction for a monopole of mass 400 GeV.

Out of 130,000 candidate events, no monopole trigger events were found, from
which the limit quoted above was extracted.  They believe they can push
the limit on masses up another 100 GeV with additional Run II data.

\subsection{Comments on CDF experiment}

One might ask how can CDF claim stronger limits than we do based on less
than 1/4 of the integrated
luminosity of our experiment.  The answer, I believe, is that
our experimental limit is extremely conservative.  They may have underestimated
the systematic effect of huge uncertainties in the production mechanism, while
at the same time they claim our limit is dependent on the trapping model,
which it is not.  Undoubtedly, the $\rmd E/\rmd x$ signature of monopoles
is much less well understood than the clear-cut electromagnetic signature
of an induction detector.  This is not to denigrate the utility of this
measurement, but to emphasize that the limits so obtained are subject to large,
relatively uncontrolled, uncertainties.

\section{Conclusions}
One magnetic monopole, carrying magnetic charge $g$, will result in the
quantization of electric charge throughout the universe,
\be
e=\frac{m'\hbar c}g,
\ee
where $m'$ is a half-integer, 
\be
m'=0,\pm\frac12,\pm1,\pm\frac32,\pm2,\pm\frac52,\dots.
\ee
That electric charge is quantized in integer multiples of the electron charge
(or integer multiples of the quark charge) is an overwhelming fact, which
does not possess a simple explanation.
This, perhaps, is the most compelling argument in favor of the existence of
magnetic charge.  A second argument is the greater symmetry (duality)
imparted to
Maxwell's equations, and to classical and quantum electrodynamics, if both
electric and magnetic charges are present.  Thus from phenomenological
and theoretical bases, the arguments in favor of the existence of magnetic
charge and for dual QED are at least as strong as those for supersymmetry.
Unfortunately, like for the latter, there is not a real shred of observational
evidence in favor of magnetic charge, although here it is far less embarrassing
to be in that situation, since the most likely mass range for magnetically
charged particles is not far from the Planck scale. 

In this review we have concentrated on the theory of point Dirac monopoles
or Schwinger dyons, starting from the classical scattering, through 
the nonrelativistic quantum mechanical description, to the quantum field
theory of such objects.  For lack of space we have only briefly referred to
the classical monopoles that arise from the solution of non-Abelian gauge
theories.  From the point of view of phenomenology and the setting of
experimental limits, the point description should be adequate, since the 
structure of composite monopoles only emerges at the energy scale that
sets the mass of the particles.  (An exception, of course, occurs with
limits based on that structure, such as the catalysis of proton decay.)
In addition, our concern has been chiefly with the quantum description,
which has been only roughly sketched for composite monopoles.

We close, as did Schwinger in his provocative article \cite{Schwinger:1969ib},
by quoting from Faraday: ``Nothing is too wonderful to be true, if it be
consistent with the laws of nature, and in such things as these, experiment
is the best test of such consistency.''

\ack
I thank the Physics Department of Washington University for its hospitality
during most of the writing of this review.  My work has been supported in
part by the US Department of Energy.  I want to thank my many collaborators,
especially J. Schwinger, L. DeRaad, Jr., G. Kalbfleisch, and L. Gamberg, for 
allowing me to describe here our joint work.

\section*{References}
\bibliography{monopole}

\begin{thebibliography}{100}

\bibitem{online}
May 28, 2005.
\newblock http://www.languagehat.com/archives/001914.php.

\bibitem{peregrinus}
Pierre de~Maricourt.
\newblock {On the Magnet, letter to Siger de Foucaucourt, (1269)}.
\newblock In {\em The Letter of Petrus Peregrinus on the Magnet}, New York,
  1904. McGraw-Hill.
\newblock Translated by Brother Arnold.

\bibitem{poincare:1896}
{H. Poincar\'e}.
\newblock {Remarques sur une exp\'erience de M. Birkeland}.
\newblock {\em Compt. Rendus}, 123:530, 1896.

\bibitem{thomson:1904}
J.~J. Thomson.
\newblock {\em Electricity and Matter}.
\newblock Scribners, New York, 1904.
\newblock Lectures given at Yale University.

\bibitem{thomson:1904a}
J.~J. Thomson.
\newblock On momentum in the electric field.
\newblock {\em Philos. Mag.}, 8:331--356, 1904.

\bibitem{faraday}
M.~Faraday.
\newblock {\em Experimental Researches in Electricity}.
\newblock Taylor, London, 1844.

\bibitem{heaviside}
O.~Heaviside.
\newblock {\em Electromagnetic Theory}, volume~1.
\newblock Benn, London, 1893.

\bibitem{curie}
P.~Curie.
\newblock On the possible existence of magnetic conductivity and free
  magnetism.
\newblock {\em S\`eances Soc. Phys. (Paris)}, pages 76--77, 1894.

\bibitem{Dirac:1931kp}
P.~A.~M. Dirac.
\newblock Quantised singularities in the electromagnetic field.
\newblock {\em Proc. Roy. Soc. Lond.}, A133:60--72, 1931.

\bibitem{gm-tot}
R.~Kane.
\newblock {\em Erkenntnis}, 24:115, 1986.

\bibitem{Schwinger:1969ib}
J.~Schwinger.
\newblock A magnetic model of matter.
\newblock {\em Science}, 165:757--761, 1969.

\bibitem{wuyang}
T.~T. Wu and C.~N. Yang.
\newblock Some solutions of the classical isotopic gauge field equations.
\newblock In H.~Mark and S.~Fernbach, editors, {\em Properties of Matter Under
  Unusual Conditions}, page 349, New York, 1969. Wiley.

\bibitem{'tHooft:1974qc}
G.~'t~Hooft.
\newblock Magnetic monopoles in unified gauge theories.
\newblock {\em Nucl. Phys.}, B79:276--284, 1974.

\bibitem{Polyakov:1974ek}
A.~M. Polyakov.
\newblock Particle spectrum in quantum field theory.
\newblock {\em JETP Lett.}, 20:194--195, 1974.
\newblock [Pisma Zh. Eksp. Teor. Fiz. 20:430-433,1974].

\bibitem{nambu}
Y.~Nambu.
\newblock Strings, monopoles, and gauge fields.
\newblock {\em Phys. Rev. D}, 10:4262, 1974.

\bibitem{Julia:1975ff}
B.~Julia and A.~Zee.
\newblock Poles with both magnetic and electric charges in nonabelian gauge
  theory.
\newblock {\em Phys. Rev.}, D11:2227--2232, 1975.

\bibitem{Wu:1975vq}
T.~T. Wu and C.~N. Yang.
\newblock Some remarks about unquantized nonabelian gauge fields.
\newblock {\em Phys. Rev.}, D12:3843--3844, 1975.

\bibitem{rl}
A.~S. Goldhaber and W.~P. Trower.
\newblock Resource letter {MM1}: Magnetic monopoles.
\newblock {\em Am. J. Phys.}, 58:429, 1990.

\bibitem{Schwinger:1976fr}
J.~Schwinger, K.~A. Milton, W.-y. Tsai, {L. L. DeRaad, Jr.}, and D.~C. Clark.
\newblock Nonrelativistic dyon dyon scattering.
\newblock {\em Ann. Phys. (N.Y.)}, 101:451, 1976.

\bibitem{Kalbfleisch:2000iz}
G.~R. Kalbfleisch et~al.
\newblock Improved experimental limits on the production of magnetic monopoles.
\newblock {\em Phys. Rev. Lett.}, 85:5292--5295, 2000.

\bibitem{Kalbfleisch:2003yt}
G.~R. Kalbfleisch, W.~Luo, K.~A. Milton, E.~H. Smith, and M.~G. Strauss.
\newblock Limits on production of magnetic monopoles utilizing samples from the
  {D0 and CDF} detectors at the {Tevatron}.
\newblock {\em Phys. Rev.}, D69:052002, 2004.

\bibitem{Alvarez:1963zp}
L.~W. Alvarez.
\newblock Design of an electromagnetic detector for {Dirac} monopoles.
\newblock 1963.
\newblock {UCRL-AGM-470}.

\bibitem{Alvarez:1970zu}
L.~W. Alvarez, P.~H. Eberhard, R.~R. Ross, and R.~D. Watt.
\newblock Search for magnetic monopoles in the lunar sample.
\newblock {\em Science}, 167:701--703, 1970.

\bibitem{Eberhard:1971re}
P.~H. Eberhard, R.~R. Ross, L.~W. Alvarez, and R.~D. Watt.
\newblock Search for magnetic monopoles in lunar material.
\newblock {\em Phys. Rev.}, D4:3260, 1971.

\bibitem{Ross:1973it}
R.~R. Ross, P.~H. Eberhard, L.~W. Alvarez, and R.~D. Watt.
\newblock Search for magnetic monopoles in lunar material using an
  electromagnetic detector.
\newblock {\em Phys. Rev.}, D8:698, 1973.

\bibitem{Eberhard:1975en}
P.~H. Eberhard, R.~R Ross, J.~D. Taylor, L.~W. Alvarez, and H.~Oberlack.
\newblock Evidence at the $1/10^{18}$ probability level against the production
  of magnetic monopoles in proton interactions at 300 {GeV}/c.
\newblock {\em Phys. Rev.}, D11:3099--3104, 1975.

\bibitem{CE}
J.~Schwinger, {L. L. DeRaad, Jr.}, K.~A. Milton, and W.-y. Tsai.
\newblock {\em Classical Electrodynamics}.
\newblock {Perseus Books/Westview Press}, New York, 1998.

\bibitem{bethe44}
H.~A. Bethe.
\newblock Theory of diffraction by small holes.
\newblock {\em Phys. Rev.}, 66:163, 1944.

\bibitem{er}
K.~A. Milton and J.~Schwinger.
\newblock {\em Electromagnetic Radiation: Variational Methods, Waveguides, and
  Acclerators}.
\newblock Springer-Verlag, Berlin, 2006.

\bibitem{thomson:1909}
J.~J. Thomson.
\newblock {\em Elements of the Mathematical Theory of Electricity and
  Magnetism}.
\newblock Cambridge University Press, Cambridge, 4th edition, 1909.

\bibitem{thomson:1937}
J.~J. Thomson.
\newblock {\em Recollections and Reflections}.
\newblock Macmillan, New York, 1937.

\bibitem{birkeland}
M.~Birkeland.
\newblock Sur un spectre des rayons cathodiques.
\newblock {\em Compt. Rendus}, 123:492, 1896.

\bibitem{Schwinger:1975ww}
J.~Schwinger.
\newblock Magnetic charge and the charge quantization condition.
\newblock {\em Phys. Rev.}, D12:3105, 1975.

\bibitem{Wu:1976qk}
T.~T. Wu and C.~N. Yang.
\newblock {Dirac's} monopole without strings: Classical {Lagrangian} theory.
\newblock {\em Phys. Rev.}, D14:437--445, 1976.

\bibitem{Wu:1976ge}
T.~T. Wu and C.~N. Yang.
\newblock {Dirac} monopole without strings: Monopole harmonics.
\newblock {\em Nucl. Phys.}, B107:365, 1976.

\bibitem{Kazama:1976sr}
Y.~Kazama and C.~N. Yang.
\newblock Existence of bound states for a charged spin-1/2 particle with an
  extra magnetic moment in the field of a fixed magnetic monopole.
\newblock {\em Phys. Rev.}, D15:2300, 1977.

\bibitem{Wu:1977qk}
T.~T. Wu and C.~N. Yang.
\newblock Some properties of monopole harmonics.
\newblock {\em Phys. Rev.}, D16:1018--1021, 1977.

\bibitem{Yang:1977qv}
C.~N. Yang.
\newblock Generalization of {Dirac's} monopole to {SU(2)} gauge fields.
\newblock {\em J. Math. Phys.}, 19:320, 1978.

\bibitem{Kazama:1977fm}
Y.~Kazama, C.~N. Yang, and A.~S. Goldhaber.
\newblock Scattering of a {Dirac} particle with charge ${Ze}$ by a fixed
  magnetic monopole.
\newblock {\em Phys. Rev.}, D15:2287, 1977.

\bibitem{Yang:1978td}
C.~N. Yang.
\newblock {SU(2)} monopole harmonics.
\newblock {\em J. Math. Phys.}, 19:2622, 1978.

\bibitem{Yang:1979tx}
C.~N. Yang.
\newblock Fiber bundles and the physics of the magnetic monopole.
\newblock Presented at Chern Symposium, Berkeley, Calif., Jun 1979.

\bibitem{gold}
A.~S. Goldhaber.
\newblock Role of spin in the monopole problem.
\newblock {\em Phys. Rev.}, 140:B1407, 1965.

\bibitem{Goldhaber:1977xw}
A.~S. Goldhaber.
\newblock {Dirac} particle in a magnetic field: Symmetries and their breaking
  by monopole singularities.
\newblock {\em Phys. Rev.}, D16:1815, 1977.

\bibitem{Milton:1976jq}
K.~A. Milton and L.~L. {DeRaad, Jr.}
\newblock Strings and gauge invariance.
\newblock {\em J. Math. Phys.}, 19:375, 1978.

\bibitem{zumino}
B.~Zumino.
\newblock In A.~Zichichi, editor, {\em Strong and Weak Interactions--Present
  Problems}, page 771, New York. Academic.

\bibitem{Boulware:1976tv}
D.~G. Boulware, L.~S. Brown, R.~N. Cahn, S.~D. Ellis, and C.-k. Lee.
\newblock Scattering on magnetic charge.
\newblock {\em Phys. Rev.}, D14:2708, 1976.

\bibitem{wu75}
T.~T. Wu and C.~N. Yang.
\newblock Concept of nonintegrable phase factors and global formulation of
  gauge fields.
\newblock {\em Phys. Rev. D}, 12:3845, 1975.

\bibitem{yang76}
C.~N. Yang.
\newblock In {\em Proceedings of the Sixth Hawaii Topical Conference on
  Particle Physics}, Honolulu, 1976. University of Hawaii.

\bibitem{yan67}
T.~M. Yan.
\newblock Classical theory of magnetic charge.
\newblock {\em Phys. Rev.}, 160:1182, 1967.

\bibitem{Brandt:1976hk}
R.~A. Brandt and J.~R. Primack.
\newblock Avoiding `{Dirac's} veto' in monopole theory.
\newblock {\em Phys. Rev.}, D15:1798--1802, 1977.

\bibitem{Brandt:1977ks}
R.~A. Brandt and J.~R. Primack.
\newblock Moving and removing {Dirac's} string.
\newblock {\em Phys. Rev.}, D15:1175, 1977.

\bibitem{hurst}
C.~A. Hurst.
\newblock Charge quantization and nonintegrable {Lie} algebra.
\newblock {\em Ann. Phys. (N.Y.)}, 50:51, 1968.

\bibitem{lipkin}
H.~J. Lipkin, W.~I. Weisberger, and M.~Peshkin.
\newblock Magnetic charge quantization and angular momentum.
\newblock {\em Ann. Phys. (N.Y.)}, 53:203, 1969.

\bibitem{tamm}
J.~Tamm.
\newblock Die verallgemeinerten kugelfunktionen und die wellenfunktionen eines
  electrons un feld eines magnetpoles.
\newblock {\em Z. Phys.}, 71:141--150, 1931.

\bibitem{banderet}
P.~Banderet.
\newblock On the theory of a point magnetic pole.
\newblock {\em Helv. Phys. Acta}, 19:503, 1946.

\bibitem{fierz}
M.~Fierz.
\newblock On the theory of particles with magnetic charge.
\newblock {\em Helv. Phys. Acta}, 17:27--34, 1944.

\bibitem{fordandwheeler}
K.~Ford and J.~A. Wheeler.
\newblock Application of semiclassical scattering analysis.
\newblock {\em Ann. Phys. (N.Y.)}, 7:287, 1959.

\bibitem{lapidus}
I.~R. Lapidus and J.~L. Pietenpol.
\newblock Classical interaction of an electric charge with a magnetic monopole.
\newblock {\em Am. J. Phys.}, 28:17--18, 1960.

\bibitem{nadeau}
G.~Nadeau.
\newblock Concerning the classical interaction of an electric charge with a
  magnetic monopole.
\newblock {\em Am. J. Phys.}, 28:566, 1960.

\bibitem{Prasad:1975kr}
M.~K. Prasad and C.~M. Sommerfield.
\newblock An exact classical solution for the 't {Hooft} monopole and the
  {Julia-Zee} dyon.
\newblock {\em Phys. Rev. Lett.}, 35:760--762, 1975.

\bibitem{Bogomolny:1975de}
E.~B. Bogomolny.
\newblock Stability of classical solutions.
\newblock {\em Sov. J. Nucl. Phys.}, 24:449, 1976.

\bibitem{Georgi:1972cj}
H.~Georgi and S.~L. Glashow.
\newblock Unified weak and electromagnetic interactions without neutral
  currents.
\newblock {\em Phys. Rev. Lett.}, 28:1494, 1972.

\bibitem{Preskill:1984gd}
J.~Preskill.
\newblock Magnetic monopoles.
\newblock {\em Ann. Rev. Nucl. Part. Sci.}, 34:461--530, 1984.

\bibitem{Kirkman:1981ck}
T.~W. Kirkman and C.~K. Zachos.
\newblock Asymptotic analysis of the monopole structure.
\newblock {\em Phys. Rev.}, D24:999, 1981.

\bibitem{coleman}
S.~Coleman.
\newblock In A.~Zichichi, editor, {\em The Unity of Fundamental Interactions},
  New York. Plenum.

\bibitem{Cho:1996qd}
Y.~M. Cho and D.~Maison.
\newblock Monopole configuration in {Weinberg-Salam} model.
\newblock {\em Phys. Lett.}, B391:360--365, 1997.
\newblock hep-th/9601028.

\bibitem{Cho:1997jn}
Y.~M. Cho and K.~Kimm.
\newblock Finite energy electroweak monopoles.
\newblock 1997.
\newblock hep-th/9707038.

\bibitem{Cho:1997zc}
Y.~M. Cho and K.~Kimm.
\newblock Electroweak monopoles.
\newblock 1997.
\newblock hep-th/9705213.

\bibitem{Weinberg:1982ev}
E.~J. Weinberg.
\newblock Fundamental monopoles in theories with arbitrary symmetry breaking.
\newblock {\em Nucl. Phys.}, B203:445, 1982.

\bibitem{Rajantie:2005hi}
A.~Rajantie.
\newblock Mass of a quantum 't {Hooft-Polyakov} monopole.
\newblock 2005.
\newblock hep-lat/0512006.

\bibitem{Sutcliffe:1997ec}
P.~M. Sutcliffe.
\newblock {BPS} monopoles.
\newblock {\em Int. J. Mod. Phys.}, A12:4663--4706, 1997.

\bibitem{Manton:2004tk}
N.~S. Manton and P.~Sutcliffe.
\newblock {\em Topological Solitons}.
\newblock Cambridge University Press, 2004.

\bibitem{Nair}
M.~P. Nair.
\newblock {\em Quantum Field Theory: A Modern Perspective}.
\newblock Springer, New York, 2005.

\bibitem{Khvedelidze:2005rv}
A.~Khvedelidze, A.~Kovner, and D.~McMullan.
\newblock Magnetic monopoles in $4{D}$: A perturbative calculation.
\newblock 2005.
\newblock hep-th/0512142.

\bibitem{Schwinger:1966nj}
J.~Schwinger.
\newblock Magnetic charge and quantum field theory.
\newblock {\em Phys. Rev.}, 144:1087--1093, 1966.

\bibitem{schwingermc1}
J.~Schwinger.
\newblock Electric and magnetic-charge renormalization. {I}.
\newblock {\em Phys. Rev.}, 151:1048, 1966.

\bibitem{schwingermc2}
J.~Schwinger.
\newblock Electric and magnetic-charge renormalization. {II}.
\newblock {\em Phys. Rev.}, 151:1055, 1966.

\bibitem{Schwinger:1968rq}
J.~Schwinger.
\newblock Sources and magnetic charge.
\newblock {\em Phys. Rev.}, 173:1536--1544, 1968.

\bibitem{Zwanziger:1968rs}
D.~Zwanziger.
\newblock Quantum field theory of particles with both electric and magnetic
  charges.
\newblock {\em Phys. Rev.}, 176:1489--1495, 1968.

\bibitem{Zwanziger:1970hk}
D.~Zwanziger.
\newblock Local {Lagrangian} quantum field theory of electric and magnetic
  charges.
\newblock {\em Phys. Rev.}, D3:880, 1971.

\bibitem{Brandt:1977be}
R.~A. Brandt, F.~Neri, and D.~Zwanziger.
\newblock Lorentz invariance of the quantum field theory of electric and
  magnetic charge.
\newblock {\em Phys. Rev. Lett.}, 40:147, 1978.

\bibitem{brandtandneri}
R.~Brandt, F.~Neri, and D.~Zwanziger.
\newblock {\em Phys. Rev. D}, 19:1153, 1979.

\bibitem{Blagojevic:1985sh}
{Blagojevi\'c, M. and Senjanovi\'c, P.}
\newblock The quantum field theory of electric and magnetic charge.
\newblock {\em Phys. Rept.}, 157:233, 1988.

\bibitem{Gamberg:1999hq}
L.~P. Gamberg and K.~A. Milton.
\newblock Dual quantum electrodynamics: Dyon-dyon and charge-monopole
  scattering in a high-energy approximation.
\newblock {\em Phys. Rev.}, D61:075013, 2000.

\bibitem{Dirac:1948um}
P.~A.~M. Dirac.
\newblock The theory of magnetic poles.
\newblock {\em Phys. Rev.}, 74:817--830, 1948.

\bibitem{psf1}
J.~Schwinger.
\newblock {\em Particles, Sources, and Fields}, volume~1.
\newblock Addison-Wesley, New York, 1970.
\newblock republished by Perseus Books, 1998.

\bibitem{dir55}
P.~A.~M. Dirac.
\newblock {\em Can. J. Phys.}, 33:650, 1955.

\bibitem{sis87}
A.~N. Sisakyan, N.~B. Skachkov, I.~L. Solovtsov, and O.~Yu. Shevchenko.
\newblock Gauge invariant approach and infrared behavior of spinor propagator.
\newblock {\em Theor. Math. Phys.}, 78:185, 1989.
\newblock [Teor. Mat. Fiz. 78:258-266, 1989].

\bibitem{Deans:1981qs}
W.~Deans.
\newblock Quantum field theory of {Dirac} monopoles and the charge quantization
  condition.
\newblock {\em Nucl. Phys.}, B197:307, 1982.

\bibitem{Schwinger:1951ex}
J.~Schwinger.
\newblock On the {Green's} functions of quantized fields. 1.
\newblock {\em Proc. Nat. Acad. Sci.}, 37:452--455, 1951.

\bibitem{Schwinger:1951hq}
J.~Schwinger.
\newblock On the {Green's} functions of quantized fields. 2.
\newblock {\em Proc. Nat. Acad. Sci.}, 37:455--459, 1951.

\bibitem{Mandelstam:1962mi}
S.~Mandelstam.
\newblock Quantum electrodynamics without potentials.
\newblock {\em Ann. Phys.}, 19:1--24, 1962.

\bibitem{Zumino:1959wt}
B.~Zumino.
\newblock Gauge properties of propagators in quantum electrodynamics.
\newblock {\em J. Math. Phys.}, 1:1--7, 1960.

\bibitem{bb62}
I.~Bialynicki-Birula.
\newblock On the gauge covariance of quantum electrodynamics.
\newblock {\em J. Math. Phys.}, 3:1094, 1962.

\bibitem{zin86}
J.~Zinn-Justin.
\newblock {\em Quantum Field Theory and Critical Phenomena}.
\newblock Oxford University Press, Oxford, 3rd edition, 1996.

\bibitem{Schwinger:1960qe}
J.~Schwinger.
\newblock Brownian motion of a quantum oscillator.
\newblock {\em J. Math. Phys.}, 2:407--432, 1961.

\bibitem{som63}
C.~Sommerfield.
\newblock On the definition of currents and the action principle in field
  theories of one spatial dimension.
\newblock {\em Ann. Phys. (NY)}, 26:1, 1963.

\bibitem{jon65}
K.~Johnson.
\newblock In S.~Deser and K.~Ford, editors, {\em Lectures on Particles and
  Field Theory}, page~1, Englewood Cliffs, 1965. Prentice-Hall.

\bibitem{sym54}
K.~Symanzik.
\newblock {\em Z. Naturforsch.}, 9A:809, 1954.

\bibitem{fri90}
H.~M. Fried.
\newblock {\em Functional Methods and Eikonal Models}.
\newblock Editions Fronti\`{e}res, Gif-sur-Yvette Cedex, 1990.

\bibitem{Nachtmann:1991ua}
O.~Nachtmann.
\newblock Considerations concerning diffraction scattering in quantum
  chromodynamics.
\newblock {\em Ann. Phys. (N.Y.)}, 209:436--478, 1991.

\bibitem{Korchemsky:1993hr}
G.~P. Korchemsky.
\newblock On near forward high-energy scattering in {QCD}.
\newblock {\em Phys. Lett.}, B325:459--466, 1994.

\bibitem{Korchemskaya:1994qp}
I.~A. Korchemskaya and G.~P. Korchemsky.
\newblock High-energy scattering in {QCD} and cross singularities of wilson
  loops.
\newblock {\em Nucl. Phys.}, B437:127--162, 1995.

\bibitem{Gellas:1998sh}
G.~C. Gellas, A.~I. Karanikas, and C.~N. Ktorides.
\newblock Worldline approach to eikonals for {QED} and linearized quantum
  gravity and their off mass shell extensions.
\newblock {\em Phys. Rev.}, D57:3763--3776, 1998.

\bibitem{Karanikas:1998tn}
A.~I. Karanikas and C.~N. Ktorides.
\newblock Worldline approach to forward and fixed angle fermion fermion
  scattering in {Yang-Mills} theories at high energies.
\newblock {\em Phys. Rev.}, D59:016003, 1999.

\bibitem{Quiros:1976fk}
M.~Quiros.
\newblock An eikonal expansion in quantum electrodynamics.
\newblock {\em Helv. Phys. Acta}, 49:849--862, 1976.

\bibitem{Fried:1996uv}
H.~M. Fried and Y.~Gabellini.
\newblock Non-{Abelian} eikonals.
\newblock {\em Phys. Rev.}, D55:2430--2440, 1997.

\bibitem{Bloch:1937pw}
F.~Bloch and A.~Nordsieck.
\newblock Note on the radiation field of the electron.
\newblock {\em Phys. Rev.}, 52:54--59, 1937.

\bibitem{fri65}
G.~W. Erickson and H.~M. Fried.
\newblock Lepton scattering amplitudes in two model field theories.
\newblock {\em J. Math. Phys.}, 6:414, 1965.

\bibitem{Fried:1971tz}
H.~M. Fried.
\newblock Construction of a complete eikonal representation.
\newblock {\em Phys. Rev.}, D3:2010--2013, 1971.

\bibitem{Levy:1969cr}
M.~Levy and J.~Sucher.
\newblock Eikonal approximation in quantum field theory.
\newblock {\em Phys. Rev.}, 186:1656--1670, 1969.

\bibitem{Levy:1970yn}
M.~Levy and J.~Sucher.
\newblock Asymptotic behavior of scattering amplitudes in the relativistic
  eikonal approximation.
\newblock {\em Phys. Rev.}, D2:1716--1723, 1970.

\bibitem{Abarbanel:1969ek}
H.~D.~I. Abarbanel and C.~Itzykson.
\newblock Relativistic eikonal expansion.
\newblock {\em Phys. Rev. Lett.}, 23:53, 1969.

\bibitem{Dittrich:1970vv}
W.~Dittrich.
\newblock Equivalence of the {Dirac} equation to a subclass of {Feynman}
  diagrams.
\newblock {\em Phys. Rev.}, D1:3345--3348, 1970.

\bibitem{Dittrich:1972ya}
W.~Dittrich.
\newblock Relativistic eikonal propagators and high-energy scattering in {QED}.
\newblock {\em Nucl. Phys.}, B45:290--302, 1972.

\bibitem{gottfried}
K.~Gottfried.
\newblock {\em Quantum Mechanics}.
\newblock Addison-Wesley, Reading, Massachusetts, 1966.

\bibitem{Itzykson:1980rh}
C.~Itzykson and J.~B. Zuber.
\newblock {\em Quantum Field Theory}.
\newblock {McGraw-Hill}, New York, 1980.

\bibitem{Urrutia:1978kq}
L.~F. Urrutia.
\newblock Zeroth order eikonal approximation in relativistic charged
  particle-monopole scattering.
\newblock {\em Phys. Rev.}, D18:3031, 1978.

\bibitem{Ore:1975my}
{Ore, F. R., Jr.}
\newblock Lorentz invariance of charge-monopole scattering.
\newblock {\em Phys. Rev.}, D13:2295, 1976.

\bibitem{Fried:1995cm}
H.~M. Fried and Y.~M. Gabellini.
\newblock Phase averaging and generalized eikonal representations.
\newblock {\em Phys. Rev.}, D51:890--905, 1995.

\bibitem{Fried:1995zr}
H.~M. Fried and Y.~M. Gabellini.
\newblock Special variant of the {Fradkin} representation.
\newblock {\em Phys. Rev.}, D51:906--918, 1995.

\bibitem{Fried:1995zx}
H.~M. Fried, Y.~Gabellini, and B.~H.~J. McKellar.
\newblock Exact and approximate fermion {Green's} functions in {QED} and {QCD}.
\newblock {\em Phys. Rev.}, D51:7083--7096, 1995.

\bibitem{fra66}
E.~S. Fradkin.
\newblock Application of functional methods in quantum field theory and quantum
  statistics (ii).
\newblock {\em Nucl. Phys.}, 76:588, 1966.

\bibitem{Shajesh:2005dy}
K.~V. Shajesh and K.~A. Milton.
\newblock Quantum mechanics using {Fradkin's} representation.
\newblock 2005.
\newblock hep-th/0510103.

\bibitem{Laperashvili:1999pu}
L.~V. Laperashvili and H.~B. Nielsen.
\newblock Dirac relation and renormalization group equations for electric and
  magnetic fine structure constants.
\newblock {\em Mod. Phys. Lett.}, A14:2797, 1999.

\bibitem{Laperashvili:2000np}
L.~V. Laperashvili and H.~B. Nielsen.
\newblock Phase transition in the {Higgs} model of scalar fields with electric
  and magnetic charges.
\newblock {\em Int. J. Mod. Phys.}, A16:2365--2390, 2001.

\bibitem{Laperashvili:2002xv}
L.~V. Laperashvili, D.~A. Ryzhikh, and H.~B. Nielsen.
\newblock Monopoles near the {Planck} scale and unification.
\newblock {\em Int. J. Mod. Phys.}, A18:4403--4442, 2003.

\bibitem{Laperashvili:2002jc}
L.~V. Laperashvili, H.~B. Nielsen, and D.~A. Ryzhikh.
\newblock Monopoles and family replicated unification.
\newblock {\em Phys. Atom. Nucl.}, 66:2070--2077, 2003.

\bibitem{Das:2005iv}
C.~R. Das, L.~V. Laperashvili, and H.~B. Nielsen.
\newblock Generalized dual symmetry of nonabelian theories, monopoles and
  dyons.
\newblock 2005.

\bibitem{binding}
L.~Gamberg, G.~R. Kalbfleisch, and K.~A. Milton.
\newblock Direct and indirect searches for low mass magnetic monopoles.
\newblock {\em Found. Phys.}, 30:543, 2000.

\bibitem{PDBook}
S.~{Eidelman} et~al.
\newblock {Review of Particle Physics}.
\newblock {\em {Physics Letters B}}, 592:1+, 2004.
\newblock http://pdg.lbl.gov.

\bibitem{Ahlen:1982mx}
S.~P. Ahlen and K.~Kinoshita.
\newblock Calculation of the stopping power of very low velocity magnetic
  monopoles.
\newblock {\em Phys. Rev.}, D26:2347--2363, 1982.

\bibitem{Ahlen:1982rw}
S.~P. Ahlen.
\newblock Monopole energy loss and detector excitation mechanisms.
\newblock In R.~A. Carrigan and W.~P. Trower, editors, {\em Magnetic
  Monopoles}, pages 259--290, New York, 1982. Plenum Press.
\newblock Wingspread 1982.

\bibitem{Luo:2002tm}
W.~Luo.
\newblock {\em Search for magnetic monopoles possibly produced by proton-
  antiproton collisions at the Tevatron collider}.
\newblock PhD thesis, University of Oklahoma, 2002.
\newblock {UMI-30-56946 (AAT 3056946)}.

\bibitem{Osland:1984yu}
P.~Osland and T.~T. Wu.
\newblock Monopole-fermion and dyon-fermion bound states. 1. {G}eneral
  properties and numerical results.
\newblock {\em Nucl. Phys.}, B247:421, 1984.

\bibitem{Osland:1984ys}
P.~Osland and T.~T. Wu.
\newblock Monopole-fermion and dyon-fermion bound states. 2. {W}eakly bound
  states for the lowest angular momentum.
\newblock {\em Nucl. Phys.}, B247:450, 1984.

\bibitem{Osland:1985ma}
P.~Osland and T.~T. Wu.
\newblock Monopole-fermion and dyon-fermion bound states. 3. {M}onopole-fermion
  system with $j = |q| -1/2$ and large $\kappa|q|$.
\newblock {\em Nucl. Phys.}, B256:13, 1985.

\bibitem{Osland:1985kz}
P.~Osland and T.~T. Wu.
\newblock Monopole-fermion and dyon-fermion bound states. 4. {D}yon-fermion
  system with $j = |q| -1/2$ and large $\kappa|q|$.
\newblock {\em Nucl. Phys.}, B256:32, 1985.

\bibitem{Osland:1985qw}
P.~Osland, C.~L. Schultz, and T.~T. Wu.
\newblock Monopole-fermion and dyon-fermion bound states. 5. {W}eakly bound
  states for the monopole-fermion system.
\newblock {\em Nucl. Phys.}, B256:449, 1985.

\bibitem{Osland:1985va}
P.~Osland and T.~T. Wu.
\newblock Monopole-fermion and dyon-fermion bound states. 6. {W}eakly bound
  states for the dyon-fermion system.
\newblock {\em Nucl. Phys.}, B261:687, 1985.

\bibitem{Malkus:1950yc}
W.~van~R. Malkus.
\newblock The interaction of the {Dirac} magnetic monopole with matter.
\newblock {\em Phys. Rev.}, 83:899--905, 1951.

\bibitem{Bracci:1983fe}
L.~Bracci and G.~Fiorentini.
\newblock Interactions of magnetic monopoles with nuclei and atoms: Formation
  of bound states and phenomenological consequences.
\newblock {\em Nucl. Phys.}, B232:236, 1984.

\bibitem{Bracci:1983tq}
L.~Bracci, G.~Fiorentini, and R.~Tripiccione.
\newblock On the energy loss of very slowly moving magnetic monopoles.
\newblock {\em Nucl. Phys.}, B238:167, 1984.

\bibitem{Bracci:1984zx}
L.~Bracci, G.~Fiorentini, G.~Mezzorani, and P.~Quarati.
\newblock Formation of monopole-proton bound states in the hot universe.
\newblock {\em Phys. Lett.}, B143:357, 1984.
\newblock Erratum {\bf B155}, 468 (1985).

\bibitem{Bracci:1984db}
L.~Bracci and G.~Fiorentini.
\newblock On the capture of protons by magnetic monopoles.
\newblock {\em Nucl. Phys.}, B249:519, 1985.

\bibitem{Sivers:1970zm}
D.~W. Sivers.
\newblock Possible binding of a magnetic monopole to a particle with electric
  charge and a magnetic dipole moment.
\newblock {\em Phys. Rev.}, D2:2048--2054, 1970.

\bibitem{Olaussen:1984xb}
K.~Olaussen and R.~Sollie.
\newblock Form-factor effects on nucleus-magnetic monopole binding.
\newblock {\em Nucl. Phys.}, B255:465, 1985.

\bibitem{Olaussen:1983qc}
K.~Olaussen, H.~A. Olsen, P.~Osland, and I.~Overbo.
\newblock Kazama-{Yang} monopole-fermion bound states. 1. {A}nalytic results.
\newblock {\em Nucl. Phys.}, B228:567, 1983.

\bibitem{Olaussen:1983bm}
K.~Olaussen, H.~A. Olsen, I.~Overbo, and P.~Osland.
\newblock Proton capture by magnetic monopoles.
\newblock {\em Phys. Rev. Lett.}, 52:325, 1984.

\bibitem{Walsh:1983bz}
T.~F. Walsh, P.~Weisz, and T.~T. Wu.
\newblock Monopole catalysis of proton decay.
\newblock {\em Nucl. Phys.}, B232:349, 1984.

\bibitem{Olsen:1990jm}
H.~A. Olsen, P.~Osland, and T.~T. Wu.
\newblock On the existence of bound states for a massive spin 1 particle and a
  magnetic monopole.
\newblock {\em Phys. Rev.}, D42:665--689, 1990.

\bibitem{Olsen:1990jn}
H.~A. Olsen and P.~Osland.
\newblock Bound states for a massive spin 1 particle and a magnetic monopole.
\newblock {\em Phys. Rev.}, D42:690--700, 1990.

\bibitem{Goebel:1983xf}
C.~J. Goebel.
\newblock Binding of nuclei to monopoles.
\newblock In J.~L. Stone, editor, {\em {Monopole '83}}, page 333, New York,
  1984. Plenum.
\newblock MAD/TH/146.

\bibitem{goto}
E.~Goto, H.~Kolm, and K.~Ford.
\newblock Search for ferromagnetically trapped magnetic monopoles of cosmic-ray
  origin.
\newblock {\em Phys. Rev.}, 132:387, 1963.

\bibitem{Carrigan:1978ku}
{R. A. Carrigan, Jr.}, B.~P. Strauss, and G.~Giacomelli.
\newblock Search for magnetic monopoles at the {CERN ISR}.
\newblock {\em Phys. Rev.}, D17:1754, 1978.

\bibitem{turnerrev}
M.~Turner.
\newblock Thermal production of superheavy magnetic monopoles in the early
  universe.
\newblock {\em Phys. Lett.}, B115:95, 1982.

\bibitem{price84}
P.~B. Price, S.~l.~Guo, S.~P. Ahlen, and R.~L. Fleischer.
\newblock Search for grand unified theory magnetic monopoles at a flux level
  below the {P}arker limit.
\newblock {\em Phys. Rev. Lett.}, 52:1265, 1984.

\bibitem{cabrera}
B.~Cabrera.
\newblock First results from a superconductive detector for moving magnetic
  monopoles.
\newblock {\em Phys. Rev. Lett.}, 48:1378, 1982.

\bibitem{cabrera2}
B.~Cabrera.
\newblock Upper limit on flux of cosmic ray monopoles obtained with a three
  loop superconductive detector.
\newblock {\em Phys. Rev. Lett.}, 51:1933, 1983.

\bibitem{Jeon:1995rf}
H.~Jeon and M.~J. Longo.
\newblock Search for magnetic monopoles trapped in matter.
\newblock {\em Phys. Rev. Lett.}, 75:1443--1446, 1995.

\bibitem{jeonerr}
H.~Jeon and M.~J. Longo.
\newblock Search for magnetic monopoles trapped in matter.
\newblock {\em Phys. Rev. Lett.}, 76:159, 1996.
\newblock Erratum.

\bibitem{ambrosio}
{M. Ambrosio \etal}.
\newblock Magnetic monopole search with the {MACRO} detector at {G}ran {S}asso.
\newblock {\em Phys. Lett.}, B406:249, 1997.

\bibitem{Price:1975zt}
P.~B. Price, E.~K. Shirk, W.~Z. Osborne, and L.~S. Pinsky.
\newblock Evidence for detection of a moving magnetic monopole.
\newblock {\em Phys. Rev. Lett.}, 35:487--490, 1975.

\bibitem{Alvarez:1975gm}
L.~W. Alvarez.
\newblock Analysis of a reported magnetic monopole.
\newblock In {\em Int. Conf. on Lepton and Photon Interactions}, Stanford,
  1975.

\bibitem{gp}
G.~Giacomelli and L.~Patrizii.
\newblock Magnetic monopole searches.
\newblock 2003.
\newblock hep-ex/0302011.

\bibitem{Giacomelli:2005xz}
G.~Giacomelli and L.~Patrizii.
\newblock Magnetic monopole searches.
\newblock 2005.
\newblock hep-ex/0506014.

\bibitem{Bertani:1990tq}
M.~Bertani et~al.
\newblock Search for magnetic monopoles at the {Tevatron} collider.
\newblock {\em Europhys. Lett.}, 12:613--616, 1990.

\bibitem{Abulencia:2005hb}
A.~Abulencia et~al.
\newblock Direct search for {Dirac} magnetic monopoles in $p \bar p$ collisions
  at $s^{1/2} = 1.96$ {TeV}.
\newblock 2005.

\bibitem{Giacomelli:1984gq}
G.~Giacomelli.
\newblock Magnetic monopoles.
\newblock {\em Riv. Nuovo Cim.}, 7N12:1--111, 1984.

\bibitem{Preskill:1979zi}
J.~Preskill.
\newblock Cosmological production of superheavy magnetic monopoles.
\newblock {\em Phys. Rev. Lett.}, 43:1365, 1979.

\bibitem{Turner:1982ag}
M.~S. Turner, E.~N. Parker, and T.~J. Bogdan.
\newblock Magnetic monopoles and the survival of galactic magnetic fields.
\newblock {\em Phys. Rev.}, D26:1296, 1982.

\bibitem{Parker:1970xv}
E.~N. Parker.
\newblock The origin of magnetic fields.
\newblock {\em Astrophys. J.}, 160:383, 1970.

\bibitem{Rubakov:1981rg}
V.~A. Rubakov.
\newblock Superheavy magnetic monopoles and proton decay.
\newblock {\em JETP Lett.}, 33:644--646, 1981.

\bibitem{Callan:1982au}
{C. G. Callan, Jr.}
\newblock Dyon-fermion dynamics.
\newblock {\em Phys. Rev.}, D26:2058--2068, 1982.

\bibitem{Ambrosio:2002qu}
M.~Ambrosio et~al.
\newblock Search for nucleon decays induced by {GUT} magnetic monopoles with
  the {MACRO} experiment.
\newblock {\em Eur. Phys. J.}, C26:163--172, 2002.

\bibitem{Ambrosio:2002qq}
M.~Ambrosio et~al.
\newblock Final results of magnetic monopole searches with the {MACRO}
  experiment.
\newblock {\em Eur. Phys. J.}, C25:511--522, 2002.

\bibitem{Price:1988ge}
P.~B. Price.
\newblock Limits on contribution of cosmic nuclearites to galactic dark matter.
\newblock {\em Phys. Rev.}, D38:3813--3814, 1988.

\bibitem{Ghosh:1990ki}
D.~Ghosh and S.~Chatterjea.
\newblock Supermassive magnetic monopoles flux from the oldest mica samples.
\newblock {\em Europhys. Lett.}, 12:25--28, 1990.

\bibitem{Lazarides:1986rt}
G.~Lazarides, C.~Panagiotakopoulos, and Q.~Shafi.
\newblock Magnetic monopoles from superstring models.
\newblock {\em Phys. Rev. Lett.}, 58:1707, 1987.

\bibitem{Kephart:2001ix}
T.~W. Kephart and Q.~Shafi.
\newblock Family unification, exotic states and magnetic monopoles.
\newblock {\em Phys. Lett.}, B520:313--316, 2001.

\bibitem{DeRujula:1994nf}
A.~{De Ruj\'ula}.
\newblock Effects of virtual monopoles.
\newblock {\em Nucl. Phys.}, B435:257--276, 1995.

\bibitem{acciarri}
M.~Acciarri \etal.
\newblock Search for anomalous ${Z} \to \gamma \gamma \gamma$ events at {LEP}.
\newblock {\em Phys. Lett.}, B345:609, 1995.

\bibitem{Ginzburg:1982fk}
I.~F. Ginzburg and S.~L. Panfil.
\newblock The possibility of observation of heavy {Dirac-Schwinger} magnetic
  pole.
\newblock {\em Sov. J. Nucl. Phys.}, 36:850, 1982.

\bibitem{Ginzburg:1998vb}
I.~F. Ginzburg and A.~Schiller.
\newblock Search for a heavy magnetic monopole at the {Tevatron} and {LHC}.
\newblock {\em Phys. Rev.}, D57:6599--6603, 1998.

\bibitem{Ginzburg:1999ej}
I.~F. Ginzburg and A.~Schiller.
\newblock The visible effect of a very heavy magnetic monopole at colliders.
\newblock {\em Phys. Rev.}, D60:075016, 1999.

\bibitem{Abbott:1998mw}
B.~Abbott et~al.
\newblock A search for heavy pointlike {Dirac} monopoles.
\newblock {\em Phys. Rev. Lett.}, 81:524--529, 1998.

\bibitem{Graf:1991xe}
S.~Graf, A.~Schaefer, and W.~Greiner.
\newblock Mass limit for {Dirac}-type magnetic monopoles.
\newblock {\em Phys. Lett.}, B262:463--466, 1991.

\bibitem{Heisenberg:1935qt}
W.~Heisenberg and H.~Euler.
\newblock Consequences of {Dirac's} theory of positrons.
\newblock {\em Z. Phys.}, 98:714--732, 1936.

\bibitem{weisskopf36}
V.~Weisskopf.
\newblock {\em Kgl. Danske Videnskab. Selskabs. Mat.-fys. Medd.}, 14(6), 1936.

\bibitem{Schwinger:1951nm}
J.~Schwinger.
\newblock On gauge invariance and vacuum polarization.
\newblock {\em Phys. Rev.}, 82:664--679, 1951.

\bibitem{psf2}
J.~Schwinger.
\newblock {\em Particles, Sources, and Fields}, volume~2.
\newblock Addison-Wesley, New York, 1973.
\newblock republished by Perseus Books, 1998.

\bibitem{Dong:1992hg}
F.-X. Dong, X.-d. Jiang, and X.-j. Zhou.
\newblock Total cross-section for the scattering of photon by photon via the
  ${W}$ loops.
\newblock {\em Phys. Rev.}, D47:5169--5172, 1993.

\bibitem{Dong:1992fa}
F.-x. Dong, X.-d. Jiang, and X.-j. Zhou.
\newblock The ${C}$ coefficients of the amplitudes of ${Z} \to 3\gamma$ and
  $\gamma \gamma \to \gamma \gamma$.
\newblock {\em J. Phys.}, G19:969--978, 1993.

\bibitem{Jikia:1993tc}
G.~Jikia and A.~Tkabladze.
\newblock Photon-photon scattering at the photon linear collider.
\newblock {\em Phys. Lett.}, B323:453--458, 1994.

\bibitem{lp}
E.~Lifshitz and L.~Pitayevski.
\newblock {\em Relativistic Quantum Theory}, volume~2.
\newblock Pergamon, Oxford, 1974.

\bibitem{Bordag:1998sw}
M~Bordag and J~Lindig.
\newblock Radiative correction to the {Casimir} force on a sphere.
\newblock {\em Phys. Rev.}, D58:045003, 1998.

\bibitem{Bordag:1983hk}
M.~Bordag and D.~Robaschik.
\newblock Quantum field theoretic treatment of the {Casimir} effect.
  {F}initeness of the {C}asimir force up to second order of perturbation
  theory.
\newblock {\em Ann. Phys. (N.Y.)}, 165:192, 1985.

\bibitem{ritus}
V.~I. Ritus.
\newblock {\em Zh.\ Eksp.\ Teor.\ Fiz.}, 69:151, 1975.
\newblock [Sov.\ Phys.-JETP 42, 774 (1976)].

\bibitem{Reuter:1996zm}
M.~Reuter, M.~G. Schmidt, and C.~Schubert.
\newblock Constant external fields in gauge theory and the spin 0, 1/2, 1 path
  integrals.
\newblock {\em Ann. Phys. (N.Y.)}, 259:313--365, 1997.

\bibitem{Fliegner:1997ra}
D.~Fliegner, M.~Reuter, M.~G. Schmidt, and C.~Schubert.
\newblock Two-loop {Euler-Heisenberg Lagrangian} in dimensional regularization.
\newblock {\em Theor. Math. Phys.}, 113:{289--300, 1442--1451}, 1997.

\bibitem{Price:1987py}
P.~B. Price, Guo-Xiao Ren, and K.~Kinoshita.
\newblock Search for highly ionizing particles at the {F}ermilab proton
  anti-proton collider.
\newblock {\em Phys. Rev. Lett.}, 59:2523--2526, 1987.

\bibitem{Price:1990in}
P.~B. Price, Gui-Rui Jing, and K.~Kinoshita.
\newblock High luminosity search for highly ionizing particles at the
  {Fermilab} collider.
\newblock {\em Phys. Rev. Lett.}, 65:149--152, 1990.

\bibitem{Ahlen:1978jy}
S.~P. Ahlen.
\newblock Stopping power formula for magnetic monopoles.
\newblock {\em Phys. Rev.}, D17:229--233, 1978.

\bibitem{bargerandollson}
V.~Barger and M.~G. Ollson.
\newblock {\em Classical Electricity and Magnetism}.
\newblock Allyn and Bacon, Boston, 1967.

\bibitem{cousins}
G.~J. Feldman and R.~D. Cousins.
\newblock Unified approach to the classical statistical analysis of small
  signals.
\newblock {\em Phys. Rev. D}, 57:3873, 1998.

\bibitem{Aktas:2004qd}
A.~Aktas et~al.
\newblock A direct search for stable magnetic monopoles produced in positron
  proton collisions at {HERA}.
\newblock {\em Eur. Phys. J.}, C41:133--141, 2005.

\bibitem{mandelstam:1976}
S.~Mandelstam.
\newblock Vortices and quark confinement in non-{Abelian} gauge theories.
\newblock {\em Phys. Rep.}, 23:245, 1976.

\bibitem{Polyakov:1976fu}
A.~M. Polyakov.
\newblock Quark confinement and topology of gauge groups.
\newblock {\em Nucl. Phys.}, B120:429--458, 1977.

\bibitem{'tHooft:1981ht}
G.~'t~Hooft.
\newblock Topology of the gauge condition and new confinement phases in
  nonabelian gauge theories.
\newblock {\em Nucl. Phys.}, B190:455, 1981.

\bibitem{Ahlen:1976jw}
S.~P. Ahlen.
\newblock Monopole track characteristics in plastic detectors.
\newblock {\em Phys. Rev.}, D14:2935--2940, 1976.

\end{thebibliography}
\end{document}